%% file: main.tex
\documentclass[a4paper,11pt]{article}
\pdfoutput=1 

\usepackage{jheppub} 

\usepackage[T1]{fontenc} 
\usepackage{multirow}
\usepackage{graphicx}
\usepackage{appendix}
\usepackage{tikz}
\usepackage[compat=1.1.0]{tikz-feynman}
\usepackage{tikz-3dplot}
\usepackage{axodraw2}
\usepackage{verbatim}
\usepackage{pgfplots}
\usepackage{xspace}
\usepackage{placeins}
\usepackage{cancel}
\usetikzlibrary{calc}
\usetikzlibrary{arrows.meta}
\usetikzlibrary{decorations.markings}
\usetikzlibrary{decorations.pathmorphing}
\usetikzlibrary{shapes.geometric}
\pgfdeclarelayer{nodelayer}
\pgfdeclarelayer{edgelayer}
\pgfsetlayers{main,edgelayer,nodelayer}
\tikzstyle{none}=[]
\input{all.tikzstyles}

\input{declarations}

\title{
    \vspace{-4cm}\hfill {\small  IPPP/26/04 \\ \vspace{-0.3cm} \hfill }\vspace{3cm}\\ 
   Regionally Optimal Contour Deformations with Neural Networks
   }

\author[a]{S. Jones,}
\author[a,1]{D. Ma\^{\i}tre\note{Corresponding author.},}
\author[a]{A. Olsson}

\affiliation[a]{Institute for Particle Physics Phenomenology, Durham University, Durham DH1 3LE, UK}

\emailAdd{stephen.jones@durham.ac.uk}
\emailAdd{daniel.maitre@durham.ac.uk}
\emailAdd{anton.olsson@durham.ac.uk}

\abstract{
In this article, we explore the selection of contour deformations for the numerical evaluation of Feynman integrals after sector decomposition.
In existing codes, the contour of integration is determined heuristically for each individual phase-space point by sampling the integrand.
In this work, we introduce two methods for selecting a valid contour deformation as a function of the kinematic variables in an entire phase-space region using only an initial sampling or training step.
In addition to ensuring the validity of the contour, our method also minimises the variance of the resulting integrand, making it optimal for Monte Carlo integration. We investigate the suitability of these contours when used for Randomised Quasi-Monte Carlo integration and show that although the optimal contour for such integration rules depends on the specific generating vector used, one of our methods typically yields contours with better performance than existing heuristic methods. 
The \textit{a priori} knowledge of the integration path obtained in this work can be used to improve the speed of conventional integration methods or be leveraged for integration using neural networks, where, crucially, it removes the need to retrain the neural network for each phase-space point.
The techniques described in this work can be adapted to other problems where a non-trivial integration path has to be chosen subject to a set of constraints.
}
\begin{document}
\maketitle
\flushbottom
\section{Introduction}

In perturbative approaches to Quantum Field Theories (QFTs) beyond tree-level, the precise definition of the theory requires a prescription for how the contour of integration should be chosen for loop integrals when singularities are encountered along the real axis of integration.
The choice of contour is connected to preserving the causality of the theory, the standard textbook approach is given by the Feynman $i \delta$ prescription.
This prescription introduces a small imaginary part to each propagator of the theory in momentum space
\begin{align}
P_i(q_i,m_i) = \frac{1}{q_i^2 - m_i^2 + i \delta},
\end{align}
which defines the contour of integration. The loop integrals are then evaluated in the limit $\delta \rightarrow 0^+$.

After parameterising the loop integrals, e.g. using Schwinger or Feynman parameters, and performing the integration over the loop momenta, the Feynman $i \delta$ prescription manifests itself as a prescription for the contour of integration for the remaining parameter integrals; see, e.g. Ref.~\cite{Eden:1966dnq}.
In fact, taking the prescription for the contour of integration in parameter space as a starting point can be advantageous for understanding causality and crossing symmetries of the scattering matrix~\cite{Hannesdottir:2022bmo}.
From a practical viewpoint, the presence of singularities on the real axis of the integration parameters poses an additional challenge for evaluating loop integrals.
One strategy is to fix $\delta$ to a positive constant, which alters the integrand, then repeat the integration for different values of $\delta$ in order to extrapolate to $\delta=0$. 
Methods for handling integrals with finite, non-zero $\delta$ and extracting the value in the infinitesimal limit have been employed successfully in various calculations~\cite{deDoncker:2004bf,Yuasa:2011ff,deDoncker:2017gnb,Baglio:2020ini,deDoncker:2024lmk,Bhattacharya:2025egw}.
Alternatively, in automated approaches to the direct integration of Feynman integrals, using a contour deformation which maintains the original integrand, is often the method of choice~\cite{Borowka:2014aaa,Borowka:2015mxa,Borowka:2017idc,Borowka:2018goh,Jahn:2020tpj,Heinrich:2021dbf,Heinrich:2023til,Smirnov:2013eza,Smirnov:2015mct,Smirnov:2021rhf,Borinsky:2023jdv}. 
Using Cauchy's theorem, the integration over the real line can be replaced by an appropriately chosen contour integral in the complex plane. 
The new contour must be picked in accordance with the $i\delta$ prescription without crossing additional poles in the complex plane~\cite{Soper:1998ye,Soper:1999xk,Binoth:2005ff,Nagy:2006xy,Anastasiou:2006hc,Anastasiou:2007qb,Lazopoulos:2007ix,Lazopoulos:2007bv,Anastasiou:2008rm,Gong:2008ww,Becker:2010ng,Mizera:2021icv,Hannesdottir:2022bmo,Borinsky:2023jdv}.

Although the procedure for finding a valid contour deformation, where it exists\footnote{The usual procedure for obtaining a contour can fail for integrals with a leading Landau singularity, see for example Ref~\cite{Gardi:2024axt}.}, is algorithmic, the resulting contours depend on a set of tunable deformation parameters, $\lambda$, (more generally functions) which must be chosen to obtain a valid contour and whose values can impact the variance of the resulting integrand.
The challenge, therefore, is to pick values for these deformation parameters which both result in a valid contour and minimise the variance of the resulting integrand.
In existing implementations of contour deformation, a pre-sampling and/or regression step is used to select valid values of the deformation parameters for each phase-space point (i.e. for each value of the kinematic invariants).
The potential for improving the efficiency of the integration of contour-deformed Feynman integrals was explored in Ref.~\cite{Winterhalder:2021ngy}, where normalising flows were used to improve the variance of the integrand.

In this article, we present a method for obtaining valid values of the deformation parameters for a range of kinematic invariants, which also results in a low variance of the resulting integrand.
This is achieved by training a neural network on the integrand both as a function of the Feynman parameters and the external kinematic invariants.
The advantage of our approach is that the potentially expensive neural network training does not need to be repeated for each choice of kinematic parameters, and instead the trained network can be used to obtain valid, low-variance contours for many phase-space points.
We focus on integrands obtained using sector decomposition~\cite{Binoth:2000ps}, as implemented in a variety of numerical packages~\cite{Borowka:2014aaa,Borowka:2015mxa,Borowka:2017idc,Borowka:2018goh,Jahn:2020tpj,Heinrich:2021dbf,Heinrich:2023til,Smirnov:2013eza,Smirnov:2015mct,Smirnov:2021rhf}.


Machine Learning has been used in other works to improve numerical integration 
\cite{Janssen:2025zke,Bothmann:2020ywa,Gao:2020zvv,Heimel:2022wyj,Gao:2020vdv,Bendavid:2017zhk,Klimek:2018mza,Bishara:2019iwh,Otten:2019hhl,Bothmann:2025lwg,Jinno:2022sbr}, in these works, the integrand is considered fixed, and the sampling is adapted to improve the convergence of the numerical integration. The method described in this article can be applied to select an integration contour (and therefore fix the integrand) before using any of these integration improvement methods.

This article is structured as follows. In Section~\ref{sec:background}, we review the theoretical background. 
It is described how contour deformation is applied to parametric Feynman integrals, and we specify the type of deformations that are investigated. We also provide a brief summary of numerical integration techniques, including Quasi-Monte Carlo and neural network integration. Finally, we also describe how contour deformation is implemented in \pysecdec~\cite{Borowka:2014aaa,Borowka:2015mxa,Borowka:2017idc,Borowka:2018goh,Jahn:2020tpj,Heinrich:2021dbf,Heinrich:2023til}, a publicly available program for the evaluation of dimensionally regulated parameter integrals via sector decomposition, as well as how imaginary parts of individual sector integrals can appear at non-physical values of the kinematics as a result of sector decomposition. 
In Section~\ref{sec:optimal_contour}, we discuss the equivalence of optimal contours between networks trained regionally and networks trained on individual phase-space points. Section~\ref{sec:training} provides the details on the neural network architecture and how the training is performed. In Sections~\ref{sec:bubble1L} and \ref{sec:elliptic2L}, the one-loop bubble and two-loop box integrals are considered as examples. In Section~\ref{sec:discussion}, we discuss the general results we obtained for the two previously mentioned examples. In Section~\ref{sec:conclusion}, we summarise our main conclusions.

\section{Background}\label{sec:background}

We review the necessary background for computing Feynman Integrals in the Minkowski (physical) regime using Feynman parameters.
In Section~\ref{sec:contour}, we present an overview of the problem of selecting a contour of integration for Feynman integrals.
In Section~\ref{sec:montecarlo}, we briefly summarise the key features of Monte Carlo and Randomised Quasi-Monte Carlo integration. 
In Section~\ref{sec:neural_integration}, we describe a method for computing integrals using a neural network to approximate the primitive.
We recap the selection of the integration contour in the public \pysecdec package in Section~\ref{sec:pysecdec}. 
Finally, in Section~\ref{sec:thresholds}, we highlight how the application of sector decomposition can complicate the threshold structure of the integrand.

\subsection{Contour deformation for Feynman integrals}
\label{sec:contour}

A general dimensionally regulated Feynman integral can be written in parametric form,
\begin{equation}
I(\mathbf{s}) = \frac{\left(-1\right)^{\nu}\Gamma\left(\nu-L D/2\right)}{\prod_{i=1}^{N+1}\Gamma\left(\nu_{i}\right)}\lim_{\delta\to0^+}\!\int\limits_{\mathbb{R}_{\geq0}^{N+1}}\prod\limits_{i=1}^{N+1}\mathrm{d}x_{i} x_{i}^{\nu_{i}-1}\frac{\mathcal{U}(\mathbf{x})^{\nu-(L+1)D/2}}{\left(\mathcal{F}(\mathbf{x};\mathbf{s})-i\delta\right)^{\nu-LD/2}}\, \delta(1-\sum\nolimits_i x_i),
    \label{eq:fp2}
\end{equation}
where $\mathcal{U}(\mathbf{x})$ and $\mathcal{F}(\mathbf{x};\mathbf{s})$ are the first and second Symanzik polynomials, they are both functions of the Feynman parameters, with the latter also depending on the kinematic invariants. We use the symbols $\mathbf{s}$ and $\mathbf{x}$ to collect all kinematic invariants and integration parameters, and we note that $\mathbf{x}$ contains $N+1$ elements before integration over the $\delta$ functional and $N$ after its integration, while $\mathbf{s}$ always contains $M$ elements. The number of space-time dimensions is denoted by $D=4-2\epsilon$, $\nu_i$ are the powers of the Feynman propagators and $\nu = \sum\nolimits_i \nu_i$.
The $\delta > 0$ is a prescription for the analytic continuation of special functions, such as logarithms, that appear in the integration result. 
Singularities originating from the integration boundaries cannot be resolved with contour deformation but can be extracted by applying a sector decomposition, which transforms the integral, $I(\mathbf{s})$, into a linear combination of sector integrals, $I_k(\mathbf{s})$, such that\footnote{There are also subtraction terms present that regulate the isolated endpoint divergences. These consist of integrands evaluated at 0 for some subset of the integration variables, and we suppress these terms for ease of notation.},
\begin{equation}
\label{eq:sector_integral}
    I(\mathbf{s}) \to \sum_{k=0} c_{k}\, I_{k}(\mathbf{s}), \quad I_k(\mathbf{s}) = \int_0^1 dx_1\cdots dx_N \, \frac{U_k(\mathbf{x})^a}{\left(F_k(\mathbf{x};\mathbf{s})-i\delta\right)^b},
\end{equation}
where the $U_k(\mathbf{x})$ and $F_k(\mathbf{x};\mathbf{s})$ are the sector decomposed Symanzik polynomials in the $k$-th sector\footnote{In the remainder of this paper, we will often drop the sector indices and consider a generic sector integral.}, and the $\epsilon$-structure has been factorised and absorbed into the coefficients $c_k$, together with all prefactors of Eq.~\eqref{eq:fp2}, such that the sector integrals $I_k(\mathbf{s})$ have only integrable singularities and are suitable for numerical integration. Note that after sector decomposition, the integration domain has been remapped to the unit hypercube $[0,1]^N$. Since the $F_k(\mathbf{x};\mathbf{s})$ can still vanish within the integration domain, a contour deformation is required for numerical integration. A general contour deformation is constructed by shifting each integration variable by a small imaginary part,
\begin{equation}
\label{eq:z_definition}
    x_i \rightarrow z_i = x_i+i\tau_i(\mathbf{x};\mathbf{s})\;,
\end{equation}
where $\tau$ is a real function vanishing at the boundary of the $x$ hypercube.
Provided the integration contour vanishes at the boundaries and does not enclose any poles nor cross any branch cuts, Cauchy's theorem can be applied to express a sector integral as,
\begin{equation}
\label{eq:sector_integral_cd}
    I_k(\mathbf{s}) = \int_C dz_1\cdots dz_N \, \frac{U_k^a(\mathbf{z})}{\left(F_k(\mathbf{z};\mathbf{s})-i\delta\right)^b} = \int_0^1 dx_1\cdots dx_N \,\frac{U_k(\mathbf{z})^a}{\left(F_k(\mathbf{z};\mathbf{s})-i\delta\right)^b }\det\left(\frac{\partial \mathbf{z}}{\partial \mathbf{x}}\right),
\end{equation}
where in the second equality we have remapped the integration variables back from the complex $z_i$ to the real $x_i \in [0,1]$.

\begin{figure}
    \centering
    \includegraphics[scale=0.7]{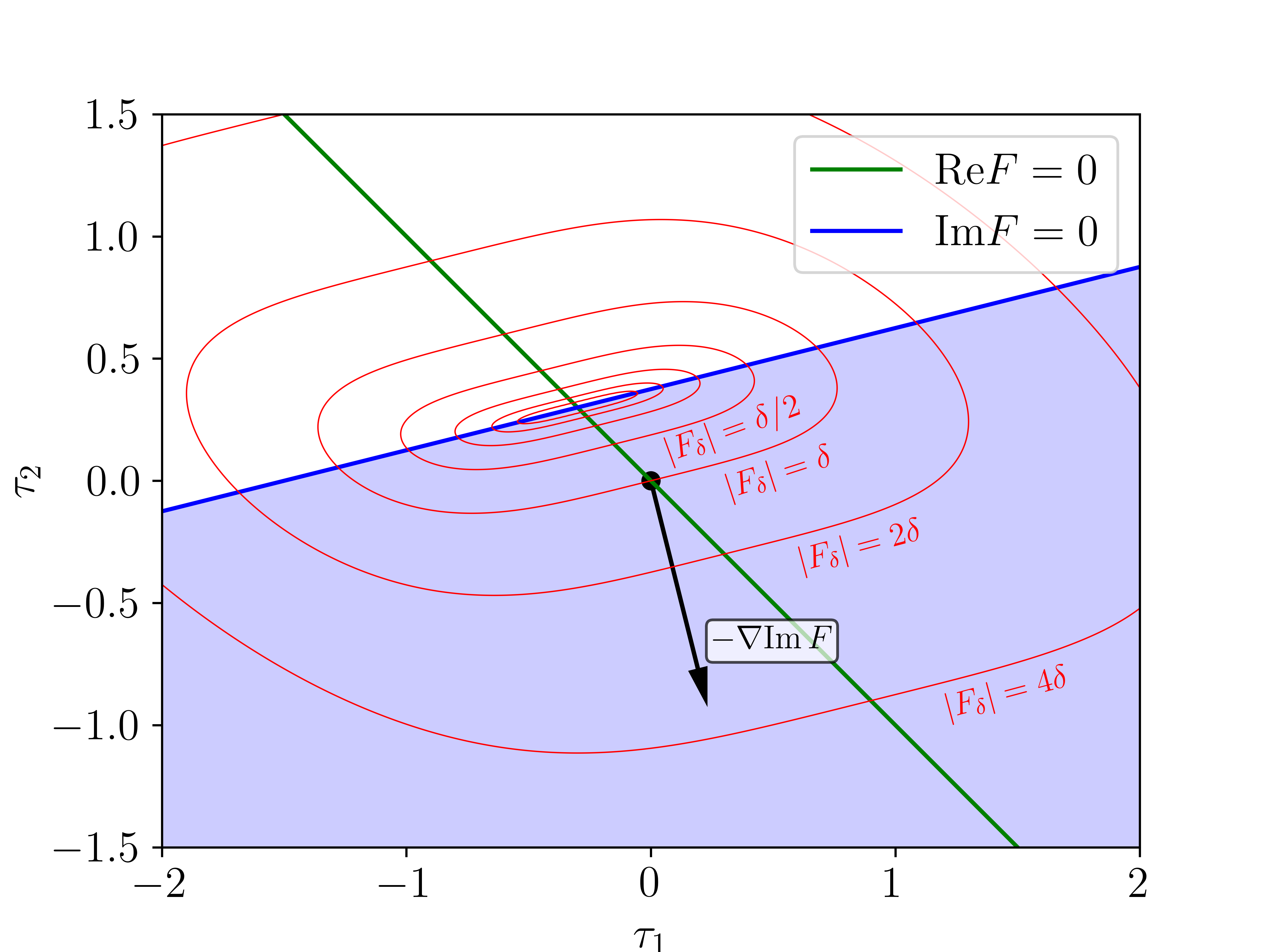}
    \caption{The plane spanned by the imaginary components of $z_1$ and $z_2$ at values of $x_1$ and $x_2$ such that $F(x_1,x_2)=0$. We consider the first sector (red) in Section~\ref{sec:thresholds} with $\delta=3/2$. The green line and the blue line represent where the real and imaginary parts of $F_\delta=0$. At $F_\delta=0$ the integrand has a divergence. The red lines are contours of constant $F_\delta$. With no deformation, the integration path is represented by the black dot. It is on the ${\rm Re} F_\delta=0$ by our choice of location for the real parts $x_1$, $x_2$. As explained in the text, the best direction to deform to avoid the zero of $F_\delta$ for $\delta \rightarrow 0$ is in the inverse direction of the gradient of $F$, shown here with the black arrow.}
    \label{fig:enter-label}
\end{figure}

The sign and magnitude of the deformations $\tau_i$ are important for the numerical efficiency of the integration. Too small a deformation leads to large peaks subject to large cancellations. Too large deformations can lead to the path crossing a pole or a branch cut, which yields the wrong result. 

To develop an intuition for the choice of a good deformation, we examine a two-dimensional example $ F_\mathrm{\delta} = F(x_1, x_2) - i\delta$. 
The integrand has a pole where $F_\delta=F-i\delta=0$. 
We now fix the values of the real part $x_1, x_2$ of $z_1$, $z_2$ to a location where the unregulated, unshifted $F$ polynomial vanishes, $F(x_1,x_2)=0$, and consider the plane spanned by the imaginary parts $\rm{Im}(z_1)=\tau_1$ and $\rm{Im}(z_2)=\tau_2$. 
The undeformed path goes through the $\tau_1=\tau_2=0$ point, at which the regulated $F_\delta$ polynomial has the value $F_\delta=-i\delta$.

When we shift the integration contour, we are moving the location of the intersection of this plane with the integration path from the origin to a different location. To maintain the same value of the integral, we have to do so without hitting a pole. The poles are located at the intersection of the ${\rm Im} F_\delta = 0$ and the ${\rm Re} F_\delta = 0$ surfaces. Since we know the original contour has ${\rm Im} F_\delta = -\delta$ we can guarantee staying away from the poles by requiring that the imaginary part of $F$ stays negative continuously as we deform the path. 
The safest direction to move into is the one that decreases the value of the imaginary part of $F$ the most, that is, following the negative gradient of the imaginary part of $F$. 
If we expand $F$ around $\tau_1=0$ and $\tau_2=0$ we obtain, 
\begin{equation}
    F(z_1,z_2) = F(x_1, x_2) + i\tau_1 \left.\frac{\partial F(z_1,z_2)}{\partial z_1}\right|_{\tau_1,\tau_2=0} +i\tau_2\left.\frac{\partial F(z_1,z_2)}{\partial z_2}\right|_{\tau_1,\tau_2=0} + \mathcal{O}(\tau_1^2,\tau_2^2, \tau_1\tau_2)\;,
\end{equation}
which means that by choosing, 
\begin{equation}
\tau_1\propto -\left.\frac{\partial F(z_1,z_2)}{\partial z_1}\right|_{\tau_1,\tau_2=0} \qquad \mbox{and} \qquad 
\tau_2\propto -\left.\frac{\partial F(z_1,z_2)}{\partial z_2}\right|_{\tau_1,\tau_2=0} , 
\end{equation}
and keeping the deformation small enough, we can avoid moving the contour across a pole.

Note that the deformation is only required in the vicinity of locations where the real part of $F$ vanishes. 
In practice, it is usually easier to define one continuous deformation along the entire integration range, even in regions where there is no risk of approaching a pole (actually introducing a deformation there creates this risk if the deformations are large enough).

We consider deformations of the form
\begin{equation}\label{eq:guided_deformation}
z_i = x_i - i \lambda_i x_i (1-x_i)\left.\frac{\partial F(\mathbf{z})}{\partial z_i}\right|_{\boldsymbol{\tau}=0} .
\end{equation}
The only free parameters are the positive constants $\lambda_i$ that are chosen such that the sign of the imaginary part of the $F$ polynomial does not change. This is the type of deformation which is implemented in \pysecdec, where the $\lambda_i$ are non-negative real numbers.

The first modification we investigate in this article is to promote the constant $\lambda$ to a function of the kinematic parameters $s_i$ and model this function with a neural network. The idea is to fit a function $\lambda(s_1,...,s_M)$ such that the variance of the integral is reduced and the sign of ${\rm Im}(F)$ stays negative.
The function for $\lambda(\mathbf{s})$ is implemented as a neural network (though it could be any other type of interpolating function) and trained to minimise the loss, 
\begin{align}
L= L_{\rm var} + C\, \int\limits d\mathbf{x}\,d\mathbf{s}\;\left[{\rm ReLu} \left({\rm Im}F(\mathbf{z}(\mathbf{x};\mathbf{s});\mathbf{s})\right)\right]^2 ,
\end{align}
where $L_{\rm var}$ is described in Section~\ref{sec:optimal_contour},  and the second term, containing the ${\rm ReLu}$ function, ${\rm ReLu}(x)={\rm max}(x,0)$, penalises choices of $\lambda$ that lead to a wrong sign in the imaginary part of $F$. 
In practice, a large value of $C$ and a safety margin ensure that the condition is imposed strictly.

The second modification we investigate is by introducing more flexibility to the deformation by promoting $\lambda(s_1,...,s_M) \to \lambda(x_1, ..., x_N;s_1,...,s_M) = \lambda(\mathbf{x};\mathbf{s})$. 
The function $\lambda(\mathbf{x};\mathbf{s})$ is much more flexible than the heuristic-lambda scheme $\lambda(\mathbf{s})$, and can negotiate situations where one would like a small $\lambda$ to reduce the variance of the integrand in one part of the integration region, but a large value of $\lambda$ to avoid large values of the integrand close to a pole. 
We refer to this kind of deformation as $\lambda(\mathbf{x};\mathbf{s})$ deformations.

We will use a neural network to find a suitable function $\lambda(x_1, ..., x_N;s_1,...,s_M)$. The Jacobian of the transformation can easily be computed by taking the appropriate derivatives of the neural network. Our implementation could leverage the infrastructure developed in Ref. \cite{Maitre:2022xle} to calculate these derivatives. 

\subsection{Monte Carlo and Randomised Quasi-Monte Carlo integration}
\label{sec:montecarlo}

The contour deformed sector integrals of Eq.~\eqref{eq:sector_integral_cd} can be evaluated numerically with Monte Carlo (MC) methods. In classical MC integration, an estimate is computed as,
\begin{equation}
\label{eq:MC}
    I[f] =\int_0^1 d\mathbf{x} \, f(\mathbf{x})\approx \frac{1}{n} \sum_{i=0}^{n-1} f(x_{1,i},...,x_{N,i}),
\end{equation}
where $f(x_1,...,x_N)$ is the integrand and the $(x_{1,i},...,x_{N,i})$ are $n$ uniformly random sample points. The randomness of the samples allows an unbiased estimate of the variance of the integral $I$ to be obtained in terms of the variance of the integrand $\sigma(f)$,
\begin{equation}
    \sigma^2(I) \approx \frac{\sigma^2(f)}{n} =\frac{I(f^2) - I(f)^2}{n}.
\end{equation}
Examining this equation, we see that for MC integration the standard deviation provides an unbiased estimate of the error of the integral, it scales as $\sigma(I) \propto n^{-1/2}$.

A more performant method of numerically evaluating sector integrals is based on Quasi-Monte Carlo (QMC) integration~\cite{Li:2015foa,Borowka:2018goh}. The idea is to choose the $n$ sample points deterministically from low-discrepancy sequences. These integration rules can provide a 
better scaling of the integration error of the integral for large $n$.

The RQMC method which we will employ in this work is the Rank-1 Shifted Lattice Rule, where the integral estimate is computed as, 
\begin{equation}
\label{eq:QMC}
    I[f] \approx \bar{Q}_{n,m}[f] \equiv \frac{1}{m} \sum_{k=0}^{m-1} Q_n^{(k)}[f], \quad Q_n^{(k)}[f] \equiv \frac{1}{n} \sum_{i=0}^{n-1} f\left(\left\{ \frac{i\mathbf{w}}{n} + \mathbf{\Delta}_k \right\} \right),
\end{equation}
where $\mathbf{\Delta}_k$ are shift vectors, $\mathbf{w} \in \mathbb{Z}^N$ is a generating vector
with elements coprime to $n$, and the curly brackets indicate taking the fractional part of each component of the vector inside. The elements of $\mathbf{\Delta}_k$ are random numbers, which allows for an unbiased error to be estimated as
\begin{equation}
    \sigma^2_{n,m}[f] \approx \frac{1}{m(m-1)} \sum_{k=0}^{m-1} (Q_n^{(k)}[f] - \bar{Q}_{n,m}[f])^2. \label{eq:qmc_variance} 
\end{equation}
In practice, only around $10-20$ random shifts are necessary to obtain a reliable estimate~\cite{Borowka:2018goh}.

An improvement to the QMC quadrature rule in Eq.~\eqref{eq:QMC} is the median lattice construction~\cite{Goda_2022,Heinrich:2023til}. 
It is motivated by the observation that lattices constructed from generating vectors with elements chosen from the set $\{1 \leq w \leq n-1 \, | \, \text{gcd}(w,n)=1\}$ are often good choices, but are known to sometimes lead to so-called unlucky lattices. These are lattices that randomly lead to a much worse error estimate than other lattices of similar size. It has been shown that picking lattices based on $\textit{median}(Q_{n,\mathbf{w}_1}[f], \dots, Q_{n,\mathbf{w}_r}[f])$, where each estimate $Q_{n,\mathbf{w}_i}[f]$ is constructed from a unique generating vector $\mathbf{w}_i$, substantially alleviates this problem, while preserving the convergence rate of the default rule. 
Furthermore, the construction of median lattice rules does not require knowledge of the relevant function space or optimal weights.
In practice, selecting the median lattice based on $\sim 13$ estimates tends to lead to good performance \cite{Heinrich:2023til}. In this paper, we use QMC integration based on median lattice rules to benchmark our choices of contour deformations for sector-decomposed Feynman integrals.

QMC integration works best for integrands that belong to certain weighted function spaces, such as the Korobov function space for periodic functions with smoothness $\alpha$. 
Our sector-decomposed integrands are typically smooth but, in general, not periodic~\cite{Li:2015foa,Borowka:2018goh}. They can be periodized by applying a Korobov transformation. This transformation is defined through a reparametrisation of the integration variables specified through a suitably normalised weight function,
\begin{align}
&\omega = A t^\alpha (1-t)^\beta,&
&\int_0^1 dt\, \omega(t) = 1,&
\end{align}
where $\alpha$ and $\beta$ are parameters of the transformation and $A$ is a normalisation constant.
The precise variable transformation is then given by
\begin{align}
x(t) = \int_0^t dt^\prime\, \omega(t^\prime).
\end{align}
Inserting into the integral definition, we obtain
\begin{equation}
\int_0^1 dx\, f(x) = \int_0^1 dt\, \omega(t) f(x(t)). 
\end{equation}
For the work presented in this article, unless otherwise specified, we use the weight function and transformation obtained by setting $\alpha=\beta=3$, namely,
\begin{equation}
\omega(t) = 140\,t^3(1-t)^3\;, \quad x = t^4 (35 - 84 t + 70 t^2 - 20t^3) 
\end{equation}
for each of the $x_i$ variables. This choice makes the integrand vanish at the $x_i$ boundaries.

\subsection{Numerical integration through a neural network}\label{sec:neural_integration}

Another integration strategy is that explored in Ref.~\cite{Maitre:2022xle}, where the idea is to fit a neural network such that its derivative matches the integrand, so that the undifferentiated neural network is a fit to the primitive of the integrand. This yields a simple evaluation of the integral as a combination of the primitive approximation evaluated at the boundaries of integration. We target contour deformed sector integrals as in Eq.~\eqref{eq:sector_integral_cd}, which we can write in the form,
\begin{equation}\label{eq:integral}
I_{\mu}(s_1,...,s_M) = \int_0^1\, dx_1  \cdots dx_N \;f_{\mu}(x_1,...,x_N;s_1,...,s_M)\;,
\end{equation}
where $\mu$ enumerates components of the integrand, such as real and imaginary parts, or related sector integrands.
We use a neural network
\begin{equation}
\mathcal{N_\mu}(x_1,...,x_N;s_1,...,s_M)
\end{equation}
which takes $N+M$ input parameters and has as one output node for each component of $I_\mu$. 
The neural network is trained to minimise the loss,
\begin{equation}\label{eq:loss}
L(x_1,\ldots,x_N; s_1,\ldots,s_M)=\sum\limits_{\mu}{\rm MSE}\left(f_\mu(x_1,...,x_N;s_1,...,s_M), \frac{d\mathcal{N}_\mu(x_1,...,x_N;s_1,...,s_M)}{dx_1 \cdots dx_N}\right)\,,
\end{equation}
where MSE is the mean squared error. In the case where $f$ is complex, $\mathcal{N}$ has two components, one for the real and one for the imaginary part, and the loss is the mean modulus of the difference between the complex integrand and the complex derivative of the network.   
After training, we obtain our estimate for the value of the integral as,
\begin{equation}\label{eq:boundarysum}
    I_\mu(s_1,...,s_M) \simeq \mathcal{I}_\mu(s_1,...,s_M)=\sum\limits_{b_1,...,b_n=0,1} (-1)^{n-\sum b_i}\mathcal{N}_\mu( b_1,...,b_N;s_1,...,s_M)\,.
\end{equation}
For this method to work for integrands necessitating a contour deformation, the form of the deformation has to be a known function of the kinematic parameters, $s_i$. 
It cannot, for example, be defined as a phase-space-dependent iterative process. The deformations we introduce in this article satisfy this requirement.

\subsection{Selecting a contour deformation in \pysecdec} 
\label{sec:pysecdec}

The integration and contour deformation procedures described in Sections~\ref{sec:contour}--\ref{sec:neural_integration} can be directly applied to finite Feynman integrals.
In practice, we find that the variance of finite integrals (after a change of variables to remove integrable singularities) is usually very high. For this reason, we prefer to consider sector-decomposed integrals. 
As a consequence of this, we can also apply all of these techniques to divergent integrals.

\pysecdec is a program for the numerical integration of dimensionally regulated parameter integrals. 
It implements various sector decomposition routines, which cast Feynman integrals in the form of Eq.~\eqref{eq:sector_integral}. 
It can obtain numerical results in the Minkowski regime by implementing a contour deformation. 
The deformation is defined as in Eq.~\eqref{eq:guided_deformation} and currently the $\lambda_i$ are initialised based on the heuristic choice,
\begin{equation}
\label{eq:pySecDecDeformation}
\lambda_i = \min \left(
    \left| x_i (1-x_i) \frac{\partial F(\mathbf{x})}{\partial x_i}\right|^{-1} \right)
\end{equation}
estimated using a set of presamples \cite{Winterhalder:2021ngy}. If the initial values are such that either the sign is wrong or the contour encloses poles, all $\lambda_i$ are iteratively reduced by a factor of $0.9$ until a valid deformation is found. Additionally, \pysecdec enforces $\lambda_i \in (10^{-6}, 1)$ to avoid numerical instabilities. While this approach works quite well in practice, there are several reasons why this might not lead to the optimal contour:
\begin{itemize}
\item not all $\lambda_i$ need to be reduced if the contour is not valid, reducing it along only one coordinate might be sufficient. 
\item the procedure starts with values of $\lambda_i$ that are close to the largest values thought empirically valid, but it is not always the case that the largest deformations lead to lower variance. Indeed, in some cases, no deformation at all is best, and the above procedure would not necessarily detect it.  
\end{itemize}
Moreover, \pysecdec recycles no knowledge of what good choices for the $\lambda_i$ are between integrations for different phase-space points. The integration wastes time evaluating presamples if the initial values are far away from producing a valid contour. 
The efficiency of \pysecdec would therefore be improved if the presampling step could be replaced by a function that immediately provides a set of valid $\lambda_i$ that integrate efficiently.

With these motivations in mind, we have extended the \textsc{Disteval} integrator in \pysecdec to accept user-defined starting values for the deformation parameters. If the specified values are found to produce an invalid contour, they will be decreased iteratively as described above, which means bad choices still produce correct results. The user-defined $\lambda_i$ are, however, not limited to the range $(10^{-6},1)$.

The feature is implemented as an input argument \texttt{initial\_deformation\_parameters} that accepts a dictionary of tuples of lists, where each entry is on the form \\ \texttt{(<integral\_name>,<sector\_name>):[$\lambda_1, \dots, \lambda_N$]}. The \texttt{<integral\_name>} is a string specified in the generation file and \texttt{<sector\_name>} is a string of the form \texttt{sector\_k\_order\_i} where \texttt{k} is the sector index and \texttt{i} is the order in the $\epsilon$ expansion (\texttt{n} is used to denote negative numbers). The names of all sectors can be found in the \texttt{disteval/<integral\_name>.json} file in the \pysecdec output directory. It is possible to specify deformation parameters for a subset of the sectors. 
In that case, any sectors left out will have the $\lambda_i$ determined by the presampling procedure described above. The feature will be made available in the next public release of \pysecdec, which also includes an example input file for the \texttt{box1L} integral.

\subsection{Thresholds and sector decomposition}
\label{sec:thresholds}

After sector decomposition, the individual sectors of a Feynman integral may develop imaginary parts starting from kinematic values strictly above the physical threshold of the integral.
This behaviour can complicate the optimisation of the contour deformation for a region of phase-space, as, after sector decomposition, the individual sector integrands can develop an imaginary part and/or discontinuities in their derivatives at values of the kinematic invariants that do not correspond to the physical thresholds.
Furthermore, this behaviour can differ for each sector of the integral, meaning that the phase-space region for which a contour is optimal can differ for each sector.

\begin{figure}[h!]
\centering
\begin{tikzpicture}
\input{figs/triangle_fd}
\end{tikzpicture}
\vspace{-0.8cm}
\caption{One-loop massive triangle integral.}
\label{fig:triangle_fd}
\end{figure}

To understand this effect, let us consider the sector decomposition of a one-loop triangle integral with three massive propagators, depicted in Figure~\ref{fig:triangle_fd},
\begin{align}
\mathcal{I}(\mathbf{s}) = 
-\Gamma(1+\epsilon) 
\int_{\mathbb{R}^3_{\ge0}} \mathrm{d} x_1 \, \mathrm{d} x_2 \,\mathrm{d} x_3  \,
\frac{\mathcal{U}(\mathbf{x})^{-1+2\epsilon}}{(\mathcal{F}(\mathbf{x};\mathbf{s})- i \delta )^{1+\epsilon}} \delta(1-x_3).
\label{eq:triangle_integral}
\end{align}
The Symanzik polynomials are given by,
\begin{align}
\mathcal{U}(\mathbf{x}) &= x_1 + x_2 + x_3, \\
\mathcal{F}(\mathbf{x};\mathbf{s}) &= -s x_1 x_3 + m^2 (x_1 + x_2 + x_3)^2.
\end{align}
Here, $m$ is the mass of the internal propagator and $p^2 = s$ is a Mandelstam invariant corresponding to the off-shellness of one external leg.

The physical thresholds of a Feynman integral are encoded in the variety defined by $\mathcal{F}(\mathbf{x};\mathbf{s})=0$ (i.e. the zero-locus of $\mathcal{F}(\mathbf{x};\mathbf{s})$).
Factoring out the squared mass, $m^2$, we can write,
\begin{align}
\tilde{\mathcal{F}}(\mathbf{x};\mathbf{s}) = - \tilde{s} x_1 x_3 + (x_1 + x_2 + x_3)^2,
\end{align}
with $\tilde{s}=s/m^2$.
Integrating over $x_3$ using the delta-functional in Eq.~\eqref{eq:triangle_integral} and solving $\tilde{\mathcal{F}}(\mathbf{x};\mathbf{s})=0$ for $\tilde{s}$ we obtain,
\begin{align}
\tilde{s} = x_1^{-1}(1 + x_1 + x_2)^{2}.
\end{align}
Examining this solution, we find that the minimum non-negative value of $\tilde{s}$ for which we can encounter a zero of $\tilde{\mathcal{F}}(\mathbf{x};\mathbf{s})$ in the non-negative octant of $\mathbf{x}$ is $\tilde{s} = 4$, which is obtained for $x_1 = 1,\, x_2 = 0$.
The value $\tilde{s}=4$ corresponds to the physical threshold of the integral at which the propagating particles with associated Feynman parameters $x_1$ and $x_3$ are on-shell.

\begin{figure}[h!]
\centering
\input{figs/triangle_np}\qquad
\input{figs/triangle_nf}
\caption{Left-panel: The Newton polytope associated with a one-loop massive triangle integral. Right-panel: Regions of the $(x_1,x_2)$-coordinate plane covered by each sector of the decomposition. The solutions of $\tilde{\mathcal{F}}(\mathbf{y};\mathbf{s})=0$ are also shown for $\tilde{s}=4$ and $\tilde{s}=9$ after $x_3$ is eliminated using the delta-functional of Eq.~\eqref{eq:triangle_integral}.}
\label{fig:triangle_np}
\end{figure}

Now, let us sector decompose this integral using the geometric approach~\cite{Bogner:2007cr,Smirnov:2008aw,Kaneko:2009qx,Kaneko:2010kj,Schlenk:2016epj,Heinrich:2021dbf}.
The first step of this algorithm is to re-express the integrand in terms of local variables obtained from the normal vectors of the facets of the Newton polytope of the integrand, depicted in Figure~\ref{fig:triangle_np}.
The change of variables to the facet coordinates is given by,
\begin{align}
&x_1  \rightarrow y_1^{-1} y_3,&
&x_2 \rightarrow y_1^{-1} y_2.&
\end{align}
Inserting this into $\tilde{\mathcal{F}}(\mathbf{x};\mathbf{s})$ after integrating over $x_3$ using the delta-functional, we obtain,
\begin{align}
\tilde{\mathcal{F}}(\mathbf{y};\mathbf{s}) = y_1^{-2} \left( -\tilde{s} y_1 y_3 + (y_1 + y_2 + y_3)^2 \right). 
\end{align}
The domain of integration is now the unit hypercube.
The minimum value of $\tilde{s}$ for which we can encounter a solution of $\tilde{\mathcal{F}}(\mathbf{y};\mathbf{s})=0$ is again $\tilde{s}=4$, which is obtained for $y_1 = 1,\, y_2 = 0,\, y_3 = 1$ (which corresponds to $x_1 = 1,\, x_2=0$).
Again, this corresponds to the physical threshold of the Feynman integral.

In Figure~\ref{fig:triangle_np}, we depict the regions of the $(x_1,x_2)$-coordinate plane covered by each sector of the sector decomposition.
The three sectors that form a sector decomposition of the integral are obtained by setting $y_1 = 1$ (red), $y_2 = 1$ (yellow) and $y_3 = 1$ (green), respectively.
The coordinate $x_1 = x_2 = 1$ is the point at which the integral is divided by the geometric sector decomposition.
In the first sector, obtained by setting $y_1 = 1$, the variety of $\tilde{\mathcal{F}}(\mathbf{y};\mathbf{s})$ enters the sector at the point $y_2 =0,\, y_3 = 1$ (which corresponds to $x_1=1,\, x_2=0$) with $\tilde{s}=4$.
In the third sector, obtained by setting $y_3=1$, the variety enters at the point $y_1 = 1,\, y_2=0$ (which again corresponds to $x_1=1,\, x_2=0$) with $\tilde{s}=4$.
However, for $\tilde{s}=4$ the variety of $\tilde{\mathcal{F}}(\mathbf{y};\mathbf{s})$ does not enter the second sector, obtained by setting $y_2=1$.
Instead, we must have  $\tilde{s} \ge 9$ before an imaginary part is possible in this sector.
In summary, the integral has an imaginary part starting at $\tilde{s}=4$. Sectors I and III contribute to the imaginary part starting at $\tilde{s}=4$, whereas Sector II contributes to the imaginary part only starting from $\tilde{s}=9$.
In Figure~\ref{fig:triangle_np}, we also depict the solutions $\tilde{\mathcal{F}}(\mathbf{x};4)=0$ and $\tilde{\mathcal{F}}(\mathbf{x};9)=0$.
We can observe that the arbitrary choice of the sector decomposition point as $x_1 = x_2 = 1$ has determined the value of $\tilde{s}$ for which the imaginary part first appears in the second sector.

In the triangle example considered, we can avoid distorting the threshold behaviour after sector decomposition by choosing a different hyper-surface in the delta-functional of Eq.~\eqref{eq:triangle_integral} and/or by selecting a different sector decomposition point, which can be realised by rescaling the Feynman parameters by positive factors.
However, public codes which automatically derive a sector decomposition of dimensionally regulated parameter integrals do not currently employ techniques to avoid distorting the threshold behaviour of integrals.

In the remainder of this work, we will assume that sector decomposition is applied as in the current public implementations. 
We must therefore devise a strategy for systematically finding the potentially unphysical thresholds in each sector.
This can be achieved using constrained numerical optimisation and/or gradient descent routines.
For example, suppose we wish to find a positive $\tilde{s}$ threshold in the $k$-th sector.
First, we solve $F_k(\mathbf{y};\mathbf{s})=0$ for $\tilde{s}$, let us label this solution $f_k(\mathbf{y};\mathbf{s}\neq\tilde{s})$.
Next, we may introduce a loss function,
\begin{align}
L(\mathbf{y};\mathbf{s}) = |f_k(\mathbf{y};\mathbf{s}\neq\tilde{s})| + C\, \theta\left(-f_k(\mathbf{y};\mathbf{s}\neq\tilde{s})\right),
\end{align}
where $C$ is a large positive constant that imposes the correct sign of $\tilde{s}$ and $\theta$ is the Heaviside function.
We may then minimise this loss function subject to the constraint that $\mathbf{y}\in[0,1]^N$, i.e. each variable remains in the unit hypercube. We will apply this procedure to an example integral in Section~\ref{sec:elliptic2L}.

The presence of an additional threshold can be detected by monitoring the following quantity during training:
\begin{align}
T=\frac{1}{i(\mathbf{x};\mathbf{s})} + \sum\limits_{i=1}^{N}x_i(1-x_i)\frac{\partial F}{\partial x_i}\;.
\end{align}
When this quantity gets small, it means that both a) the integrand $i(\mathbf{x};\mathbf{s})$ diverges and b) the contour deformation vanishes and cannot therefore regulate the integral. As long as $T=0$ is avoided away from the boundary, the training can proceed in a region. Since the optimal deformation parameter is generally not smooth over the threshold at which a sector start developing an imaginary part (see for example Figures~\ref{fig:lambda_comparison_A}~and~\ref{fig:lambda_comparison_B} ), better results are obtained by separating the training regions at this threshold, even if it is not strictly required. This is the approach we have taken in this work. We show an example of this numerical detection method in Appendix~\ref{sec:threshold_finder}.

\section{Optimal contours}

In heuristics-based approaches, the deformation parameters are chosen according to a pre-defined rule (see Eq.~\eqref{eq:pySecDecDeformation}) where the main goal is to produce a valid contour that is numerically stable at most phase-space points.
In this article, we explore defining the contour in a more global way that allows us to optimise the deformation beyond mere validity. The most natural improvement of the integrand is to reduce its variance. This improves the integration error when using MC methods, is straightforward to implement as a loss function and allows for the relationship between local and regional optimality described in Section~\ref{sec:optimal_contour} below.

\subsection{Variance-based optimal contour equivalence }\label{sec:optimal_contour}

In this section, we consider how the variance of the regionally determined contour compares with that obtained by optimising on a phase-space point-by-phase-space point basis. Let us consider the integrand,
\begin{equation}
f_\lambda(x_1,...,x_N; s_1,...,s_M),
\end{equation}
where $\lambda$ parametrises the contour deformation. The $x$ parameters are the ones to integrate over, and the $s$ parameters are the ``kinematic'' parameters. 
For simplicity, we take the range of $\mathbf{x} = (x_1, \ldots, x_N)$ and $\mathbf{s} = (s_1, \ldots, s_M)$ to be in the unit hypercube, with a mapping of the unit hypercube $\mathbf{s}$ variables to the physical values between their minimum and maximum.  

In a local integration strategy, for each set of parameters $\mathbf{s}$, the best path can be found by minimising the standard deviation of the integrand,
\begin{equation}
\sigma^2_{loc}(\mathbf{s},\lambda) = \int d \mathbf{x}\, \left( f_\lambda(\mathbf{x};\mathbf{s}) - \hat{f}(\mathbf{s}) \right)^2,
\end{equation}
with respect to the path,
\begin{equation}
\lambda_{loc} = \arg \min\limits_{\lambda}  \sigma^2_{loc}(\mathbf{s},\lambda),
\end{equation}
with 
\begin{equation}
    \hat{f}(\mathbf{s}) = \int d \mathbf{x}\, f_\lambda(\mathbf{x};\mathbf{s}).
\end{equation}

To regionally determine the contour, we minimise the standard deviation of the integrand over the full physical parameter range of $\mathbf{s}$ instead of fixing its value and optimising at each phase-space point. The function to minimise is the standard deviation,
\begin{equation}
\sigma^2_{glob}(\mathbf{s}) =  \int d\mathbf{s} \int d\mathbf{x}\, \left( f_{\lambda(\mathbf{s})}(\mathbf{x};\mathbf{s}) - \bar{f} \right)^2,
\end{equation}
where,
\begin{equation}
    \bar{f} = \int d\mathbf{s} \int d\mathbf{x}\, f_\lambda(\mathbf{x};\mathbf{s}),
\end{equation}
is the average of the integral over the physical phase-space, $\mathbf{s}$. 
By definition of the contour deformation, $\bar{f}$ is independent of the choice of contour, while $\sigma^2_{glob}$ can change.

It is instructive to decompose the standard deviation above,
\begin{eqnarray}\label{eq:var}
\int d\mathbf{x} \int d\mathbf{s} \left( f_\lambda(\mathbf{x};\mathbf{s}) - \bar{f} \right)^2 &=& \int d\mathbf{x} \int d\mathbf{s}\, \left( f_\lambda(\mathbf{x};\mathbf{s}) - \hat{f}(\mathbf{s}) + \hat{f}(\mathbf{s}) - \bar{f} \right)^2 \nonumber\\
&=& \int d\mathbf{x} \int d\mathbf{s}\, \left( f_\lambda(\mathbf{x};\mathbf{s}) - \hat{f}(\mathbf{s})\right)^2  + \int d\mathbf{x} \int d\mathbf{s} \left(\hat{f}(\mathbf{s}) - \bar{f} \right)^2\nonumber\\
&&+2  \int d\mathbf{x} \int d\mathbf{s}\, \left( f_\lambda(\mathbf{x};\mathbf{s}) - \hat{f}(\mathbf{s})\right) \left(\hat{f}(\mathbf{s}) - \bar{f} \right)\nonumber\\
&=& \int d\mathbf{s}\; \sigma^2_{loc}(\mathbf{s}) + \int d\mathbf{s} \left(\hat{f}(\mathbf{s}) - \bar{f} \right)^2,
\end{eqnarray}
where we recognised in the first term the minimisation objective of the local contour optimisation and used the definition of $\hat{f}(\mathbf{s})$ to show that the mixed term vanishes.
The minimum of the regional objective is therefore not necessarily consistent with the minimum of individual local optimisations because of the influence of the second term.

If $\hat{f}(\mathbf{s})$ was known we could redefine,
\begin{equation}
f\rightarrow g(\mathbf{x},\mathbf{s})\equiv f(\mathbf{x},\mathbf{s})-\hat{f}(\mathbf{s}), 
\end{equation}
giving,
\begin{equation}
\hat {g}(\mathbf{s}) =  \int d\mathbf{x}\, g(\mathbf{x};\mathbf{s}) =   \int d\mathbf{x}\, f(\mathbf{x},\mathbf{s})-\hat{f}(\mathbf{s}) =0,
\end{equation}
and 
\begin{equation}
\bar {g} =  \int d\mathbf{s}\int d\mathbf{x}\, g(\mathbf{x};\mathbf{s}) =   \int d\mathbf{s} \int d\mathbf{x}\,  f(\mathbf{x};\mathbf{s})-\hat{f}(\mathbf{s}) =0.
\end{equation}

With this modification, the second term in Eq.~\eqref{eq:var} vanishes, while the first remains unchanged as,
\begin{equation}
\sigma^2_{loc}(\mathbf{s}) \rightarrow \int d\mathbf{x}\, \left( g_\lambda(\mathbf{x};\mathbf{s}) - \hat{g}(\mathbf{s}) \right)^2 = \int d\mathbf{x} \left( f_\lambda(\mathbf{x};\mathbf{s}) - \hat{f}(\mathbf{s}) - 0 \right)^2 = \sigma^2_{loc}(\mathbf{s}).
\end{equation}
After modification, the optimal contours in the local and regional approaches would be the same. Of course, the value $\hat{f}(\mathbf{s})$ is the value of the integration and is not known, but an approximation for it would lead to optimal contours for the regional method being close to the optimal paths for the local method. 

In practice, we augment our training with a network $m(\mathbf{s})$ that attempts to emulate $\hat{f}(\mathbf{s})$. 
We consider the loss,
\begin{equation}
L = \int d\mathbf{s}\,d\mathbf{x}\left( f_\lambda(\mathbf{x};\mathbf{s}) - \bar{f} - (m(\mathbf{s}) - \overline{m})\right)^2
\end{equation}
where $\overline{m}$ is the average of $m(\mathbf{s})$ and both the path $\lambda$ and the approximation of the average $m(\mathbf{s})$ have neural network parameters to be optimised.

Note that there is a redundancy in the definition of $m(\mathbf{s})$: any constant shift is cancelled in the average subtraction. To prevent $m(\mathbf{s})$ from drifting to large values and induce numerical noise, it is useful to add a term in the loss that pins its average to the average of $f$, such that $m(\mathbf{s})$ will be optimised to be the integral of $f$ for a fixed $\mathbf{s}$ value. After this last modification, we obtain our loss function,
\begin{equation}
L_{\rm var} =(\bar{f}-\overline{m})^2 +  \int d\mathbf{x}\, d\mathbf{s}\,  \left( f_\lambda(\mathbf{x};\mathbf{s}) - \bar{f} - (m(\mathbf{s}) - \overline{m})\right)^2 \;.
\end{equation}\label{eq:lvar}
The relative weight between the two terms could be tuned, but given that the network can easily adjust the second term to vanish by an appropriate setting of the bias in its last layer, it is not an important consideration in practice. 

\subsection{Optimality for RQMC integration}

In many practical cases, the convergence rate of MC integration is not sufficient, and RQMC methods are used instead. It would be convenient to find an optimal contour for RQMC integration, but unfortunately, we find that the optimal contour displays too large a dependence on the details of the grid used for a generically optimal contour to be determined for any RQMC integration, see Appendix~\ref{sec:minlambda}. 

In the result section, we show that, in the absence of a generic optimal contour for RQMC integration, using the variance-based optimal contour can yield good performance. It is known that this strategy will not yield the true minimum of the RQMC variance~\cite{Dick_Kuo_Sloan_2013}. However, it is aligned with the heuristics already implemented in public codes~\cite{Smirnov:2021rhf,Heinrich:2023til} and used elsewhere in the literature on obtaining improved integration contours for Feynman integrals~\cite{Winterhalder:2021ngy}.
In the remainder of this section, we describe the difficulties of optimising contours for QMC integration in a generic way.  

First, let us briefly recall the relevant properties of QMC integration, for a more detailed review, see Ref.~\cite{Dick_Kuo_Sloan_2013}.
The worst-case error for a given normed function space $H$ and point set $P_{n}$ is given by,
\begin{align}
e_{n,N} (P; H) = \sup_{||f||_H \le 1} \left|\, I[f] - Q_n[f] \,\right|,
\end{align}
where $Q_n$ is the QMC rule.
For any function $f \in H$, the integration error is bounded by the product of the worst-case error and the norm of the function in the appropriate function space,
\begin{align}
\left|\, I[f] - Q_n[f]\, \right| \le e_{n,N}(P;H)\ ||f||_H. \label{eq:qmc_worst_case_error}
\end{align}
For example, if $H$ is a Korobov function space with smoothness parameter\footnote{The smoothness parameter $\alpha$ is related to the rate of decay of the Fourier coefficients of $f$, for integer $\alpha$ it corresponds to the number of square-integrable derivatives of $f$.} $\alpha$, the worst-case error is bounded, for component-by-component lattice rules constructed with weights $\gamma_u$, which define the relative importance of each direction, by~\cite{Dick_Kuo_Sloan_2013},
\begin{align}
e_{n,N}^2(P;H) \le \left( \frac{1}{\varphi(n)} \sum_{\emptyset \neq u \subseteq \{1,\ldots,N\}} \gamma_u^\lambda ( 2 \zeta(2\alpha\lambda))^{|u|} \right)^{1/\lambda}
\end{align}
for all $\lambda \in (1/(2\alpha),1]$. Here $\zeta$ is the Riemann zeta function and $\varphi(n)$ is the Euler totient function.
Therefore, the worst-case error of the integral of any $f\in H$ scales arbitrarily close to $e_{n,N} \propto n^{-\alpha}$.
The appropriate norm for this space is given by,
\begin{align}
||f||_H = \left( \sum_{\mathbf{k} \in \mathbb{Z}^N} w_\alpha(\gamma) | \hat{f}(\mathbf{k})|^2 \right)^{1/2}, \label{eq:qmc_korobov_norm}
\end{align}
where $\hat{f}(\mathbf{k})$ is the $\mathbf{k}$-th Fourier coefficient of $f$ and $w_\alpha(\gamma) \ge 1$ is a function of weights, $\gamma$. 

Although the above analysis is useful for determining QMC point sets with good properties, using Eq.~\eqref{eq:qmc_worst_case_error} and Eq.~\eqref{eq:qmc_korobov_norm} to directly optimise the properties of $f$ can be inefficient in practice~\cite{Dick_Kuo_Sloan_2013,Goda_2022}.
The evaluation of the norm of the function, using for example Eq.~\eqref{eq:qmc_korobov_norm}, is often prohibitively expensive.
Furthermore, the worst-case error bound, Eq.~\eqref{eq:qmc_worst_case_error} provides only a biased upper bound estimate of the error, whereas an unbiased estimator would be preferable.
The worst-case error bound is also observed in practice not to be a tight error bound for a specific function, $f$, in a given function space, $H$.
This means that it can very much overestimate the true error.
Relatedly, it is not necessarily trivial to establish which function space a particular integrand belongs to and the best possible choice of weights.
A detailed analysis of the possible function spaces which can be encountered for the parameter integrals resulting from the sector decomposition and series expansion of arbitrary dimensionally regulated integrals and the optimal choice of weights has not, to our knowledge, yet appeared in literature.
This means that the correct choice of function norm $||\, f\, ||_H$ is not currently known to us in all cases.
Nevertheless, we do periodize functions and avoid introducing additional discontinuities in their derivatives in an attempt to obtain a higher smoothness and/or a better scaling behaviour. 
We will see in the examples below that this strategy is successful.

For the above reasons, considerable effort has been dedicated in the mathematical literature to the understanding of Randomised QMC (RQMC) methods.
These methods yield an unbiased estimator of the error, which is practical to obtain and typically 
smaller than the worst-case error bound.

One of the goals of this work is to investigate how the unbiased error estimator, $\sigma_{n,m}^2[f]$, 
is affected by the choice of contour used to define the integrand $f$.
The most direct approach to minimising Eq.~\eqref{eq:qmc_variance} is to evaluate $Q_n^{(k)}[f]$ using different integration contours (deformations of $f$). 
However, this is impractical as it requires evaluating $(m\times n)$ samples for each possible contour.
Another approach to minimise the variance of Eq.~\eqref{eq:qmc_variance} is to ensure that the results of each individual QMC shift, $Q_n^{(k)}[f]$, are as similar as possible theoretically, suggesting that we should attempt to minimise Eq.~\eqref{eq:qmc_worst_case_error} by minimising $||\, f\, ||_H$.
Unfortunately, we currently do not know of a practical method of achieving this, especially without knowledge of $H$.

Instead, we could turn to the classical QMC literature and consider other bounds on $| I[f] - Q_n[f]|$ that do not rely on knowledge of the function space, $H$, and which may be more practical to compute. 
The classical Koksma-Hlawka inequality states that for some point set $P=\{\mathbf{x}_i\}$ for $i=0,\ldots,n-1$,
\begin{align}
\left| \frac{1}{n} \sum_{i=0}^{n-1} f(\mathbf{x}_i) - \int_0^1 d \mathbf{x}\  f(\mathbf{x}) \right| \le D_n^*(P)\ V[f],
\end{align}
where $D_n^*(P)$ is the star discrepancy of the point set and $V[f]$ is the Hardy-Krause variation of $f$.
If $f$ has continuous mixed first partial derivatives, the variation is given by,
\begin{align}
V[f] = \sum_{\emptyset \neq u \subseteq \{1,\ldots,N\}} \int_0^1 d \mathbf{x} \left| \frac{\partial^{|u|}}{\partial \mathbf{x}_u} f(\mathbf{x}_u; \mathbf{1})\right|,
\end{align}
where $\mathbf{x}_u$ denotes the components $x_i$ of $\mathbf{x}$ with $i \in u$ and $(\mathbf{x}_u;\mathbf{1})$ indicates that the components of $\mathbf{x}$ for $j \notin u$ are replaced by 1.
Unfortunately, this classical result on the integration error suffers from many of the same drawbacks as the QMC worst-case error bound.
In particular, it is computationally expensive to compute and it provides only a weak (or very weak\footnote{For some discontinuous functions $V[f]=\infty$, yet, QMC estimates converge at a rate of at least $n^{-1/2}$.}) bound on the integral error.
We observe that for the examples we study in this work, the Koksma-Hlawka inequality bound can be an order of magnitude or more larger than the error estimate obtained using Eq.~\eqref{eq:qmc_variance}.

In this work, our approach will therefore be to determine a contour that minimises the variance.
This will provide contours that are provably optimal for Monte Carlo integration.
Additionally, we will then test these contours using RQMC integration and discuss their performance relative to the heuristic approach currently employed in public codes.
We note that if an efficient measure of the optimality of a contour in the context of RQMC, satisfying that a) it can be calculated tractably on a phase-space sample and b) that its optimisation on the $(x-s)$ space is equivalent to optimisation on individual $s$ values, see Section~\ref{sec:optimal_contour}, is devised in future work, then our proposed procedure for determining the contour as a function of the kinematic parameters could still be employed, with the loss function Eq.~\ref{eq:lvar} suitably adapted.

\section{Network design and training}\label{sec:training}

In this section, we describe the design and training of the neural networks we use to define the contour deformations. The focus of this article is to demonstrate that such an approach can work and, in addition to providing a deformation \emph{a priori}, yields improvements in the variance of the integrand. No extensive hyperparameter scan has been performed to optimise the learning rate, network size, activation function or the optimiser parameters. The networks are trained in separate regions of phase-space delimited by the thresholds of the sector decomposed integrands, as discussed in Section~\ref{sec:thresholds}, in order to avoid the appearance of divergences in the training data. In addition, the training may be further split into smaller regions to improve the training performance.

\subsection{$\lambda(\mathbf{s})$ deformation}

When the network is initialised, the deformation it represents might not represent a valid contour. To avoid this issue, we start the training with a large value of $\delta$ in Eq.~\eqref{eq:fp2}, reducing it gradually to ensure the validity of the contours throughout the training. 

This addresses the scale disparity issue on the input side. 
On the output side, the optimal values of $\lambda$ can also span several orders of magnitude but must remain positive. 
To address this, we use the neural network, $\mathcal{N}_\lambda$, to predict the logarithm of $\lambda$,
\begin{equation}\label{eq:guided_nn}
\lambda(s_1,...,s_M) = \exp\left( \mathcal{N}_{\lambda}(s_1,\ldots,s_M)\right),
\end{equation}
ensuring both positivity and a suitable scale coverage.

The approximation $m(\mathbf{s})$ of $\hat f(\mathbf{s})$ used to encourage the network to learn the locally best value of $\lambda$ is also allowed to span several orders of magnitude by fitting the modulus and phase of the approximation,
\begin{equation}\label{eq:ftilde}
m(s_1,...,s_M) = \exp\left( \mathcal{M}_{r}(s_1,...,s_M)\right) \exp\left(i\mathcal{M}_{\phi}(s_1,...,s_M)\right).
\end{equation}
The functions $\mathcal{N}_\lambda$, $\mathcal{M}_r$ and $\mathcal{M}_\phi$ are fully connected neural networks with $\tanh$ activation functions. We use 4 layers of 40 units each. 

The training is organised in two phases. The loss, given in Eq.~\eqref{eq:loss} is the same in both phases. In the first phase, the value of $\delta$ is progressively reduced by a constant factor, but the learning rate is kept constant. Once the regulator $\delta$ has reached a small value, the second phase starts, where the learning rate is reduced by a factor when the validation loss reaches a plateau. We do not reduce the learning rate in the first phase because the loss is expected to increase as the regulator gets closer to 0, this is not a sign that the learning has stalled but an expected behaviour. When the integrand over the phase space considered for training spans several orders of magnitude, it is useful to normalise the variance by the approximation of the integral $m(\mathbf{s})$, so that the loss approaches the average of the relative variance. This modification is only effective once the approximation $m(\mathbf{s})$ has been learned for the true integrand, i.e. when the regulator has converged to 0. With this modification, the loss takes the form
\begin{align}
L =\frac{L_{\rm var}}{|m(\mathbf{s})|} + C \,\left[{\rm ReLu}\left({\rm Im}F(\mathbf{z};\mathbf{s}) \right)\right]^2\;.
\end{align}
We run 3 independent trainings and keep the network which reached the lowest loss at the end of the training. 

\subsection{$\lambda(\mathbf{x};\mathbf{s})$ deformation}

For the $\lambda(\mathbf{x};\mathbf{s})$ deformations, we use a 3-layer network with 40 GELU\footnote{The activation function for the GELU (Gaussian Error Linear Units) is $\mathrm{GELU(}x) = x \,\Phi(x)$, where $\Phi(x)$ is the Cumulative Distribution Function for the Gaussian Distribution.}~\cite{hendrycks2023gaussianerrorlinearunits} units for each integration dimension. 
We chose this activation function because we need it to be continuously differentiable for the computation of the Jacobian. These derivatives are computed through the chain rule, as outlined in Ref.~\cite{Maitre:2022xle}. For the simplest example, we use the same two phases as described for the $\lambda(\mathbf{s})$ deformation. 

\section{One-loop bubble example}\label{sec:bubble1L}

\begin{figure}[h!]
\centering
\begin{tikzpicture}
\input{figs/bubble}
\end{tikzpicture}
\vspace{-0.8cm}
\caption{One-loop equal-mass bubble integral.}
\label{fig:bubble}
\end{figure}
The first example we consider is that of a scalar one-loop equal-mass bubble, see Figure~\ref{fig:bubble}. The integral is given by,
\begin{align}\label{eq:bubble}
I(\mathbf{s}) &= \Gamma\left(\epsilon\right) \int_{\mathbb{R}^2_{\geq0}}\!\!\mathrm{d}x_1\mathrm{d}x_2 \, \frac{(x_1+x_2)^{-2+2\epsilon} \, \delta\left(1-x_1-x_2\right)}{\left(-sx_1x_2+m^2\left(x_1+x_2\right)^2-i\delta\right)^{\epsilon}} = \Gamma\left(\epsilon\right) \int_0^1\;\mathrm{d}x \frac{1}{F(x,s)^{\epsilon}} \\ &= \frac{1}{\epsilon} - \gamma_{\mathrm{E}} -\int_0^1\;\mathrm{d}x \log\left(F(x,s)\right) + \mathcal{O}(\epsilon), \qquad F(x,s)= -x(1-x) s + m^2 -i\delta,
\end{align}
where $s=p^2$ is the invariant mass squared of the incoming and outgoing momenta and $m$ is the mass of the internal propagators. For $m>0$ the integral is free of endpoint singularities and does not need a sector decomposition. 
We can factor out an overall $m^2$ and consider the integral as a function of $\tilde{s}=s/m^2$. 
\begin{align}
I(\mathbf{s}) &= \frac{1}{\epsilon} - \gamma_{\mathrm{E}} -\log(m^2) - \int\limits_0^1\;\mathrm{d}x \log\left(\tilde{F}(x,\tilde{s})\right) + \mathcal{O}(\epsilon), \quad \tilde{F}(x,\tilde{s})= -x(1-x) \tilde{s} + 1 -i\delta,
\end{align}
where the last term is the numerical integration we will consider.
For $\tilde{s} > 4$, there are two singularities in the integration region, and a contour deformation is necessary. 

\begin{figure}
\centering
\includegraphics[scale=0.45]{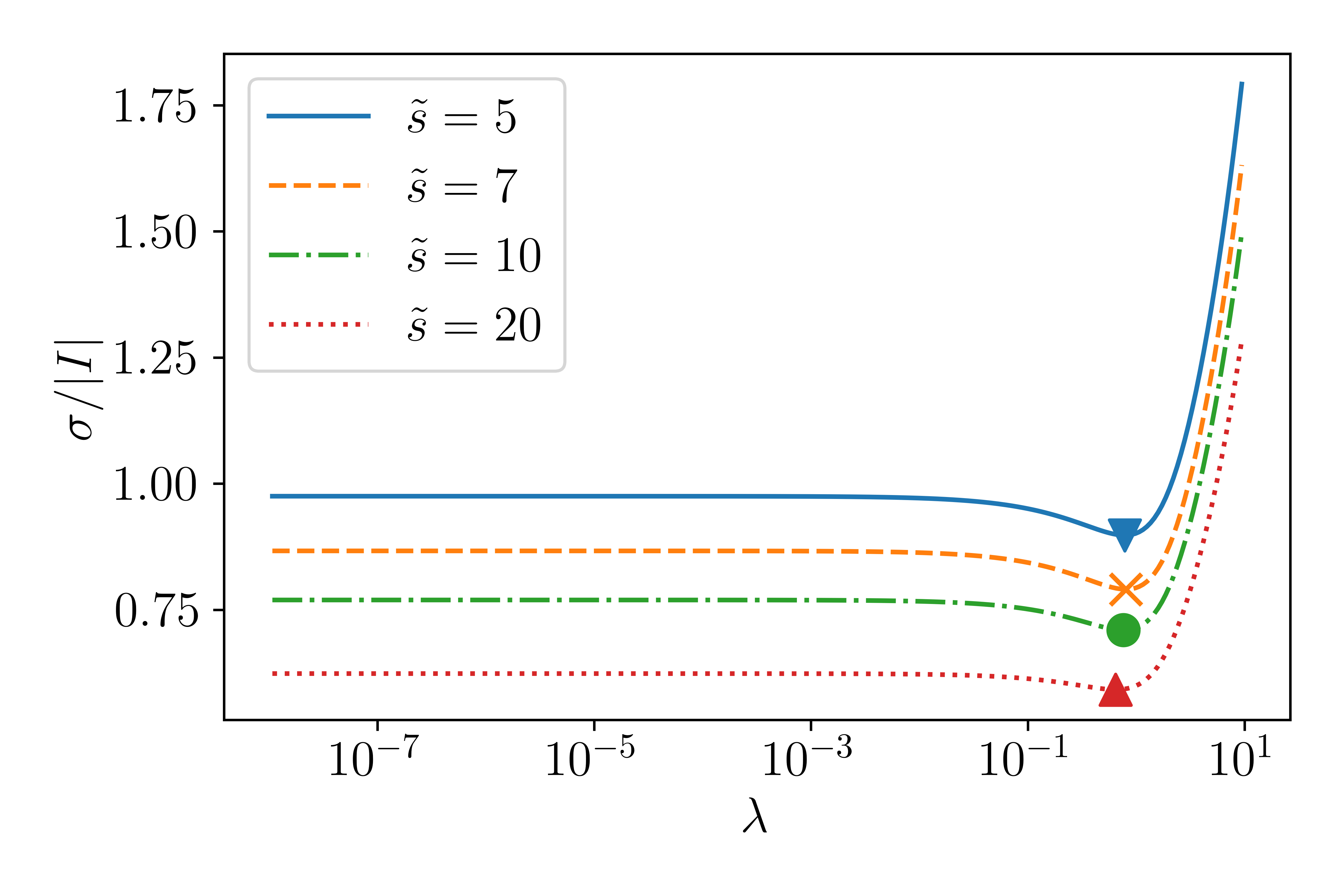}
\includegraphics[scale=0.45]{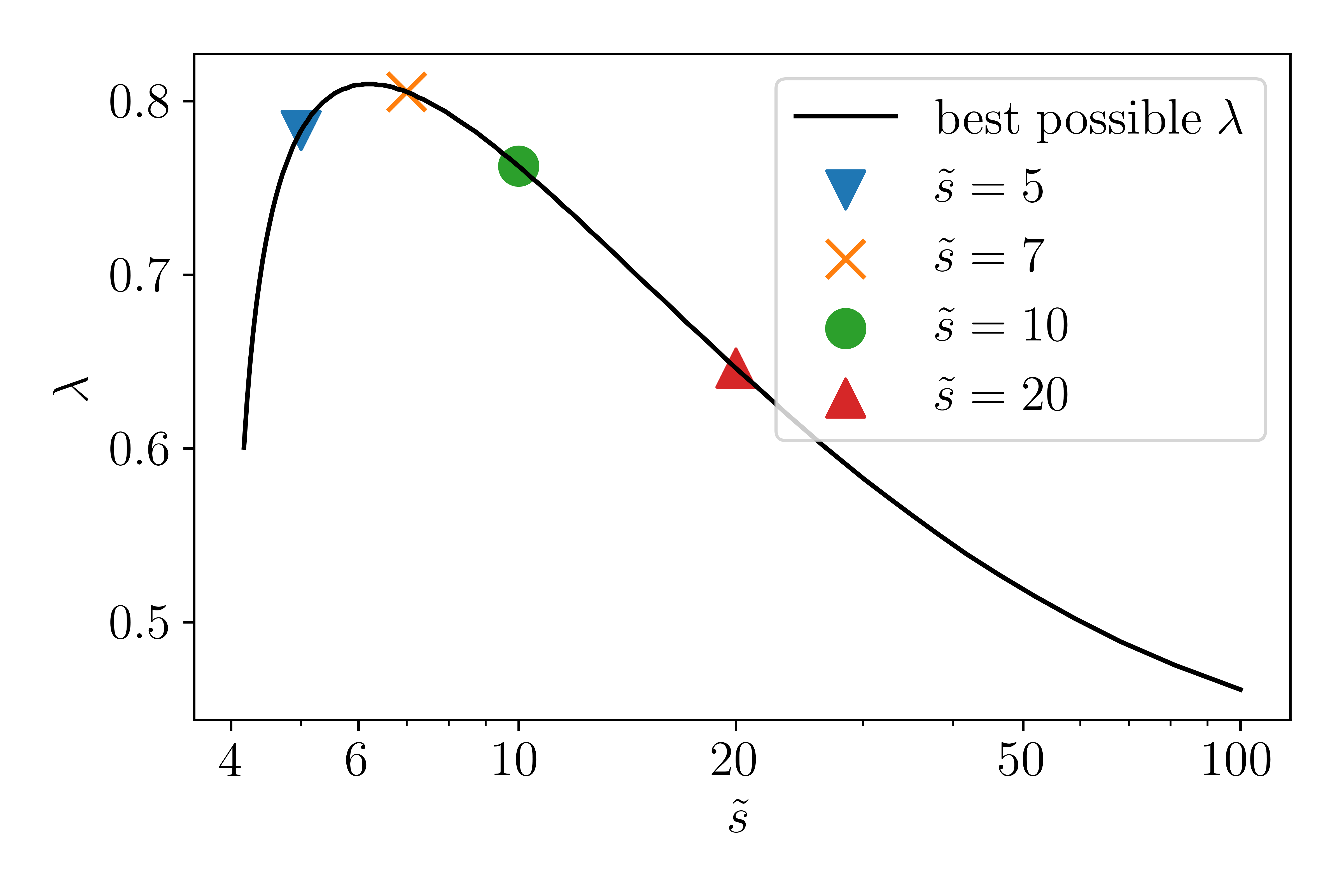}
\caption{Left-hand: variance of the integrand normalised by the integrand as a function of $\lambda$ for a selection of values of $\tilde{s}$. The markers show the position of the minimum. Right-hand pane: value of lambda that minimises the variance of the integrand. The markers correspond to the values of the minima for the curves on the left-hand side plot.}\label{fig:variance_of_lambda}
\end{figure}

For the 1-loop bubble integral the heuristic-$\lambda$ deformation of Eq.~\eqref{eq:guided_deformation} is given by
\begin{equation}
z = x - i x (1-x) \, \tilde{s} \, (2x-1)\lambda.
\end{equation}
We are ready to investigate the $\lambda$ dependence of the integral variance. The left-hand pane of Figure~\ref{fig:variance_of_lambda} shows the dependence of the variance of the integrand as a function of the deformation $\lambda$ for set values of $\tilde{s}$. 
In general, we would expect a small value of the deformation to lead to a large variance, because the integration path is close to the poles on the real axis. In this example, the variance reaches a plateau for small values of the deformation. This is because the exponent of the $F$ polynomial in the denominator of Eq.~\eqref{eq:bubble} is $\epsilon$ and therefore there is no singularity in the integrand.  

Too large a deformation also leads to an increased variance, this time due to the large added imaginary part. 
Given the simplicity of the integral, we can compute the value of the variance of the integrand as a function of $\lambda$ and numerically determine the optimal value of $\lambda$ for each value of $\tilde{s}$. The right pane of Figure~\ref{fig:variance_of_lambda} shows the value of $\lambda$ that minimises the variance of the integrand as a function of $\tilde{s}$. 
For large values of $\tilde{s}$, the poles are situated close to the boundaries of integration, and it becomes beneficial to have a larger slope, see Figure~\ref{fig:deformation_2D}.

\begin{figure}
\begin{centering}
\includegraphics[scale=0.5]{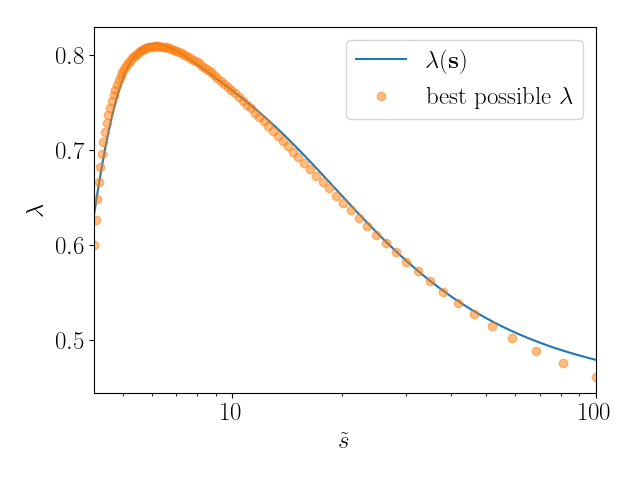}
\caption{Comparison of the optimal values of $\lambda$ compared with the values determined by the $\lambda(\mathbf{s})$ deformation neural network.}\label{fig:wide_lambda}
\end{centering}
\end{figure}

We now want to test a neural network determination of the $\lambda$ value to use. We use a single neural network for the region $4.01<\tilde{s}<100$. After training, the values of $\lambda$ obtained match the optimal ones very well, as can be seen from Figure~\ref{fig:wide_lambda}.

We could also apply the modification described in Section~\ref{sec:optimal_contour} (i.e. we can subtract the estimated value of the integrand, $m(\mathbf{s})$), but this has almost no effect since the values of $\lambda$ obtained without the modification are already very close to their optimal value. This is because the contribution to the variance coming from the variance of $\hat{f}(\mathbf{s})$ (the first term in the right-hand side of Eq~\eqref{eq:var}) is negligible compared to that coming from the integrand variance at fixed values of $\tilde{s}$ (the second term in the right-hand side of Eq~\eqref{eq:var}). 

\subsection{$\lambda(\mathbf{x};\mathbf{s})$  deformation}

\begin{figure}
\begin{centering}
\includegraphics[scale=0.5]{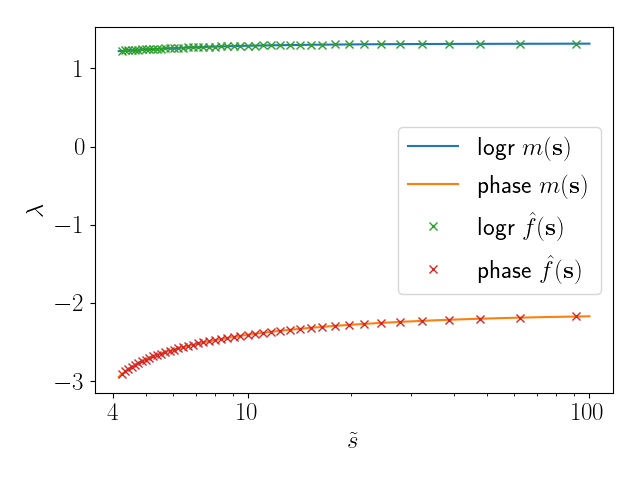}
\caption{Comparison of the true value of $\hat{f}(\mathbf{s})$ and its approximation $m(\mathbf{s})$. The values are expressed in polar coordinates, and we plot the logarithm of the radius along with the argument of the true value of the integral for fixed $\tilde{s}$ (crosses) and its approximation $m(\mathbf{s})$ (solid line).}
\label{fig:ftilde_wide}
\end{centering}
\end{figure}

We now turn to the deformation where the $\lambda_i$ are allowed to take the much more flexible form of a function of both the kinematic invariants and the integration variables; $\lambda_i := \lambda_i(x_1, ..., x_N; s_1,...,s_M)$. 
We train this deformation with the modification described in Section~\ref{sec:optimal_contour}. First, we can verify that the neural network was able to get a good approximation $m(\mathbf{s})$ of $\hat{f}(\mathbf{s})$. This is shown in Figure~\ref{fig:ftilde_wide} where we compare the neural network approximation to the true value of $\hat{f}(\mathbf{s})$ obtained through direct numerical integration. 

\begin{figure}
\includegraphics[scale=0.5]{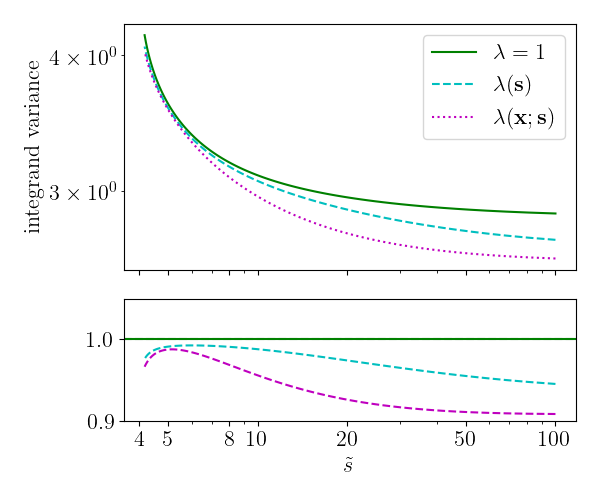}
\includegraphics[scale=0.5]{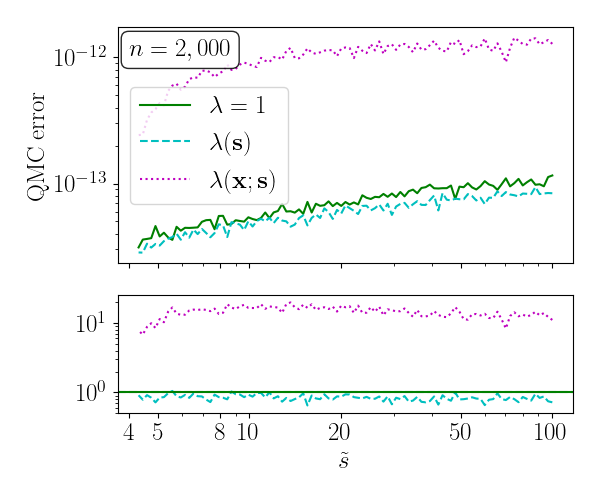}
\caption{Comparison of the variance and QMC error estimates of the heuristic-$\lambda$ deformation and the $\lambda(\mathbf{s})$ and $\lambda(\mathbf{x};\mathbf{s})$ deformations. Left pane: variance of the integrand. Right pane: QMC error estimate obtained from the two methods. }\label{fig:wide_variance_comparison}
\end{figure}

On the left-hand side of Figure~\ref{fig:wide_variance_comparison}, we compare the variance of the integrands obtained with the heuristic-$\lambda$ deformation and those obtained using the $\lambda(\mathbf{s})$ or $\lambda(\mathbf{x};\mathbf{s})$ deformations. We can see that the $\lambda(\mathbf{x};\mathbf{s})$ deformation achieves a lower variance. This is not very surprising given that they have a much less restricted functional form, and, given enough parameters, the full neural network solution should at least be able to match the $\lambda(\mathbf{s})$ deformation results by matching its functional form. On the same plot we show the variance from the deformation selected by \pysecdec, which in this case is $\lambda=1$ because of the limit on the maximum size of the deformation described in Section~\ref{sec:pysecdec}. 

\begin{figure}
\centering
\includegraphics[scale=0.45]{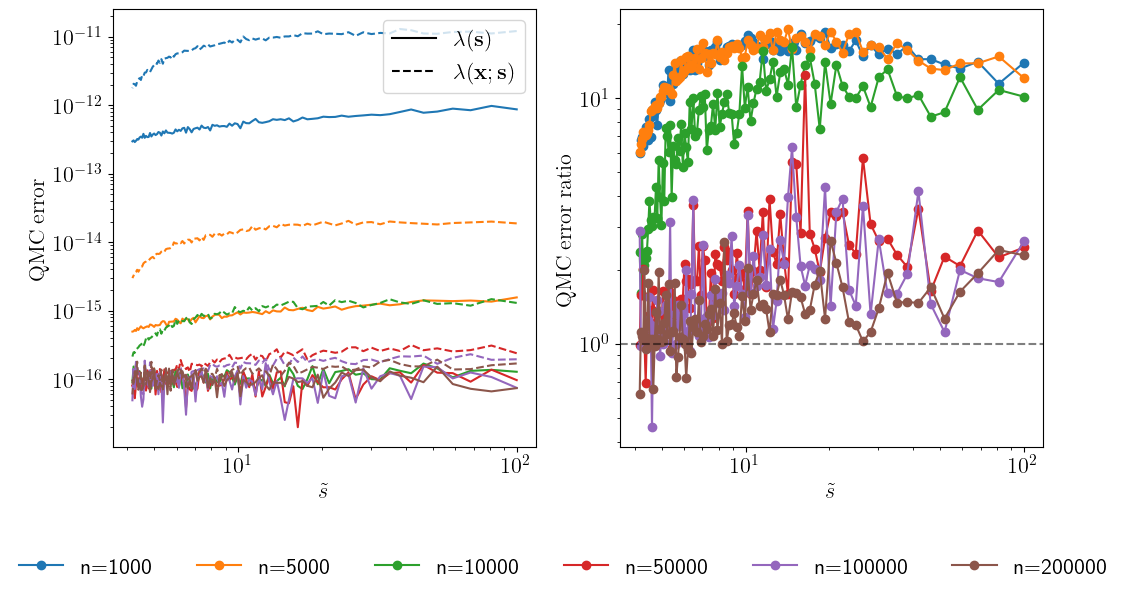}
\caption{QMC performance comparison between the $\lambda(\mathbf{s})$ deformation and the $\lambda(\mathbf{x};\mathbf{s})$ deformation as a function of the lattice size. Left pane: QMC error estimate for different lattice sizes. Dashed lines correspond to the $\lambda(\mathbf{x};\mathbf{s})$ neural network deformation and solid lines to the $\lambda(\mathbf{s})$ deformation. Right pane: The curves show the ratio of the QMC error estimates from the $\lambda(\mathbf{x};\mathbf{s})$ deformation to those from the $\lambda(\mathbf{s})$ deformation. 
}\label{fig:QMC_n_ratio}
\end{figure}

An improvement in the variance of the integrand translates directly to a better performance for a standard Monte Carlo integration. In practice, for large dimensions, a QMC is often used instead of conventional MC, and there is no guarantee that an improvement in the variance translates into an improvement in the QMC performance. If we plot the estimated error from 32 random shifts of a lattice computed with the median of means method with 21 candidates, we find the first example of a somewhat counter-intuitive result of this article: the improvement of the variance does not always translate into an improvement of the QMC performance. This is illustrated in the right-hand side of Figure~\ref{fig:wide_variance_comparison} for a lattice size of 10,000\footnote{Whenever we refer to a lattice size $N$, we mean "the first prime integer larger than $N$." }. The QMC error estimate does not change significantly when the deformation is only rescaled from $\lambda=1$ to $\lambda(\mathbf{s})$, but it changes when the functional form is altered using $\lambda(\mathbf{x};\mathbf{s})$. Plotting the $\lambda(\mathbf{x};\mathbf{s})$ and $\lambda(\mathbf{s})$ QMC error estimate for a variety of lattice sizes in Figure~\ref{fig:QMC_n_ratio} shows that the effect is not limited to a single lattice size and seems systematic, but decreases with large lattice sizes. 


\begin{figure}
\centering
\includegraphics[scale=0.5]{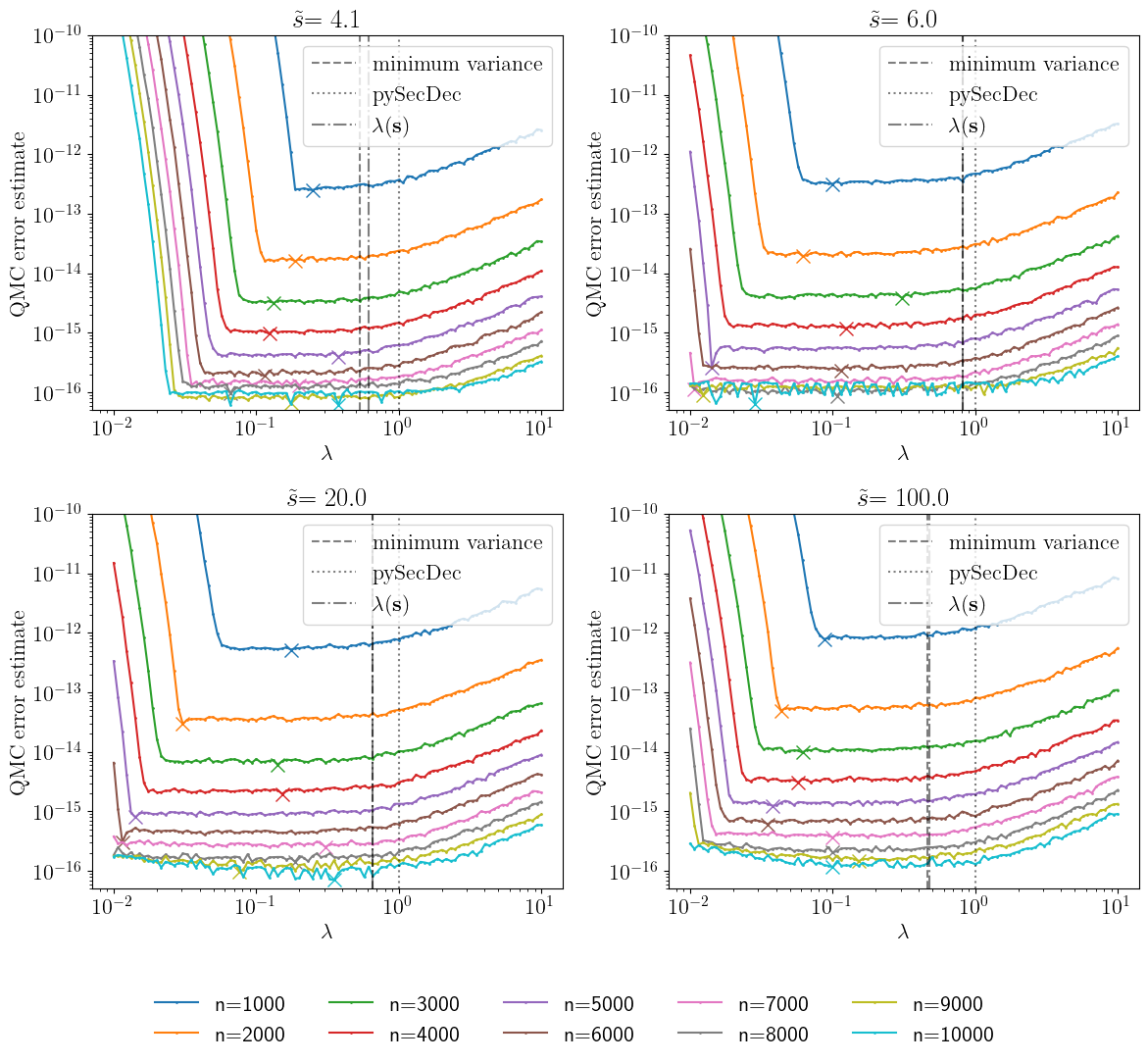}

\caption{QMC performance as a function of the deformation strength $\lambda$ for a set of lattice sizes. The dashed vertical line corresponds to the optimal value of $\lambda$ to minimise the integrand variance. The crosses show the minimum of the QMC error estimate, whose location has a strong dependence on the grid size.
}\label{fig:QMCerror_lambda_n}
\end{figure}

In Figure~\ref{fig:QMCerror_lambda_n} we plot the dependence of the QMC error estimate as a function of the lattice size. We see that the optimal contour for the purpose of minimising the QMC error depends on the lattice size. While the optimal contour from the integrand variance point of view is independent of the lattice size, this dependence means that there is no unique optimal contour for QMC integration. We note that for this integral there is a plateau of good $\lambda$ choices and both the optimal $\lambda$ choice for a given lattice and the $\lambda$ that minimises the variance are close to this plateau. A more detailed exposition of the grid size dependence of the RQMC error estimate can be found in Appendix~\ref{sec:minlambda}.

It would be interesting to investigate the behaviour of the minimum of the QMC error estimate for large lattice sizes, but as can be seen in the left-hand pane of Figure~\ref{fig:QMC_n_ratio}, lattice sizes above 50000 reach the limit of the numerical accuracy one can expect from double-precision floating point operations.

\begin{figure}[h!]
\centering
\includegraphics[scale=0.45]{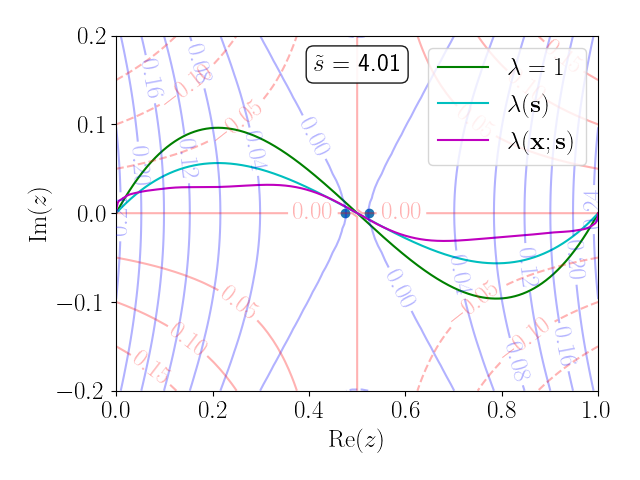}
\includegraphics[scale=0.45]{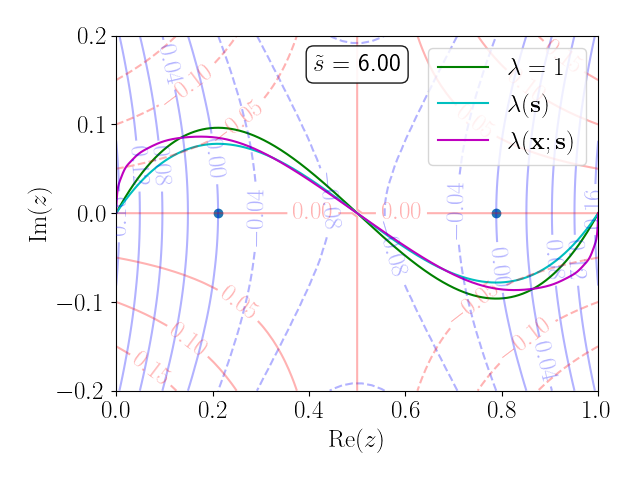}\\
\includegraphics[scale=0.45]{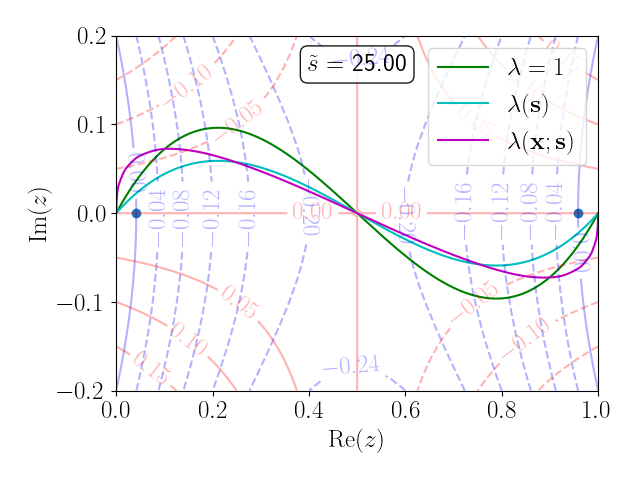}
\includegraphics[scale=0.45]{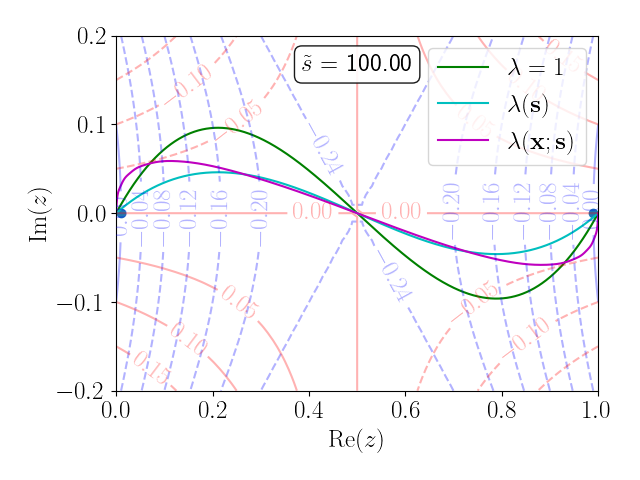}\\
\caption{Deformed path for the heuristic-$\lambda$ and neural network deformations $\lambda(\mathbf{s})$ and $\lambda(\mathbf{x};\mathbf{s})$ for three values of $\tilde{s}$: 4.01 (top left), 6 (top right), 25 (bottom left), 100 (bottom right). The faint blue and red lines show contours of constant real and imaginary part of the $F$ polynomial, respectively.
}\label{fig:deformation_2D} 
\end{figure}

For this one-dimensional integral, it can be instructive to visualise the deformations. Figure~\ref{fig:deformation_2D} shows the deformed path for three different values of $\tilde{s}$. 
The shape of the $\lambda(\mathbf{x};\mathbf{s})$ deformation visibly differs from the $\lambda(\mathbf{s})$ and heuristic-$\lambda$ deformations. For the phase-space points considered, it always has a steeper start at the integration boundaries and a smaller magnitude than that picked using the heuristic approach.
\FloatBarrier
\section{One-loop triangle example}\label{sec:triangle1L}
We consider the one-loop, equal-mass triangle integral defined by the Feynman diagram in Figure~\ref{fig:triangle_fd}. The integral can be written as\footnote{The Cheng-Wu theorem has been used to change $\delta(1-\sum\nolimits_{i=1}^N x_i) \to \delta(1-x_N)$.\label{chengwu}},
\begin{equation}
\label{eq:triangle}
    I(\mathbf{s}) = -\Gamma\left(1+\epsilon\right) \int_{\mathbb{R}^3_{\geq0}}\!\,\prod_{i=1}^{3}\mathrm{d}x_i\frac{\left(x_1+x_2+x_3\right)^{-1+2\epsilon} \delta\left(1-x_3)\right)}{\left(-p^2x_1x_2+m^2\left(x_1+x_2+x_3\right)^2-i\delta\right)^{1+\epsilon}}
\end{equation}
This integral has a physical threshold at $s=(p_1+p_2)^2=4m^2$. For the remainder of this section, we consider the integral for $m^2=1$ (or, equivalently, as a function of the rescaled parameter $s/m^2$). Applying geometric decomposition as implemented in \pysecdec yields 3 sectors\footnote{With \texttt{decomposition\_method=geometric}, see Ref. \cite{Schlenk:2016epj} for a description of the implementation.\label{geosecdec}}. To each sector integral, a Korobov transformation with $\alpha=\beta=3$ is applied to improve the error convergence of the QMC integrator \cite{Borowka:2018goh}. As described in Section~\ref{sec:thresholds}, after sector decomposition, integrals can develop an imaginary part at kinematic values that do not correspond to physical thresholds. In the present case, this occurs at the ratio $s/m^2 = 9$, in addition to the physical threshold at $s/m^2 = 4$. 
\subsection{Results}
\begin{figure}
\centering
\includegraphics[scale=0.45]{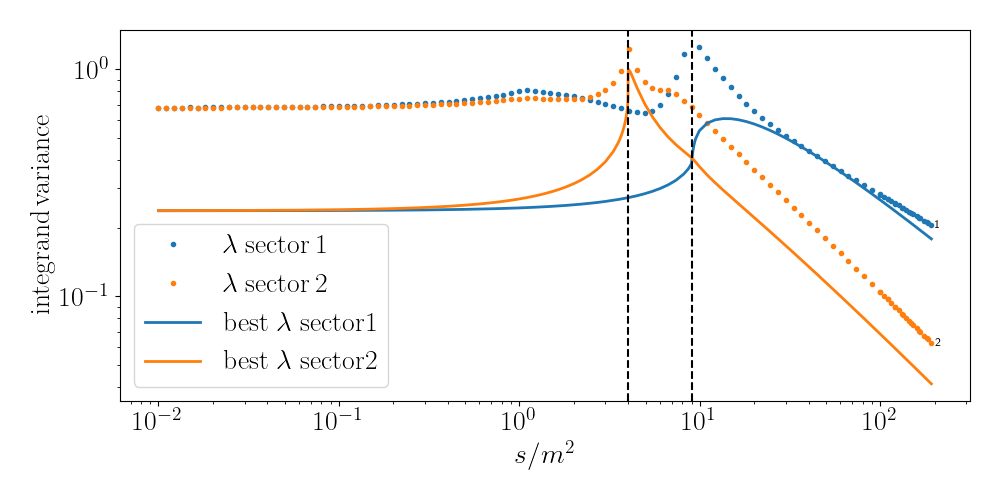}
\includegraphics[scale=0.45]{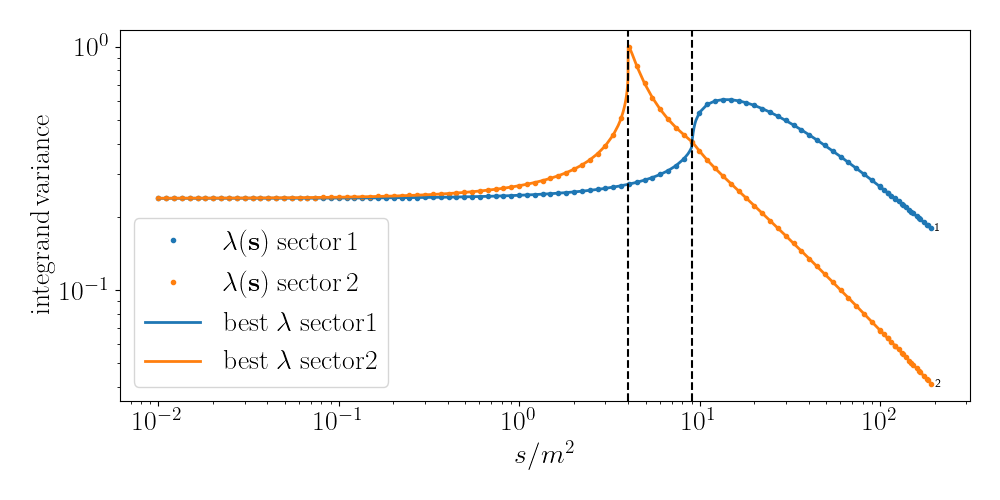}
\includegraphics[scale=0.45]{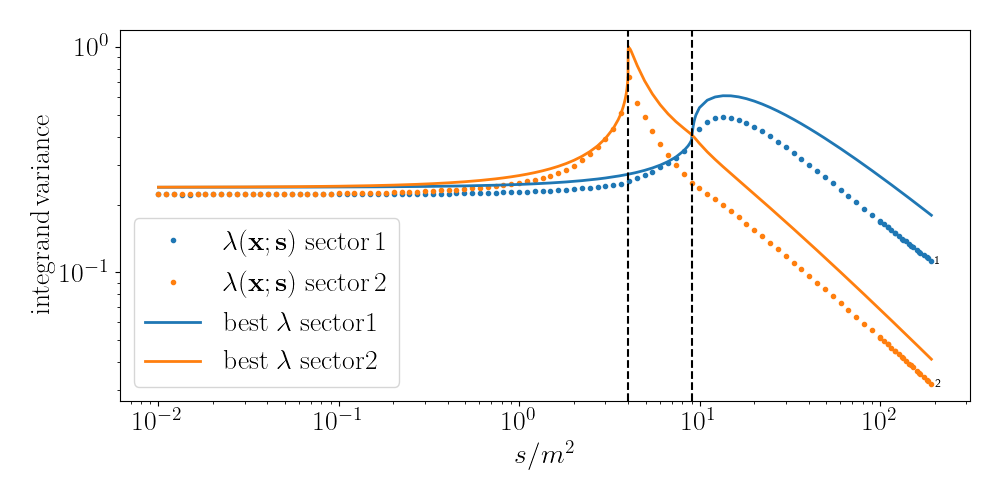}
\caption{Variance of the integrand as a function of the $s/m$ ratio for the three different types of contours. The solid line shows the value obtained by explicitly minimising the variance for each value of $s/m$.}\label{fig:triangle_std}
\end{figure}

Figure~\ref{fig:triangle_std} shows the variance for the heuristic-$\lam$, $\lams$ and $\lamsx$ deformation as a function of the ratio $s/m^2$. As a comparison, we performed an explicit optimisation of the variance on the integrand. This gives us a lower bound on the variance that can be obtained using the $\lam$ and $\lams$ deformation. The top plot of Figure~\ref{fig:triangle_std} shows that the heuristically determined deformation falls short of the optimal variance. The middle plot shows that the neural network determination matches to optimal values across the $s$ range. With the increased flexibility the $\lamsx$ deformation can reach variances below the minimal floor for an $x$-independent $\lam$.

\begin{figure}
\centering
\includegraphics[scale=0.35]{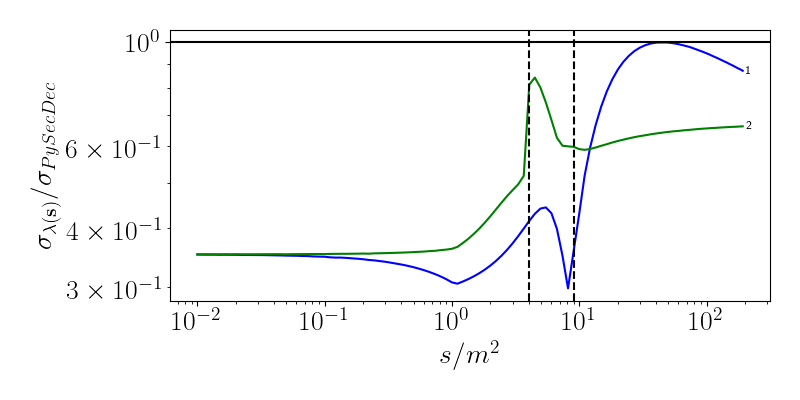}
\includegraphics[scale=0.35]{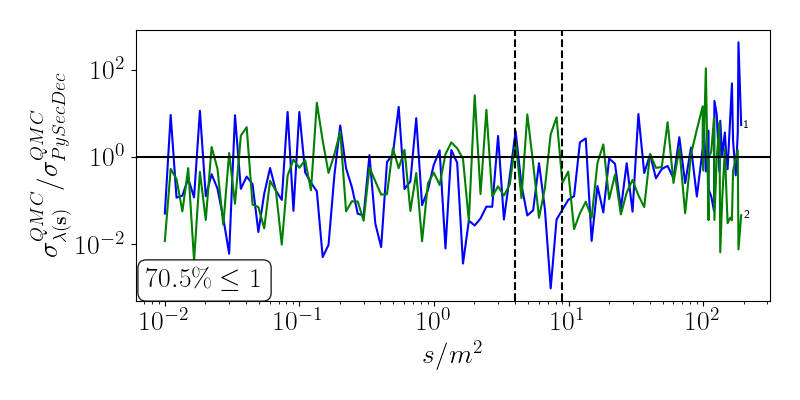}
\includegraphics[scale=0.35]{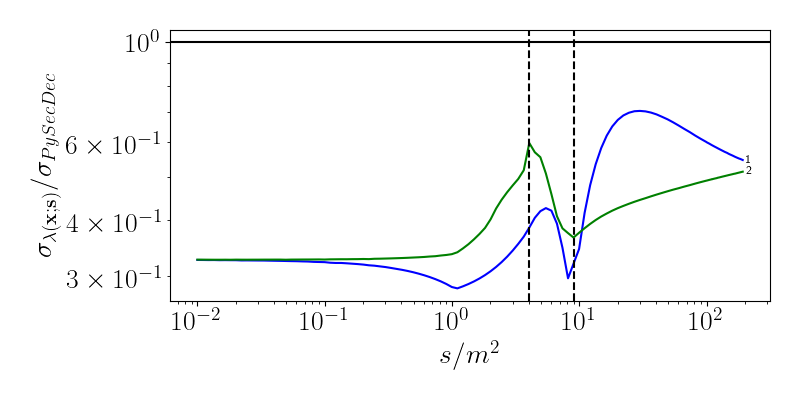}
\includegraphics[scale=0.35]{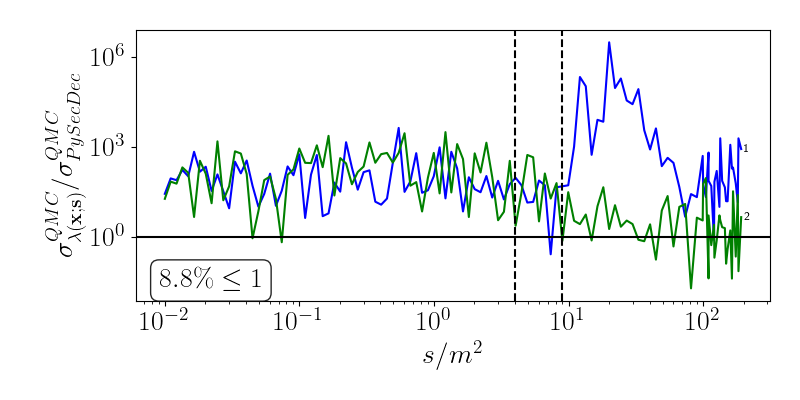}
\includegraphics[scale=0.35]{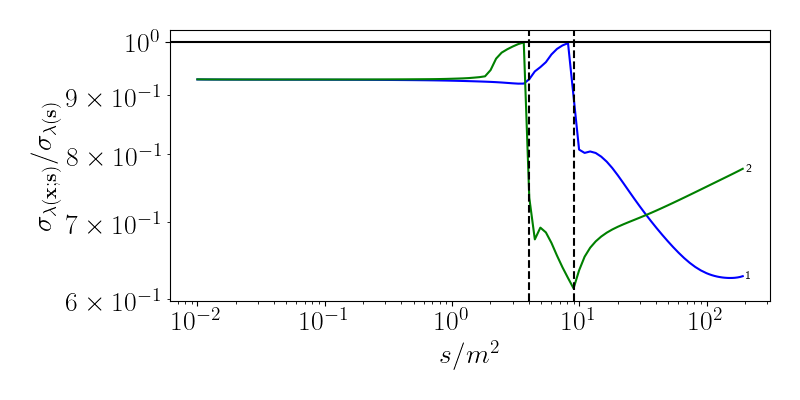}
\includegraphics[scale=0.35]{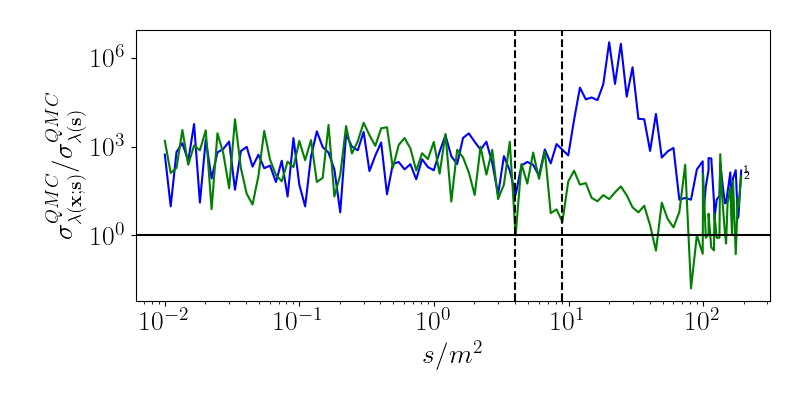}
\caption{Left side: ratios of the standard deviation of the integrand. Right side: ratios of the standard deviation of the shift ensemble for a lattice size of $2\cdot 10^4$. The upper row shows the ratios between $\lams$ and the heuristic $\lam$, the middle row the ratios between $\lamsx$ and $\lam$, and the lower row the ratios between $\lamsx$ and $\lams$.}
\label{fig:triangle_nnlambda}
\end{figure}

The upper row of plots in Figure~\ref{fig:triangle_nnlambda} shows the ratio of the variance between the $\lams$ and the $\lam$ contours as a function of the $s/m^2$ ratio on the left and the RQMC error estimate ratio is shown in the lower part of Figure~\ref{fig:triangle_nnlambda}. Across the data displayed the $\lams$ deformation achieves a better error estimate for around 70\% of the points. The middle row of plots shows the same ratios but between the $\lamsx$ and $\lam$ deformations. 

The plot shows that the variance improves from the heuristic $\lam$ to $\lams$ to $\lamsx$, as one would expect, but the QMC error estimate is best for the $\lams$ deformation.  
\FloatBarrier
\section{Two-loop box example}\label{sec:elliptic2L}
\begin{figure}[h!]
  \centering
  \input{figs/box}
  \caption{Feynman diagram for the two-loop elliptic box integral. Figure adapted from Ref. \cite{Winterhalder:2021ngy}.} 
  \label{fig:elliptic}
\end{figure}

We consider a two-loop integral defined by the Feynman diagram of Figure~\ref{fig:elliptic}. The integral can be written as\footref{chengwu},
\begin{equation}
    I(\mathbf{s}) = \Gamma(2\epsilon+2) \int_{\mathbb{R}^6_{\geq0}} \, \!\!\prod_{i=1}^{6}\mathrm{d}x_i \, \frac{\mathcal{U}(\mathbf{x})^{3\epsilon}}{\left(\mathcal{F}(\mathbf{x};\mathbf{s}) -i\delta\right)^{2\epsilon+2}} \, \delta\left(1-x_6\right),
\end{equation}
where the $\mathcal{U}$ and $\mathcal{F}$ polynomials are given by,
\begin{align}
\mathcal{U}(\mathbf{x}) = &\,\, x_3 x_4 + x_3 x_5 + x_4 x_5 + (x_4 + x_5) x_6 + x_1 (x_3 + x_5 + x_6) + 
 x_2 (x_3 + x_5 + x_6), \nonumber \\
   \mathcal{F}(\mathbf{x};\mathbf{s}) = &-((t x_4 x_5 + p_4^2 (x_3 (x_1 + x_2 + x_4) + (x_2 + x_3) x_5)) x_6) -
 s x_1 (x_3 x_5 + x_2 (x_3 + x_5 + x_6))  \nonumber \\
 &+ \, m^2 (x_1 + x_2 + x_3 + x_4 + x_6) (x_3 x_4 + x_3 x_5 + x_4 x_5 
 + (x_4 + x_5) x_6 + x_1 (x_3 + x_5 + x_6) \nonumber \\ &+ x_2 (x_3 + x_5 + x_6)). \nonumber
\end{align}
The external legs $p_1$, $p_2$ and $p_3$ are massless, and $p_4$ is massive. We consider the physical region with $p_1,p_2$ incoming and $p_3,p_4$ outgoing. This integral has physical thresholds at $s=(p_1+p_2)^2=4m^2$, $p_4^2=4m^2$ and $t=(p_1-p_3)^2=4m^2$.
For the remainder of this section, we consider the integral for $m^2=1$ (or, equivalently, as a function of the rescaled parameters $s/m^2, t/m^2$ and $p_4^2/m^2$). 
The integral has endpoint divergences and requires a sector decomposition. Applying geometric decomposition as implemented in \pysecdec yields 22 sectors\footnote{\ref{geosecdec}}. The labelling of the sectors has no intrinsic meaning\footnote{Indeed, \pysecdec might return them in a different order for each call.}, but they are consistent throughout this work. To each sector integral, a Korobov transformation with $\alpha=\beta=3$ is applied to improve the error convergence of the QMC integrator \cite{Borowka:2018goh}.

\begin{figure}[t]
\begin{centering}
\includegraphics[scale=0.5]{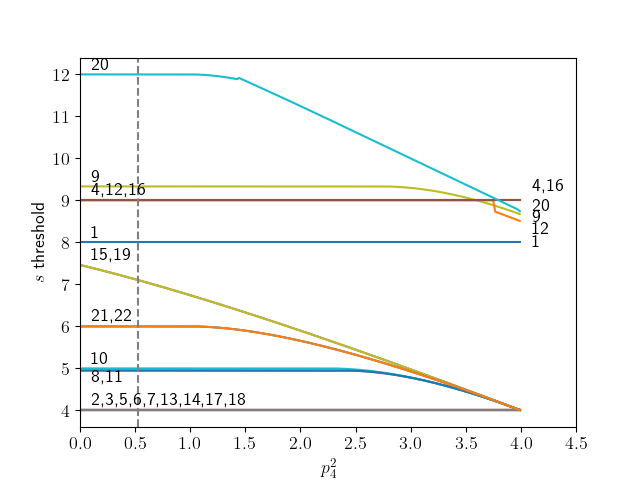}
\caption{Lowest $s$ threshold as a function of $p_4^2$. Several curves overlap for some sectors; their number(s) is displayed close to the respective curve. The vertical line corresponds to the value $p_4^2/m^2$=12/23 used for the numerical evaluation of these sectors in this work.}\label{fig:thresholds_elliptic2L}
\end{centering}
\end{figure}

As described in Section~\ref{sec:thresholds}, after sector decomposition, integrals can develop an imaginary part at kinematic values that do not correspond to physical thresholds, furthermore, these unphysical thresholds can have a complicated dependence on the kinematic invariants. 
The location of the lowest threshold in $s$, that is, the smallest value of $s$ at which a given sector develops an imaginary part, is displayed in Figure~\ref{fig:thresholds_elliptic2L}.
The location of the threshold for each sector can be determined numerically by investigating the corresponding $F$-polynomial. 
First, we insert the value of $p_4^2$ and $m^2=1$, then substitute 
$$t\rightarrow -\tau (s - p_4^2)$$ 
where $\tau\in[0,1]$ covers all allowed values of $t$. At this point, in each sector, the $F(\mathbf{y};\mathbf{s})$ polynomial is a linear function of $s$ with coefficients that are functions of the facet variables, $\mathbf{y}$, and $\tau$. 
We solve the equation $F(\mathbf{y};\mathbf{s})=0$ for $s$ and obtain
\begin{align}
s=f(\mathbf{y};\mathbf{s}\neq\tilde{s}),
\end{align}
which we numerically minimise over the unit hypercube spanned by $\mathbf{y},\tau$. 
This procedure will find the smallest value of $s$ for which $F(\mathbf{y};\mathbf{s})=0$ has a solution, and therefore the location of the threshold. 
The numerically found solutions can be verified analytically. 

In the first part of this section, we will work with a fixed value of $p_4^2/m^2=12/23$, which approximates the ratio $m_t^2/m_H^2$, where $m_t$ is the on-shell top-quark mass and $m_H$ is the on-shell Higgs boson mass. 
In the second part of this section, we will relax this constraint to facilitate comparison with previous work. 
We will investigate the deformations to evaluate it in three regions: $1<s<s_{threshold}^{(k)}$, $s_{threshold}^{(k)}<s<100$ and $100 < s < 10,000$ where the separation point, $s_{threshold}^{(k)}$, is sector dependent, and given in Table~\ref{tab:thresholds}. The deformation for each of these regions is represented by a single neural network. The separation at the threshold value is needed because the integral is not well defined at the physical threshold, and the separation at $s=100$ is set to prevent the training from spanning three orders of magnitude, which helps convergence. This illustrates the two reasons that can motivate the choice of regions on which the deformation is defined.

\begin{table}[t]
\centering
\input{tables/thresholds.tex}
\caption{Lowest $s$ thresholds for each sector, $s_{thresold}^{(k)}$. The non-integer values are algebraic functions involving the value 12/23 used as the ratio $p_4^2/m^2$.}\label{tab:thresholds}
\end{table}

For fixed $p_4^2/m^2$, the $F$ polynomial for each sector scales with at most one power of $s/m^2$ or $t/m^2$. This also means that the derivative of $F$ entering Eq.~\eqref{eq:guided_deformation} scales with $s/m^2$ and $t/m^2$. In the physical region where $s$ or $t$ are large, this leads to large deformations, to counterbalance this effect, we scale the value of $\lambda$ predicted by the network by $1/s$. In principle, the network could learn this functional dependence, but the training would be more difficult, and the issue of large initial deformation not being valid (see Section~\ref{sec:training}) would be exacerbated.

Since we are considering a wide range of scales, especially in the large-$s$ region, we first apply the transformation 
\begin{equation}
s\rightarrow \log(s)\;,\qquad t\rightarrow -\frac{t}{s}\in [0,1]
\end{equation}
before standardisation and evaluation of the neural network.

\subsection{Results}

\begin{figure}
\centering
\includegraphics[scale=0.5]{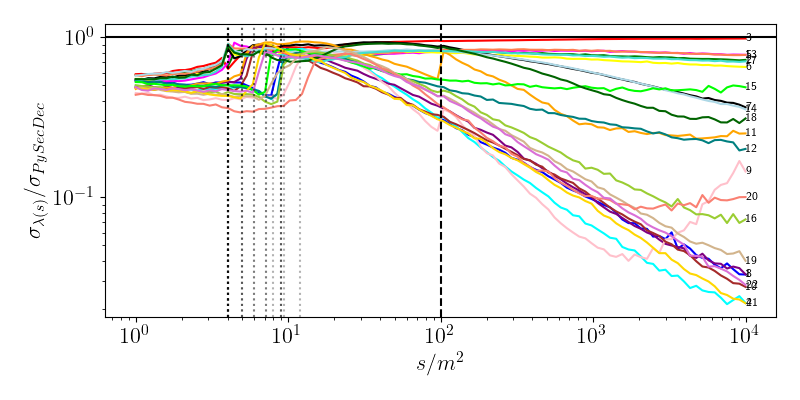}
\includegraphics[scale=0.5]{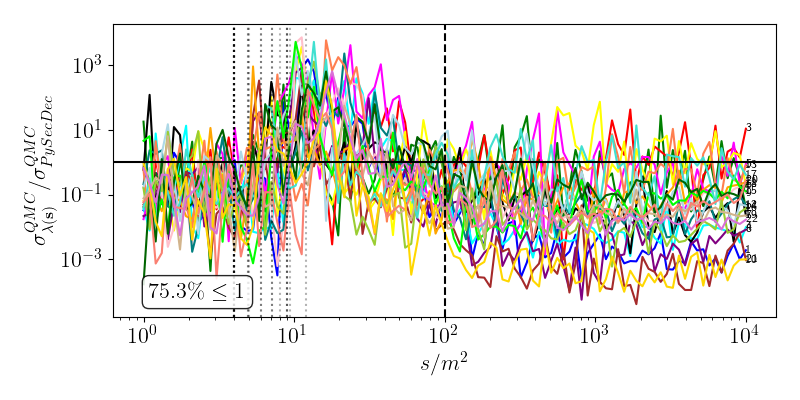}
\caption{Upper plot: Ratio of the standard deviation of the integrand for the $\lambda(s)$ deformation to the \pysecdec integrand. Lower plot: ratio of the standard deviation of the shift ensemble for a lattice size of $2\cdot 10^6$. The vertical lines show the location of the thresholds and the separation of the large $\tilde{s}$ region, where discontinuities can be expected.}\label{fig:elliptic_std}
\end{figure}

To assess the suitability of the deformation from our methods, we compare two quantities. The first is the variance of the integrand. For this, we expect the neural network methods to yield a better result, since this is what the training is aimed at. Since we minimise this variance for a range of values of the kinematic parameters, some regions may underperform the \pysecdec variance, because their contribution to the loss is too small to be prioritised in the learning process. In other words, the noise in the regions where the integrand is large prevents the network from learning the optimal choice for the regions where the integrand is much smaller. We consider a one-dimensional slice where the value of $t$ is fixed at the middle value of the allowed interval,
\begin{equation}
t = \frac{1}{2} \left(t_{\text{min}}+t_{\text{max}}\right),\; t_{\text{min}} = -s+p_4^2 \;,t_{\text{max}}=0.
\end{equation}

We run \pysecdec for this range of $s$ and $t$ and record the final value of the deformation parameters $\lambda$. We then compare the variance of the integrand and the QMC error estimate with those obtained through our neural network. 

The upper pane of Figure~\ref{fig:elliptic_std} shows the ratio of the variance of the integrand obtained from the $\lambda(\mathbf{s})$ method to \pysecdec. 
Each sector is shown as a line. We can see that for all regions and all sectors, the variance is indeed reduced. 

The second quantity we consider to assess the suitability of the NN-chosen contours is the error estimate from the QMC integration. To obtain this estimate, we follow the method described in Ref.~\cite{Goda_2022}. For a given prime $n$, we generate $n_l$ random lattices and select the random lattice with the median mean. For this lattice, we generate $n_s$ random shifts and use the mean of the results from these $n_s$ shifted lattices as the integral estimate and the standard deviation of the $n_s$ results as the error estimate.
We repeat a QMC integration with a given lattice $n_{s}$ times, each with a different random shift. The standard deviation of the results gives an estimate of the error of the QMC integration. In principle, a better-behaved integrand yields a lower error estimate. We find that reducing the variance of the integrand is not a guarantee that the QMC error estimate will be reduced. The lower pane of Figure~\ref{fig:elliptic_std} shows the ratio of the QMC error estimates between the contour chosen by the neural network and the one chosen following the \pysecdec procedure for a lattice size of 2 million points. We see that in some cases this ratio can take dramatically large values, although the integrand has a small variance.  

\begin{figure}
\begin{centering}
\includegraphics[scale=0.5]{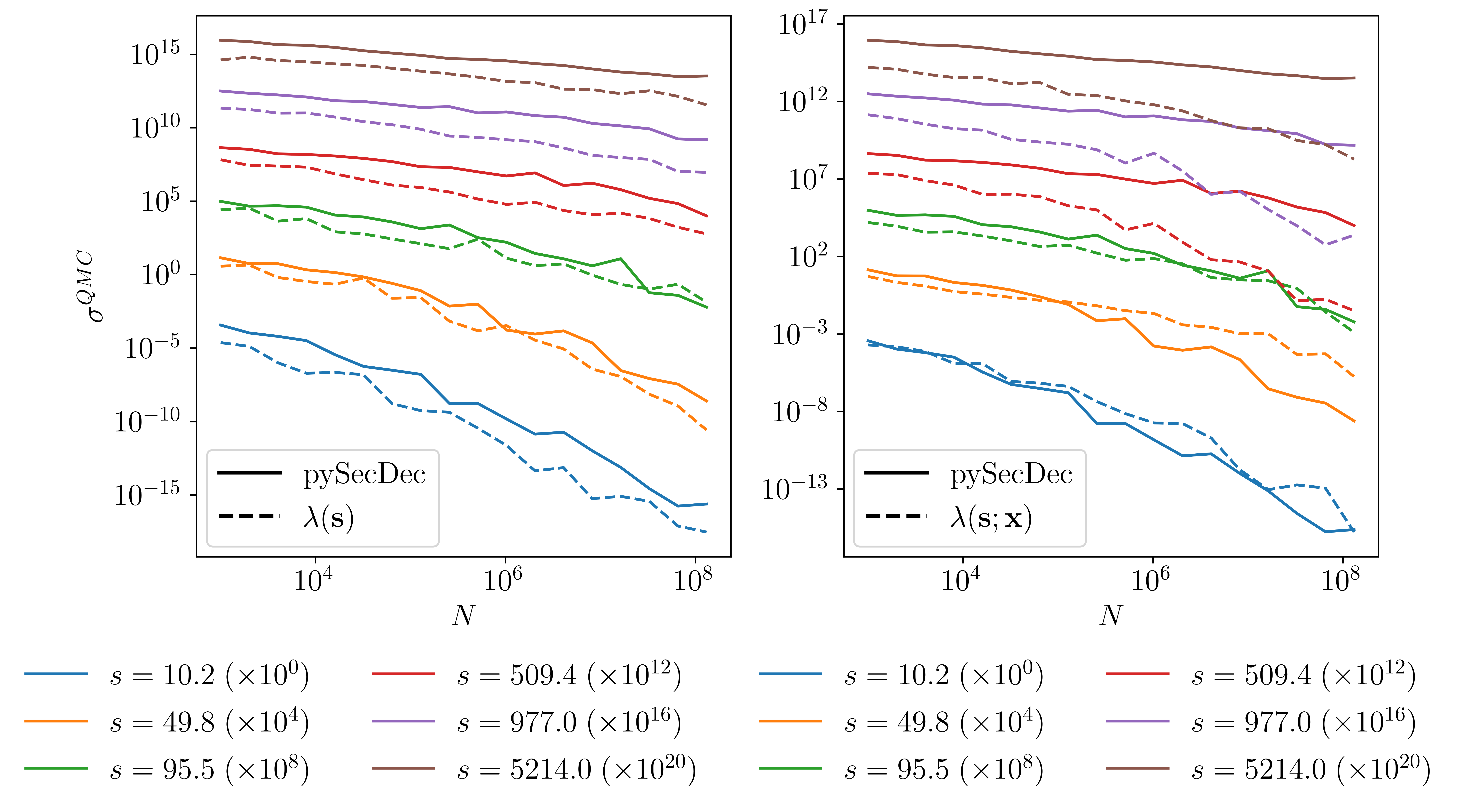}
\caption{QMC error estimate of sector 20 as a function of the lattice size for different values of~$s$. The line for each value of $s$ have been scaled by the same factor for readability.} \label{fig:lattice_size}
\end{centering}
\end{figure}

In Figure~\ref{fig:lattice_size}, we see that the scaling of the QMC error estimate with the number of lattice points is similar for both methods of choosing a contour, although the error obtained with the two methods can differ significantly. 

\begin{figure}
\includegraphics[scale=0.4]{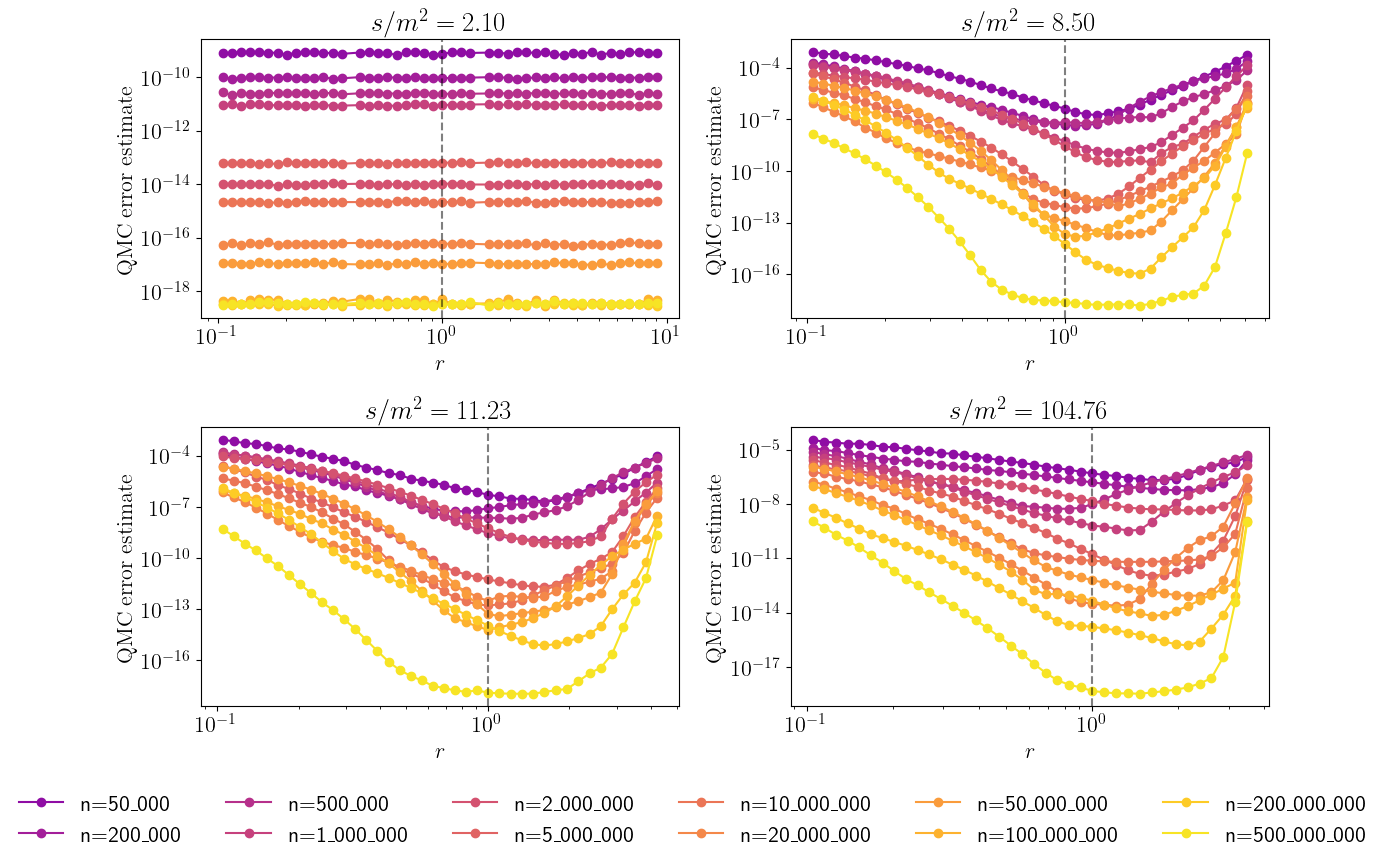}
\caption{QMC error estimate for sector 11 as a function of the rescaling of the deformation strengths $\lambda_i$ by a fixed scalar $r$ for different lattice sizes. The dashed line shows the $\lambda$ value that minimises the variance of the integrand.} \label{fig:QMC_error_scan_elliptic_2L}
\end{figure}

In Figure~\ref{fig:QMC_error_scan_elliptic_2L}, we investigate the dependence of the QMC error as a function of both the deformation parameter and the lattice size. To do so, we take a reference value of $\lambda_i$ and rescale all components by a fixed factor $r$. 
The reference value of $\lambda_i$ is obtained by numerically minimising the variance as a function of the deformation parameters $\lambda_i$. We take as a starting point the value given by the $\lambda(\mathbf{s})$ deformation found by the neural network training.
Too large values of $r$ may describe an invalid shift, so we check that for the data points plotted, the value of the integrand has not departed too much from its true value. We can see that the minimum of the QMC error estimate does not match the value that minimises the variance, but it appears to be the case that as the lattice size increases, the optimal QMC $\lambda$ converges to the value that minimises the variance. Given that there is no unique choice of $\lambda$ that would be optimal for all lattice sizes, choosing the values that minimise the variance does not appear to be a bad choice.

The curves showing the behaviour of the QMC error estimate on the rays defined above are shown in Figure~\ref{fig:QMC_error_scan_elliptic_2L}. Again, we see that the optimal contour for the purpose of QMC error estimate optimisation is dependent on the lattice size. We also note that there are many crossovers between curves corresponding to different lattice sizes. When this happens, and a curve for a lower lattice size crosses that of a larger size, this means that the QMC error estimate will grow despite an increase in lattice size. These counterintuitive situations illustrate one of the complexities of QMC integration. 

\begin{figure}
\includegraphics[scale=0.55]{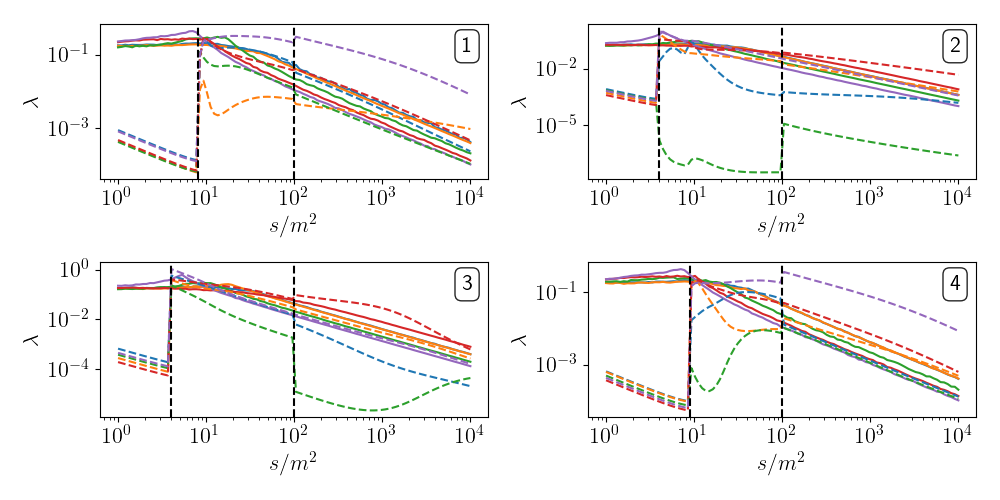}
\caption{Comparison between the $\lambda$ values for \pysecdec (solid) and the $\lambda(\mathbf{s})$ values (dashed). Curves of the same colour correspond to the deformation strength for the same coordinate. The sector number is displayed in the upper right of each plot. The vertical dashed lines show the separation between different trainings, where discontinuities can be expected. 
}\label{fig:lambda_comparison_1_4}
\end{figure}

Figure~\ref{fig:lambda_comparison_1_4} shows a comparison of the lambda values from \pysecdec and from our $\lambda(\mathbf{s}$ deformation for the first four sectors. The comparison for all sectors is shown in Appendix~\ref{sec:lambda_plots}
We observe that the $\lambda$ values selected by the neural network have a much larger range of values and that the scale of the deformations applied to each parameter differs more for the neural network approach than the heuristic approach.

\subsection{$\lambda(\mathbf{x};\mathbf{s})$ deformation}
\begin{figure}
\centering
\includegraphics[scale=0.6]{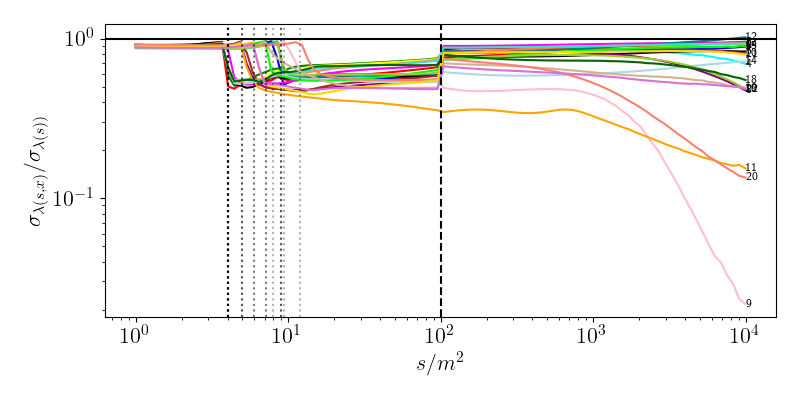}
\caption{Ratio of the $\lambda(\mathbf{x};\mathbf{s})$ deformation variance over the $\lambda(\mathbf{s})$ variance for all 22 sectors of the 2-loop elliptic integral.}\label{fig:free_over_guided_2L} 
\end{figure}

Allowing the deformation to now depend also on the sampling point $x_1,\ldots,x_N$, such that $\lambda_i := \lambda_i(x_1, ..., x_N;s_1,...,s_M)$, we can improve on the results of the previous section. Figure~\ref{fig:free_over_guided_2L} shows the ratio of the variance obtained with $\lambda(\mathbf{x};\mathbf{s})$ deformation over that obtained from the $\lambda(\mathbf{s})$ deformation. For $s<4$ the improvement is hardly perceptible, due to the fact that the $\lambda(\mathbf{s})$ deformations are already almost optimal (almost vanishing). Between the thresholds, the improvements are visible but not substantial. Above threshold, the picture is slightly mixed. There are some large improvements, especially at very large values of $s$, but at the cost of a small loss of performance at moderate $s$ values. This is likely due to limitations of the training and compromises made between different $s$ and $t$ regions. More parameters and/or more training would reduce the variance further.   

\begin{figure}
\centering
\includegraphics[scale=0.6]{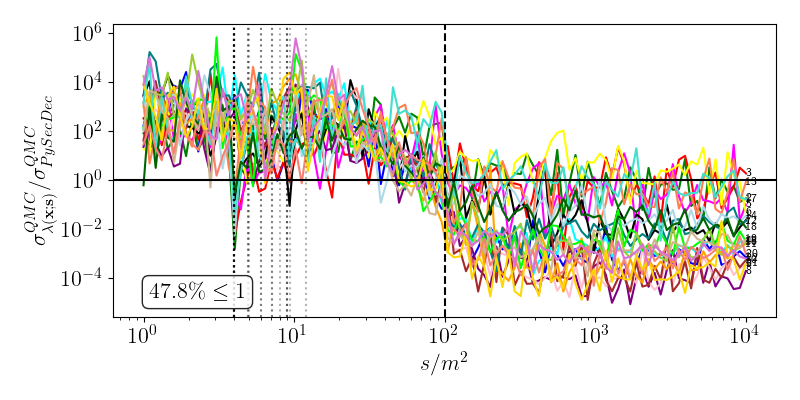}
\includegraphics[scale=0.6]{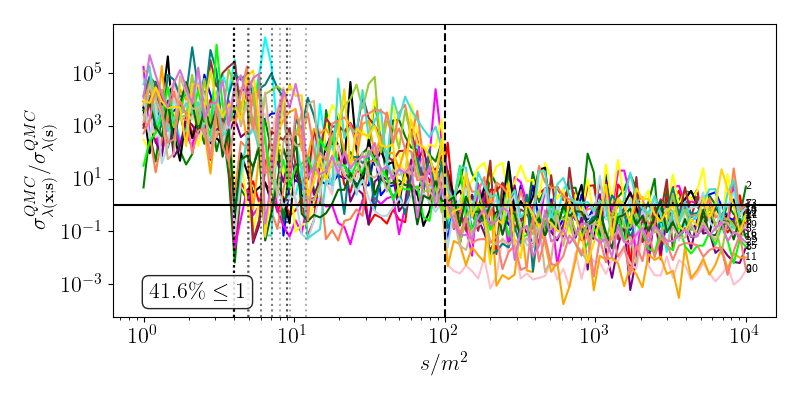}
\caption{Upper plot: Ratio of the $\lambda(\mathbf{x};\mathbf{s})$ deformation QMC error over that of the heuristic $\lambda$ for all 22 sectors of the 2-loop elliptic integral. Lower plot: Ratio of the $\lambda(\mathbf{x};\mathbf{s})$ deformation QMC error over the $\lambda(\mathbf{s})$ variance for the same sectors.}\label{fig:free_over_pysecded_2L_QMC} 
\end{figure}

Figure~\ref{fig:free_over_pysecded_2L_QMC} shows the comparison for the QMC error estimate. Once again, we see that even though the variance is improved, the QMC error estimate is not always improving. The error estimate is significantly worse below the threshold (though very low given the integral is Euclidean in this region). In the intermediate region, the error estimate is worse than either the heuristic-$\lambda$ or $\lambda(\mathbf{s})$ deformations. On the other hand, it is improving over the heuristic-$\lambda$ deformations in the large-$s$ region, and it is a further slight improvement over the $\lambda(\mathbf{s})$ deformation. 

\subsection{Comparison with alternative neural network approach}\label{sec:ramon}

In this section, we compare our approach with that of Ref.~\cite{Winterhalder:2021ngy}, where the authors investigated two \emph{local} ways of determining the deformation. The first is similar to the $\lambda(\mathbf{s})$ deformation, in that the parameters $\lambda$ of the deformation are determined by minimising the variance of the integrand, but unlike in our approach, that minimisation is repeated for each phase-space point. The second combines it with normalising flows to further improve the variance of the integrand. The training of the normalising flow is also performed for each phase-space point independently. We specifically compare their results for the ``elliptic2L'' integral and focus on their Figure 8. This integral corresponds to our sector 19 in the previous section, but with $p_4^2$ not set to the fixed value $12/23$.
With the help of the authors, we reproduce this plot in the top row of  Figure~\ref{fig:ramon_plots}.
Note that Ref.~\cite{Winterhalder:2021ngy} did not use a Korobov transformation in their analysis. 
To facilitate comparison, we also refrain from using one in this section.

\begin{figure}
\includegraphics[scale=0.5]{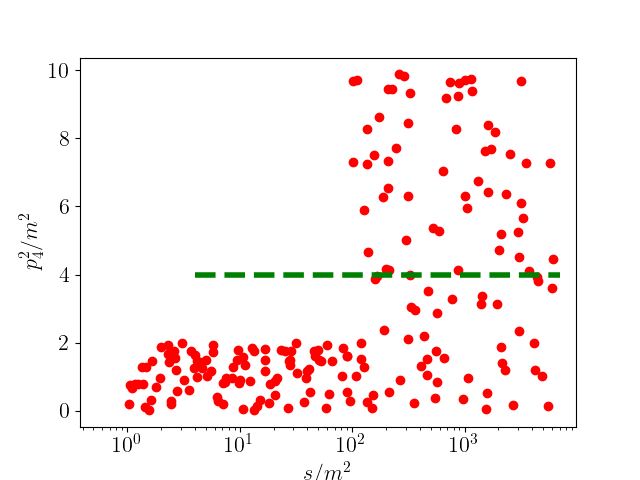}
\includegraphics[scale=0.5]{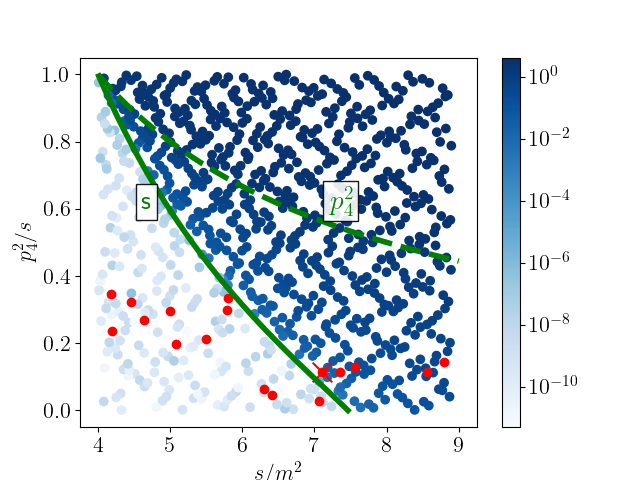}
\caption{Left-hand panel: Distribution of sampled points in Figure 8 from \cite{Winterhalder:2021ngy}. The green line shows the location of the $p_4^2=4m^2$ threshold. Right-hand panel: threshold location for Sector 1 of Ref.~\cite{Winterhalder:2021ngy}. The two green curves are the thresholds in the $s$ and $p_4^2$ channels. The colour scale shows the magnitude of the imaginary part of the integrand. The red dots represent points sampled in Ref.~\cite{Winterhalder:2021ngy} and the cross marks the location of the points giving rise to outliers in Figure~\ref {fig:ramon_plots}.
}
\label{fig:ramon_threshold}
\end{figure}

\begin{figure}
\centering
\includegraphics[scale=0.5]{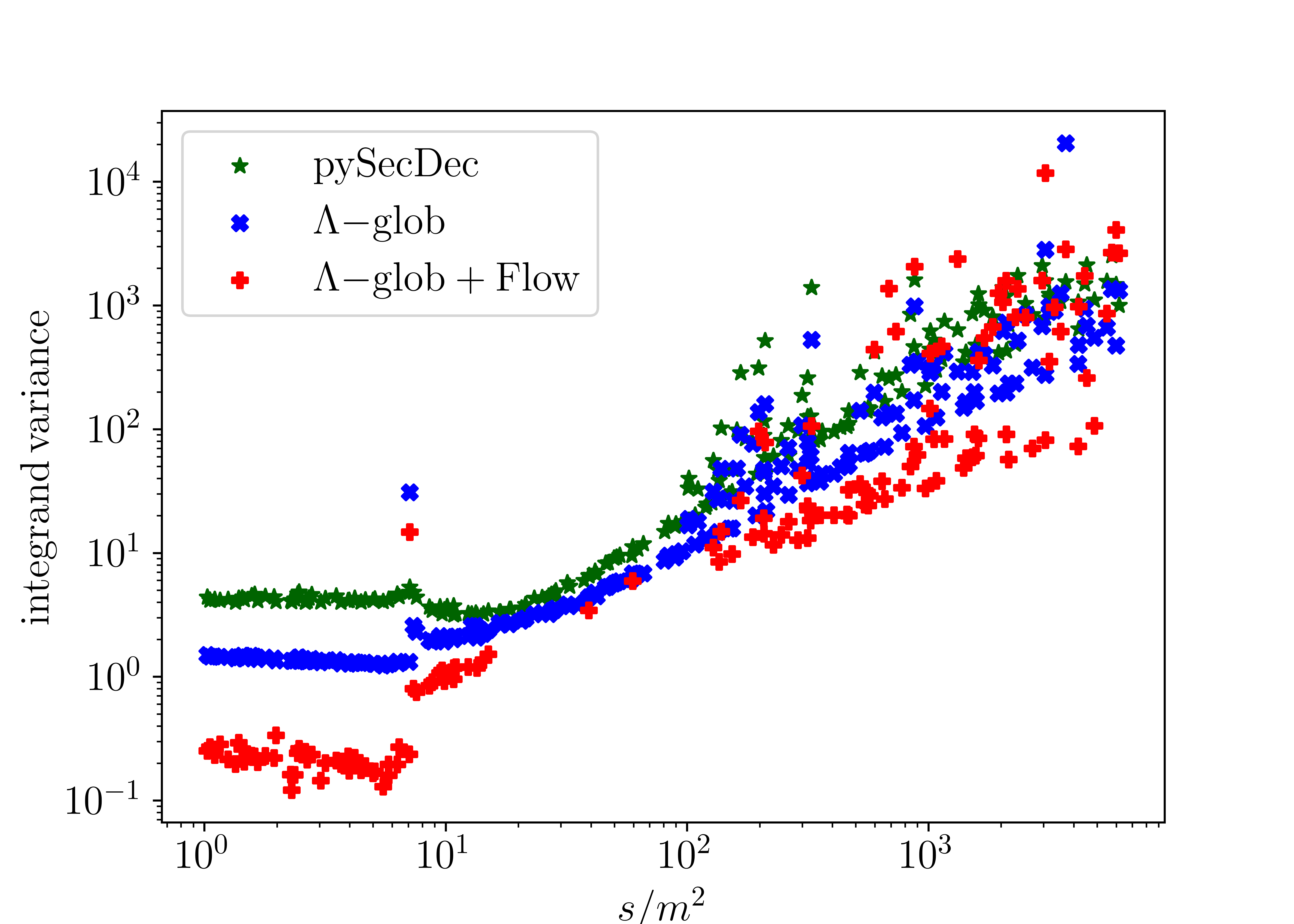}

\includegraphics[scale=0.4]{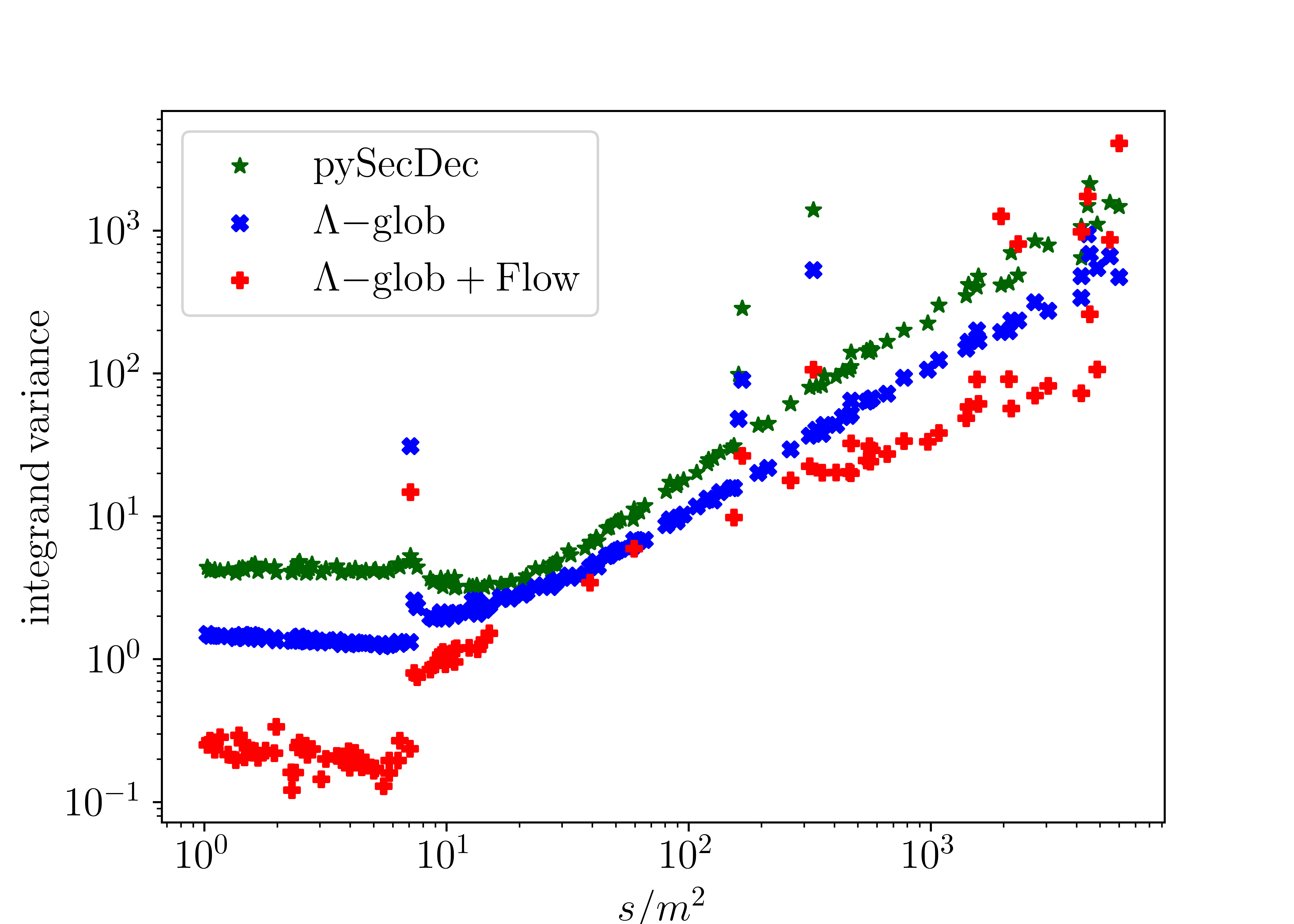}
\includegraphics[scale=0.4]{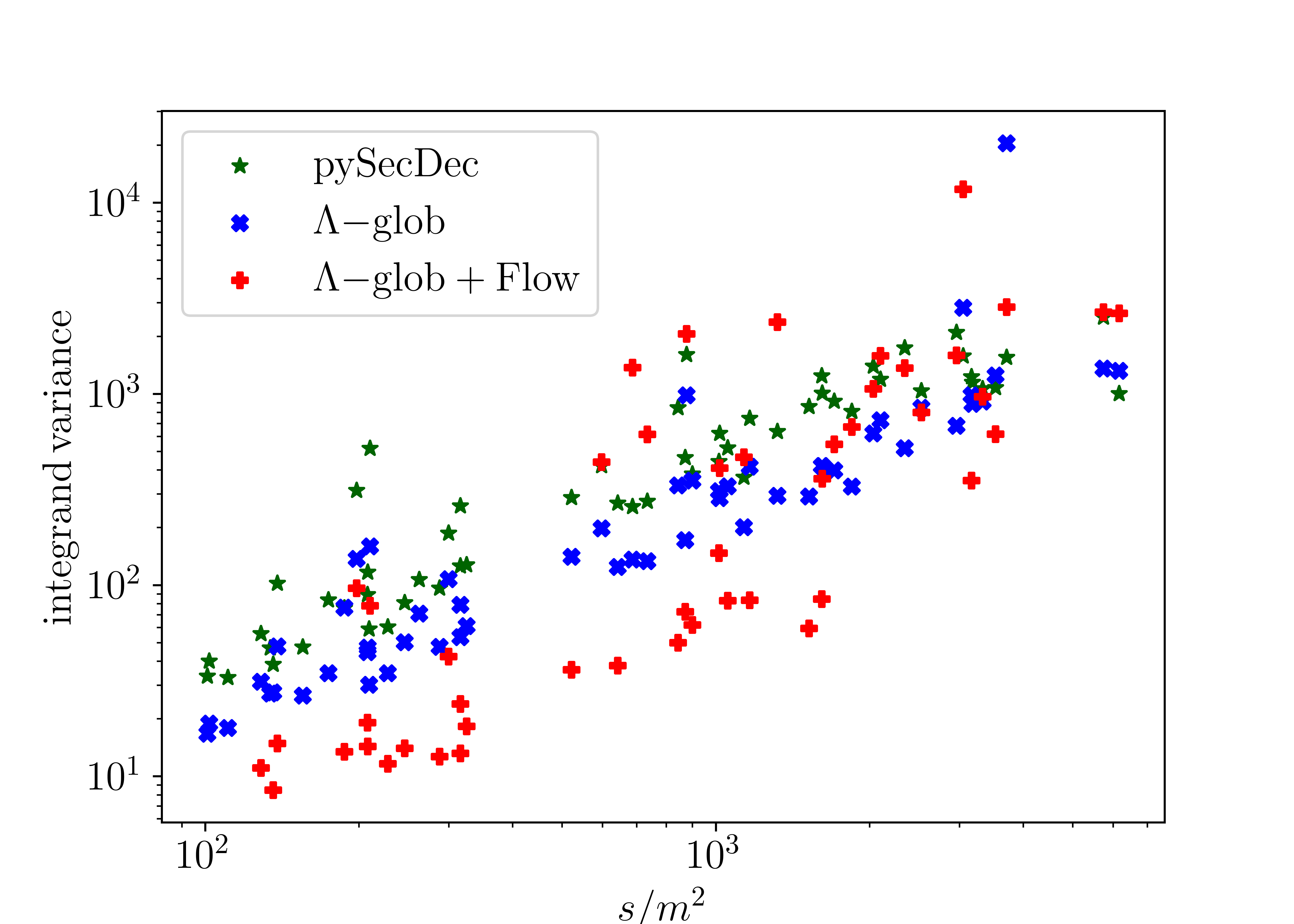}
\includegraphics[scale=0.4]{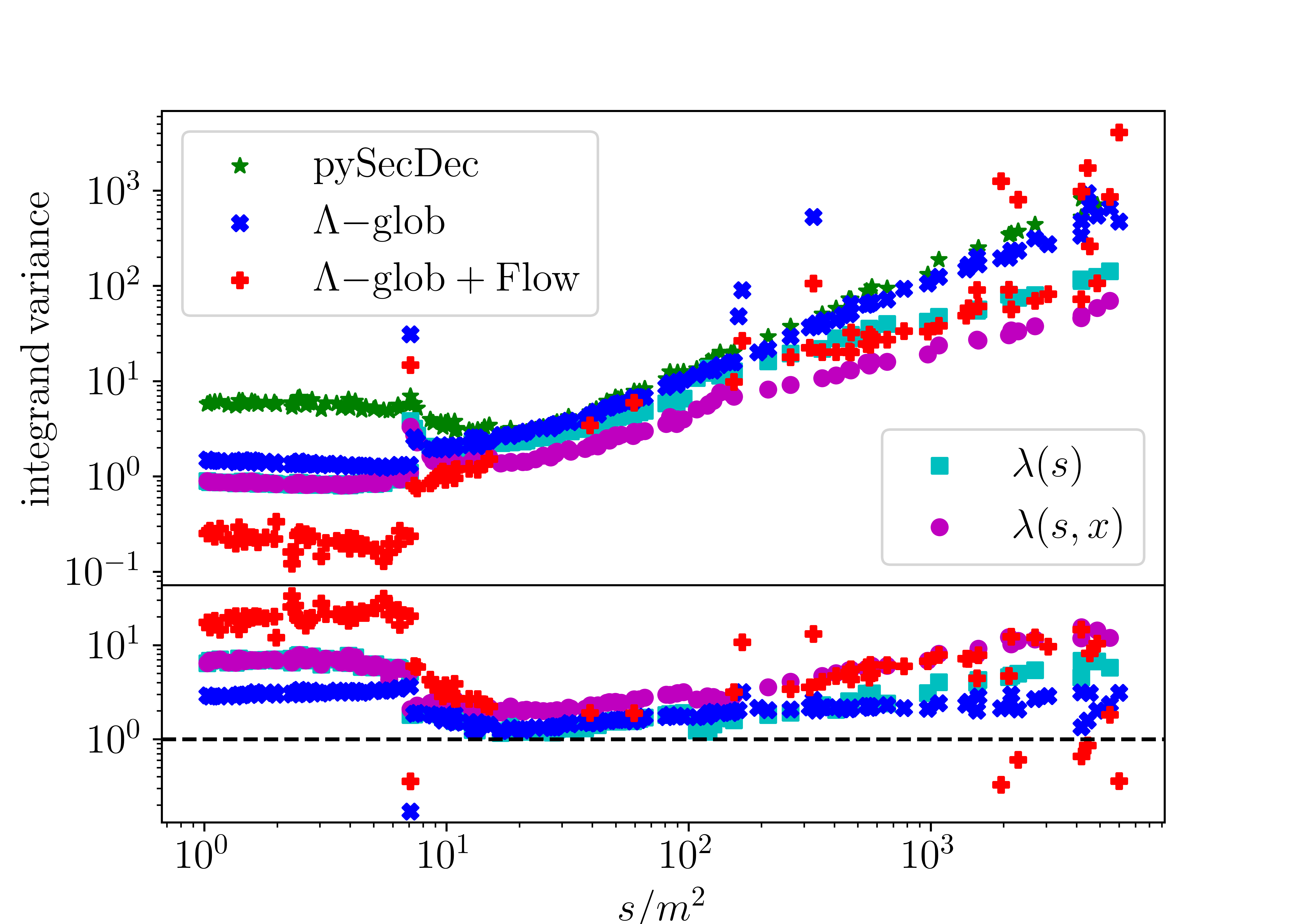}
\includegraphics[scale=0.4]{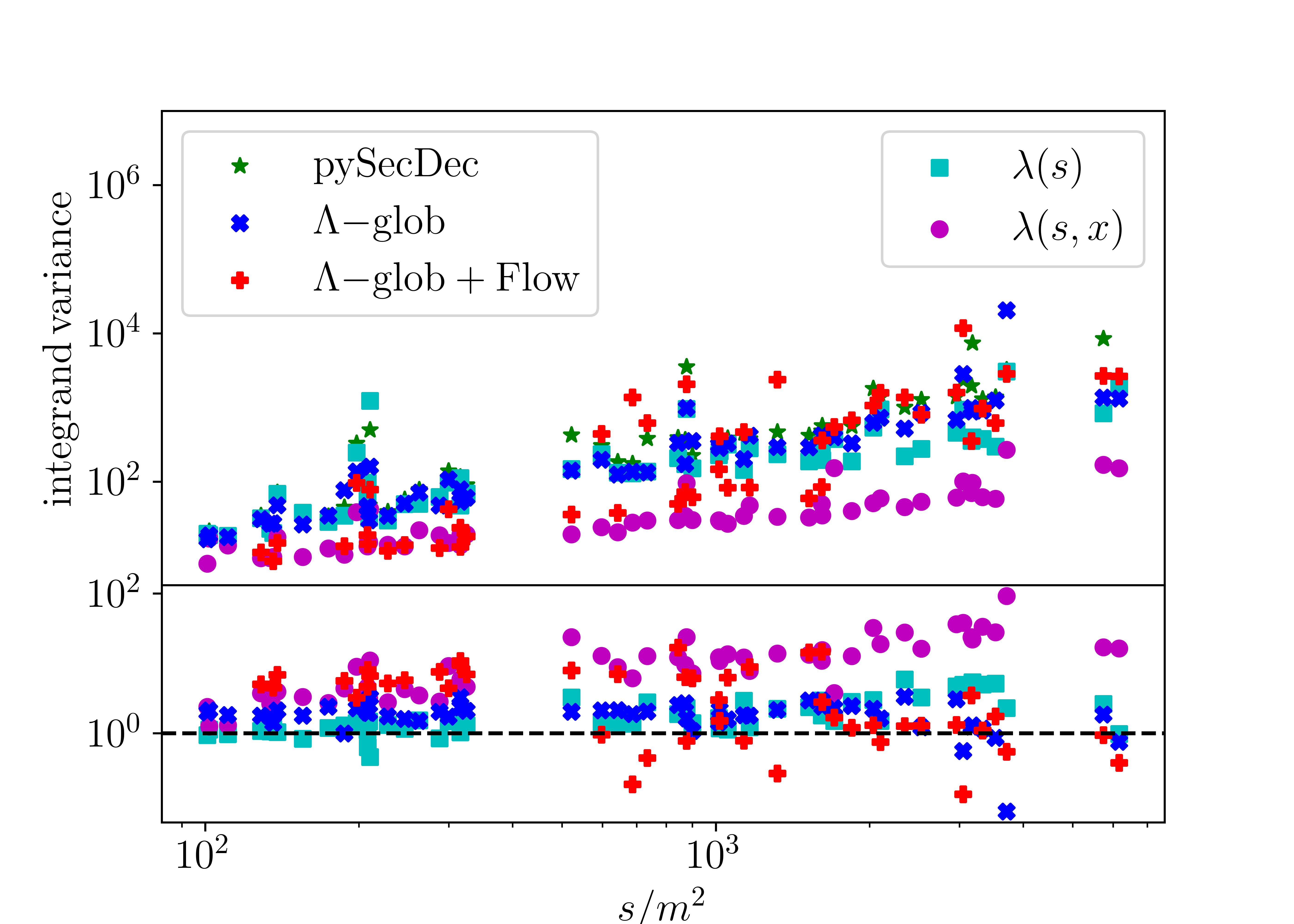}
\caption{Top row: Reproduced Figure 8 from Ref.~\cite{Winterhalder:2021ngy}. The $y$-axis has been rescaled to display the integrand variance rather than the relative error (shown in the original plot). Middle row: data from the top row separated according to the location with respect to the $p_4^2$ threshold. The right pane shows only points for which $p_4^2<4m^2$. The left pane shows those for which $p_4^2>4m^2$, this region was only sampled for values of $s>100$, see Figure~\ref{fig:ramon_threshold}. In the lower row the results using the $\lambda(\mathbf{s})$ and $\lambda(\mathbf{x};\mathbf{s})$ networks, produced in this work, are also shown. Following the choice made in Ref.~\cite{Winterhalder:2021ngy}, the ratio plot shows the variance improvement over the \pysecdec contour, therefore higher is better.}\label{fig:ramon_plots}
\end{figure}

This plot was obtained by setting $m^2=1$ and sampling points in two regions, one with moderate
$s<100$ and one with large values of $100<s<7000$. These points are displayed on the left-hand side of Figure~\ref{fig:ramon_threshold}. In the larger $s$ sample, we observe a larger spread of variances. This is due to the way the phase-space points were chosen. On the right-hand side of Figure~\ref{fig:ramon_threshold}, we plot the value of the external mass $p_4^2$ and $s$. For $s<100$, the external mass was chosen according to $p_4^2<2$, instead, for larger values of $s$, the external mass was varied in the range $p_4^2<100$. 
At the value $p_4^2=4$, a new threshold appears, which explains the different behaviour. If we restrict the plot to those points with values of $p_4^2$ below this threshold, we obtain a more continuous plot, as shown in the middle row of plots in Figure~\ref{fig:ramon_plots}. This threshold is not an issue for the method of Ref.~\cite{Winterhalder:2021ngy}, as each phase-space point is treated independently, but for our method it is important to separate them, as the optimal contour parameters cannot be expected to be continuous over the threshold, and the presence of a divergence in the middle of a training sample can cause significant issues with convergence and lead to non-optimal results.  

Figure~\ref{fig:ramon_threshold} (right-panel) shows the value of the imaginary part of the integrand. 
Phase-space points with a small value of the imaginary part (up to our numerical integration precision) correspond to the Euclidean region, and points with a non-vanishing imaginary part demonstrate the appearance of a threshold. 
The figure shows the two thresholds, the sector decomposition procedure shifts the $s=4$ threshold, while the $p_4^2=4$ threshold remains unshifted. 
The phase-space points selected in Ref.~\cite {Winterhalder:2021ngy} are shown in red. 
The point denoted by a cross, located close to the $s$ threshold, is the outlier in the plot in the top row of Figure~\ref{fig:ramon_plots}. 
This figure explains the existence and location of the outlier. In Appendix~\ref{sec:analytical_threshold} we derive an analytical form for the threshold as a function of $p_4^2$. 

Now that we have gained a better understanding of the features of the plot we wish to compare to, we add the result of our methods in the bottom row of Figure~\ref{fig:ramon_plots}. We observe that our methods yield improved performance, although they are trained once over entire phase-space regions and not repeatedly for each phase-space point.
In the Euclidean region (to the left of the green line on the right side of Figure~\ref{fig:ramon_threshold}), we observe that the $\lambda(\mathbf{s})$ and $\lambda(\mathbf{x};\mathbf{s})$ methods perform similarly to each other and better than the pySecDec heuristic choice and $\Lambda$-glob, but worse than $\Lambda$-glob + Flow, in this region a good contour can be obtained with no deformation.
Immediately above the $s$ threshold, $\lambda(\mathbf{s})$ performs similarly to $\Lambda$-glob and $\lambda(\mathbf{x};\mathbf{s})$ provides better performance, but still slightly worse than $\Lambda$-glob + Flow.
At very large values of $s$, $\lambda(\mathbf{s})$ performs similarly to $\Lambda$-glob + Flow, with $\lambda(\mathbf{x};\mathbf{s})$ providing the best performance.

\section{Discussion}\label{sec:discussion}

One could imagine using a different loss to optimise the QMC integration error estimate rather than the variance of the integrand, but there are some issues with this endeavour:
\begin{itemize}
\item The discussion of Section~\ref{sec:optimal_contour}, where the equivalence of local and regional optima is demonstrated, would need to be established for a loss function based on the QMC error estimate. 
\item The QMC loss has a strong dependence on the size of the lattice, as seen in Figure~\ref{fig:QMCerror_lambda_n} and Figure~\ref{fig:QMC_error_scan_elliptic_2L}, and it is impossible to make a choice of lambda that is simultaneously optimal for all lattice sizes. Furthermore, the QMC error is not a monotonic function of the lattice size. The error estimation of QMC integration is still not completely understood and is the subject of current research, see Ref.~\cite{owen2025errorestimationquasimontecarlo} for a recent status.
\end{itemize}

Using the methods demonstrated in this article as a replacement for the default deformation has the following potential efficiency savings in \pysecdec (and other similar programs):
\begin{itemize}
\item If a valid deformation is given without the need for pre-sampling, the first stage of the integration can be skipped.
\item The computation time associated with iterations of the integration with non-valid deformations is avoided. 
\end{itemize}
The use of the technique is only beneficial if the gains above are amortising the computing time needed to train the networks. 

We considered two types of deformations. The $\lambda(\mathbf{s})$ deformation is a lot easier to integrate into existing tools as only the new values of $\lambda_i$ have to be passed. Indeed, it is now implemented in \pysecdec. The $\lambda(\mathbf{x};\mathbf{s})$ deformation requires providing both the deformed contour and the Jacobian for each lattice point evaluated, and the mechanism to do so would need to be efficiently implemented.

\section{Conclusion}\label{sec:conclusion}

In this article, we have investigated the task of determining a valid contour deformation for a range of kinematic parameters \emph{ab initio}, that is, before numerical integration for a fixed set of kinematic parameters is started.
The ability to determine contours in this manner removes the need to find valid contours on a phase-space-by-phase-space basis, potentially improving the performance of existing tools for the evaluation of Feynman integrals in the physical Minkowski regime.
Furthermore, knowledge of a function that can return valid contours of integration regionally allows integration techniques that require the rapid evaluation of the integrand as a function of the phase-space, such as the neural network technique of Ref.~\cite{Maitre:2022xle}, to be employed.

We have demonstrated that this task can be accomplished using Neural Networks and present two techniques, one which returns the magnitude of the deformation as a function of the phase-space point only, and another that returns for each phase-space point a deformation function depending on the integration variables.
We show that by first finding the thresholds appearing in each sector of the integrand and then training a Neural Network for each kinematic region bounded by these thresholds, it is possible to construct a function that returns contours with optimal variance (i.e. contours which yield an integrand with minimal variance and which perform as well as contours optimised for a specific local phase-space point).
We observe that although such contours are not necessarily optimal for QMC methods of integration, in fact, no contour is optimal for all lattice sizes, the performance of one of the methods of selecting contours is, nevertheless, very competitive with other heuristic and local methods of determining integration contours.

The picture emerging is that the best variance is obtained for the most flexible $\lambda(\mathbf{x};\mathbf{s})$ method, but a smaller QMC error estimate is obtained using the simpler $\lambda(\mathbf{s})$. The latter represents a good compromise between training/evaluation performance across metrics and ease of practical implementation.

We compare the performance of our techniques to the previous work on determining optimal contours, presented in Ref.~\cite{Winterhalder:2021ngy}, and observe that our procedures compare favourably. 
However, in the present work, we need to train a Neural Network only once per region of phase space, rather than individually for each phase-space point.
This means that when evaluating a large number of phase-space points, the potentially substantial time required to train a Neural Network can be offset by the time-saving of not having to optimise the contour for each phase-space point.
We leave a further detailed study of the performance of our contour deformations to future work.  

\section*{Acknowledgments}
It is our pleasure to thank Vitaly Magerya for the discussion of the interplay of sector decomposition and thresholds, and Vitaly Magerya and Ramon Winterhalder for clarifications and the sharing of the raw data and code of Ref.~\cite{Winterhalder:2021ngy}. We also thank the other members of the \pysecdec collaboration for useful discussions; Gudrun Heinrich, Bakar Chargeishvili, Matthias Kerner, Vitaly Magerya and Johannes Schlenk.
This research was supported in part by the UK Science and Technology Facilities
Council under contracts ST/X000745/1 and ST/X003167/1.
SJ is additionally supported by a Royal Society University Research
Fellowship (URF/R1/201268, URF/R/251034).

\bibliography{main,manual}
\bibliographystyle{JHEP}

\appendix
\newpage
\section{Analytical threshold derivation}\label{sec:analytical_threshold}
In this section, we derive an analytic formula for the $p_4^2$ dependence of the $s$ threshold for the integrand of Section~\ref{sec:ramon}. 
We start from the F polynomial
\begin{eqnarray}F&=&m^{2} \left(x_{2} x_{5} + x_{1} x_{5} + x_{3} + x_{4} x_{5} + 1\right) \nonumber\\&&\times\left(x_{2} x_{3} + x_{2} x_{5} + x_{2} + x_{1} x_{3} + x_{1} x_{5} + x_{1} + x_{3} x_{4} + x_{3} + x_{4} x_{5} + x_{4} + 1\right) \nonumber\\&&- p^{2}_{4} \left(x_{2} x_{3} + x_{1} x_{3} + x_{1} x_{5} + x_{3} x_{4} + x_{3}\right) \nonumber\\&&- s x_{2} x_{5} \left(x_{1} x_{3} + x_{1} x_{5} + x_{1} + x_{3}\right) - t x_{4} x_{5}
\end{eqnarray}
and substitute $m^2=1$, $t=-(s-p_4^2)\tau$ and $p_4^2=\gamma s$. 
Using the numerical minimisation described in Section~\ref{sec:elliptic2L}, we observe that the minimum always occurs for $x_1=1$, $x_2=1$, $x_4=0$ and $x_5=1$, leaving us with,
\begin{align}
F=-s (\gamma +(3 \gamma +2) x_3+2)+3 x_3^2+14 x_3+15\;.
\end{align}
We now solve $F=0$ for $s$ to obtain
\begin{align}\label{eq:s_of_x3}
    s=\frac{3 x_3^2+14 x_3+15}{\gamma +3 \gamma  x_3+2 x_3+2}\;,
\end{align}
the minimum value for $s$ is obtained for 
\begin{align}
    x_3 = \frac{4 \sqrt{3} \sqrt{6 \gamma ^2+5 \gamma +1}-3 \gamma -6}{3 (3 \gamma +2)}\;.
\end{align}
Inserting back into Eq.~\eqref{eq:s_of_x3} and simplifying yields
\begin{align}
    s=\frac{4 \left(9 \gamma +2 \sqrt{3 \gamma  (6 \gamma +5)+3}+4\right)}{(3 \gamma +2)^2}\;.
\end{align}
This expression evaluates to $4+2\sqrt{3}\approx7.46$ in the limit of vanishing $p_4^2$ and coincides with the physical threshold of $s=4$ in the limit $p_4^2\rightarrow s$.  
\newpage
\section{Grid size dependence of the RQMC error estimate}\label{sec:minlambda}
In this section, we show that the RQMC error estimate and the optimal integration contour have a strong dependence both on the grid size and on the specific generating vector used for the rank-1 lattice.

Each data point in the figures of this section is obtained by identifying the deformation ($\lambda_i$) that minimises the RQMC error for a given grid size and generating vector. The optimisation was performed using the Nelder-Mead method. To limit the noise in the RQMC error in Eq.~\eqref{eq:qmc_variance}, we use 1000 shifts, which are kept fixed during the optimisation. After optimisation, we record the location of the optimal deformation parameters $\lambda_i$ and the value of the RQMC error estimate. This procedure is repeated 16 times for each grid size, each time a new  generating vector for the lattice is determined using the median method using 21 test generating vectors. 

\begin{figure}
\centering
\includegraphics[scale=0.5]{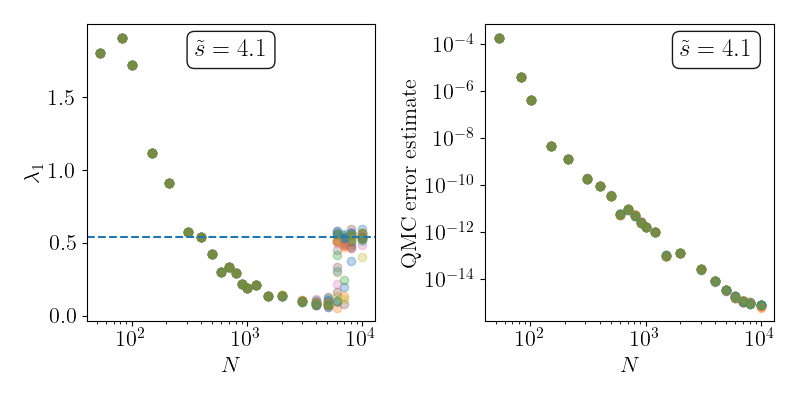}
\includegraphics[scale=0.5]{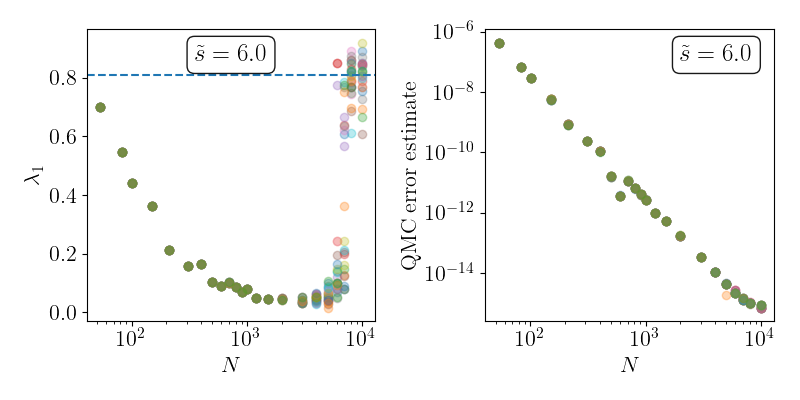}
\includegraphics[scale=0.5]{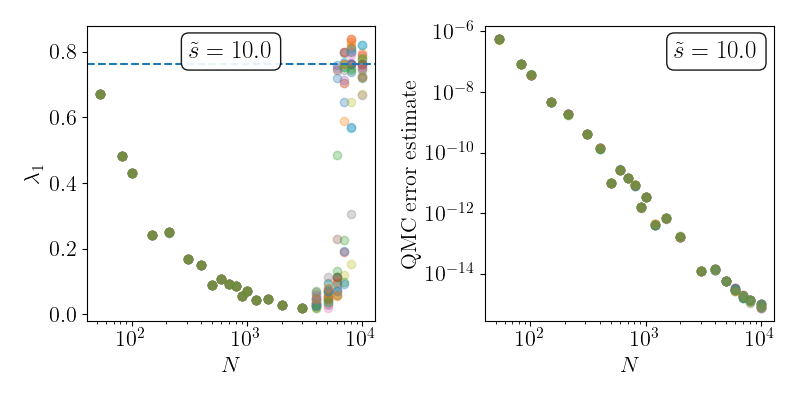}
\includegraphics[scale=0.5]{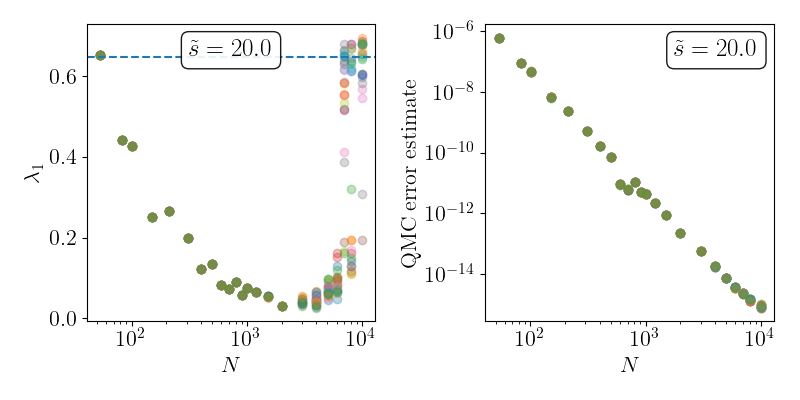}
\caption{Optimal deformation parameter for the 1-loop bubble example for different values of $\tilde{s}$. The left panels show the optimal $\lambda$, the right panels show the variance of the 1000 shift results. The dashed line on the plots on the left shows the optimal value for the variance.
}\label{fig:true_min_1loop}
\end{figure}

In Figure~\ref{fig:true_min_1loop} we show the location of the optimal deformation for various RQMC lattices as a function of the grid size for the case of the one-loop bubble integral discussed in Section~\ref{sec:bubble1L}. For moderate values of $N$, the dependence on the generating vector is very weak (all 16 points coincide), but for larger values, the dependence begins to appear. The return to the variance-optimal value at large grid sizes is because the optimisation starts at that value, and for large values of $N$ the QMC integration reaches the numerical accuracy floor of 64-bit floating point operations, as is visible in Figure~\ref{fig:QMCerror_lambda_n}.

We observe that the best choice of $\lambda_1$ strongly depends on the grid size. 
Since this integrand depends only on a single $\lambda$, there is no possibility of finding a contour described by a fixed value of $\lambda_1$ that consistently minimises the RQMC error independently of $N$.
Therefore, for the median lattice rules we use for integration, the optimal contour can not be found by minimising any quantity independent of $N$, such as the variation, $V[f]$, or a function norm, $||f||_H$.

\begin{figure}
\centering
\includegraphics[scale=0.35]{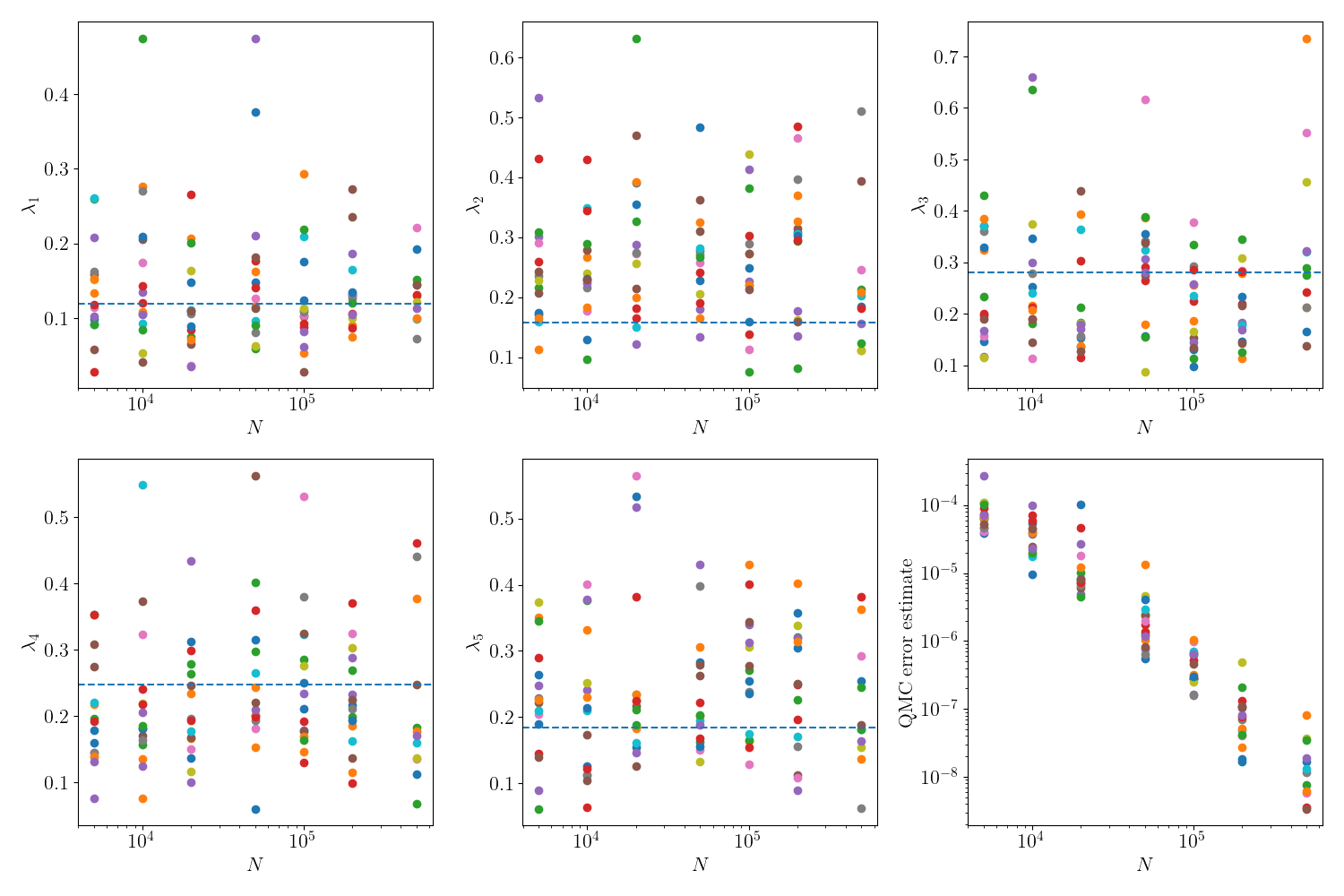}
\caption{First 5 panels: Optimal deformation location for 16 different generating vectors. The dashed line represents the value of the deformation parameters $\lambda_i$ that optimise the variance. Points of the same colour for the same value of $N$ correspond to the same generating vector and correspond to the point with that colour in the sixth panel. Points with the same colour for different values of $N$ are uncorrelated. Last panel: Value of the RQMC error estimate at the optimal location.}\label{fig:min_lambda_RQMC}
\end{figure}

We now turn to the 2-loop example discussed in Section~\ref{sec:elliptic2L}. The first 5 panels of Figure~\ref{fig:min_lambda_RQMC} show the location of the optimal deformation parameters for 16 different generating vectors as a function of the grid size. This figure shows that the locations of the optimal contour parameters $\lambda_i$ are strongly dependent on the generating vector. The last panel also shows that the error estimate exhibits a large dependence on the generating vector.     
\newpage
\section{Comparison of $\lambda_i$ values}\label{sec:lambda_plots}
\begin{figure}
\includegraphics[scale=0.55]{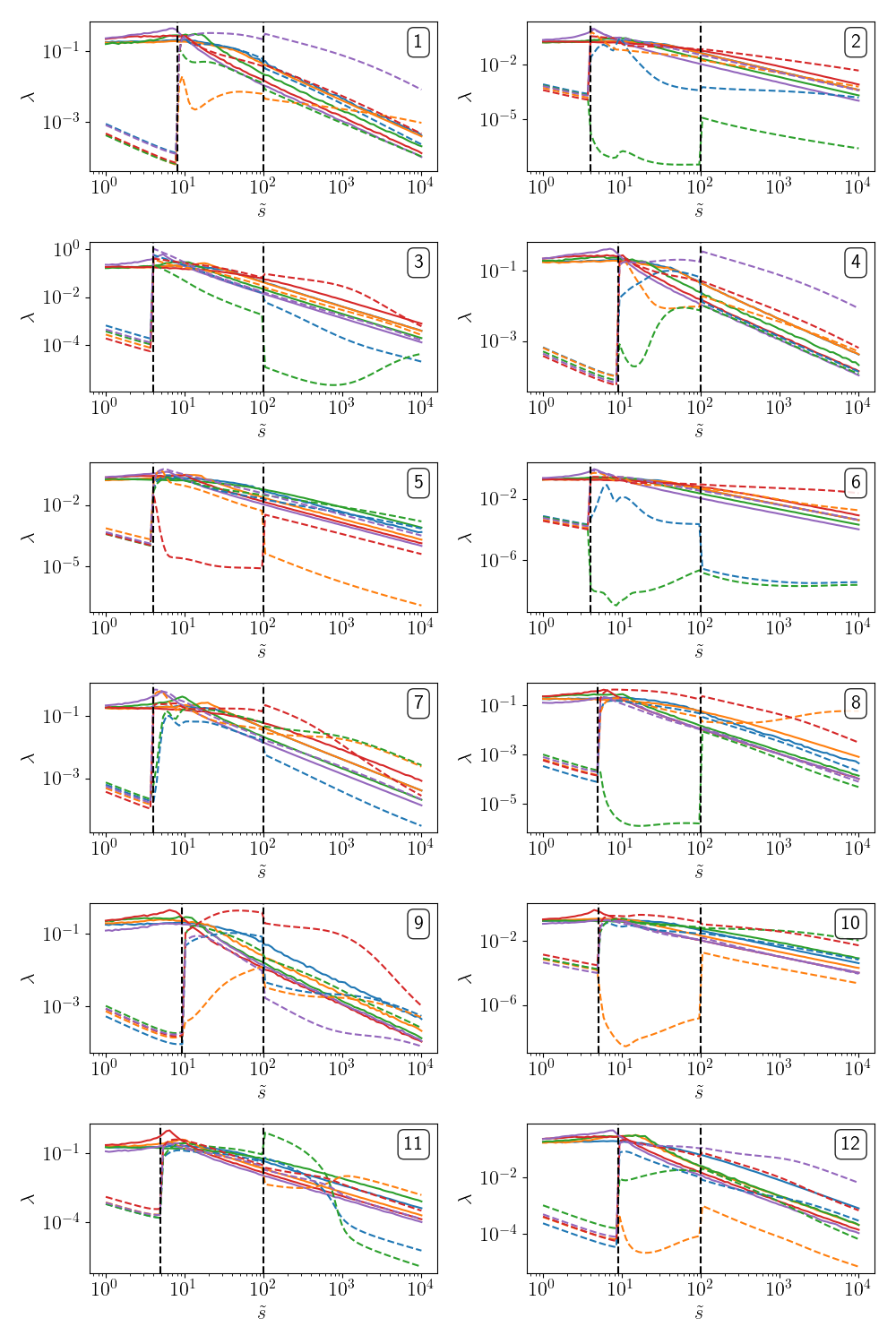}
\caption{Comparison between the $\lambda$ values for \pysecdec (solid) and the $\lambda(\mathbf{s})$ values (dashed). Curves of the same colour correspond to the deformation strength for the same coordinate. The sector number is displayed in the upper right of each plot. The vertical dashed lines show the separation between different trainings, where discontinuities can be expected. 
}\label{fig:lambda_comparison_A}
\end{figure}

\begin{figure}
\includegraphics[scale=0.55]{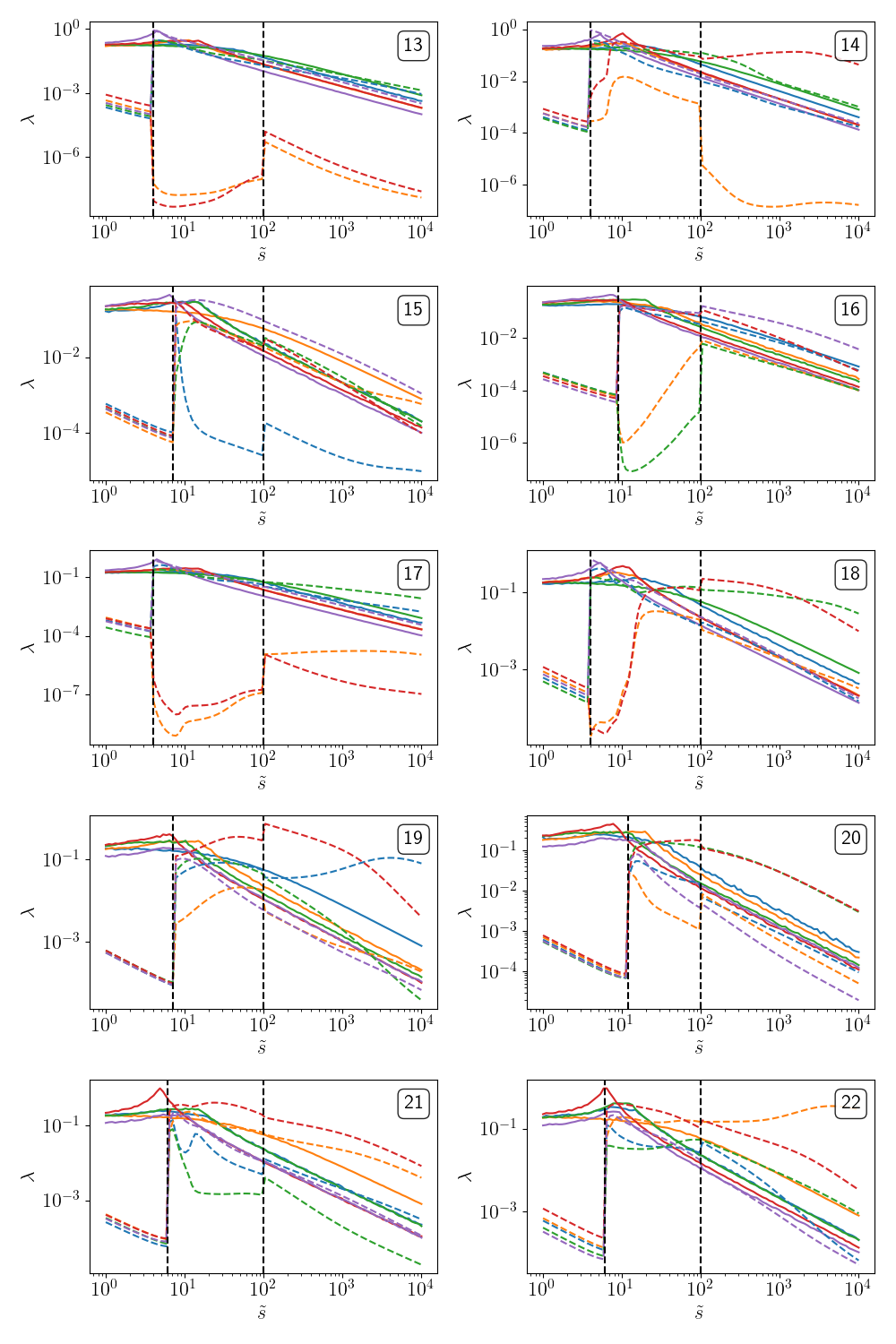}
\caption{Comparison between the $\lambda$ values for \pysecdec (solid) and the $\lambda(\mathbf{s})$ values (dashed). Curves of the same colour correspond to the deformation strength for the same coordinate. The sector number is displayed in the upper right of each plot. The vertical dashed lines show the separation between different trainings, where discontinuities can be expected. 
}\label{fig:lambda_comparison_B}
\end{figure}

Figure~\ref{fig:lambda_comparison_A} and Figure~\ref{fig:lambda_comparison_B} show a comparison of the values of $\lambda_i$ chosen by \pysecdec and those chosen by the $\lambda(\mathbf{s})$ networks.

\clearpage
\section{RQMC error validation}
We validate the fact that the RQMC error estimate Eq.~\eqref{eq:qmc_variance} is reliable by calculating the residuals
\begin{align}\label{eq:RQMC_residue}
z=\frac{e_i-t_i}{\sigma_i}
\end{align}
where $e_i$ is the mean of the $m$ integration results, $t_i$ is true reference value and $\sigma_i$ is the QMC error estimate from \ref{eq:qmc_variance}. Where the integral has more than one sector the errors of each sector is added in quadrature to obtain the error for the total integral. A reliable error estimator gives a Gaussian distribution for the residuals. The following plots show that this is the case for our three examples for all choices of deformation.

Figure~\ref{fig:bubble_residuals} shows the residuals for the one-loop bubble integral. They were obtained using the median grid method with grids of size $N=2000$, as larger grids yield results that are dominated by the floating-point precision. 

\begin{figure}
\includegraphics[scale=0.36]{figs/bubble_RQMC_validation.png}
\caption{Residuals for the one-loop bubble integral.}\label{fig:bubble_residuals}
\end{figure}

In Figure~\ref{fig:triangle_residuals} we show the residuals for the one-loop triangle example. They were obtained with a grid of size $N=20,000$ The reference values were obtained using the known analytic formula.

\begin{figure}
\includegraphics[scale=0.36]{figs/RQMC_validation.png}
\caption{Residuals for the total one-loop triangle.}\label{fig:triangle_residuals}
\end{figure}

In Figure~\ref{fig:elliptic_residuals} we show the residuals for the two-loop box example. The reference values were obtained using {\tt AMFlow}~\cite{Liu:2017jxz,Liu:2021wks,Liu:2022chg,Huang:2026rjb} with a requested precision of 20 digits. The required integration-by-parts reduction was obtained using {\tt Kira}~\cite{Maierhofer:2017gsa,Klappert:2020nbg,Lange:2025fba}.

\begin{figure}
\includegraphics[scale=0.36]{figs/elliptic_RQMC_validation.png}
\caption{Residuals for the total two-loop box.}\label{fig:elliptic_residuals}
\end{figure}

\FloatBarrier

\section{Numerical threshold detection}\label{sec:threshold_finder}
\begin{figure}
\centering
\includegraphics[scale=0.5]{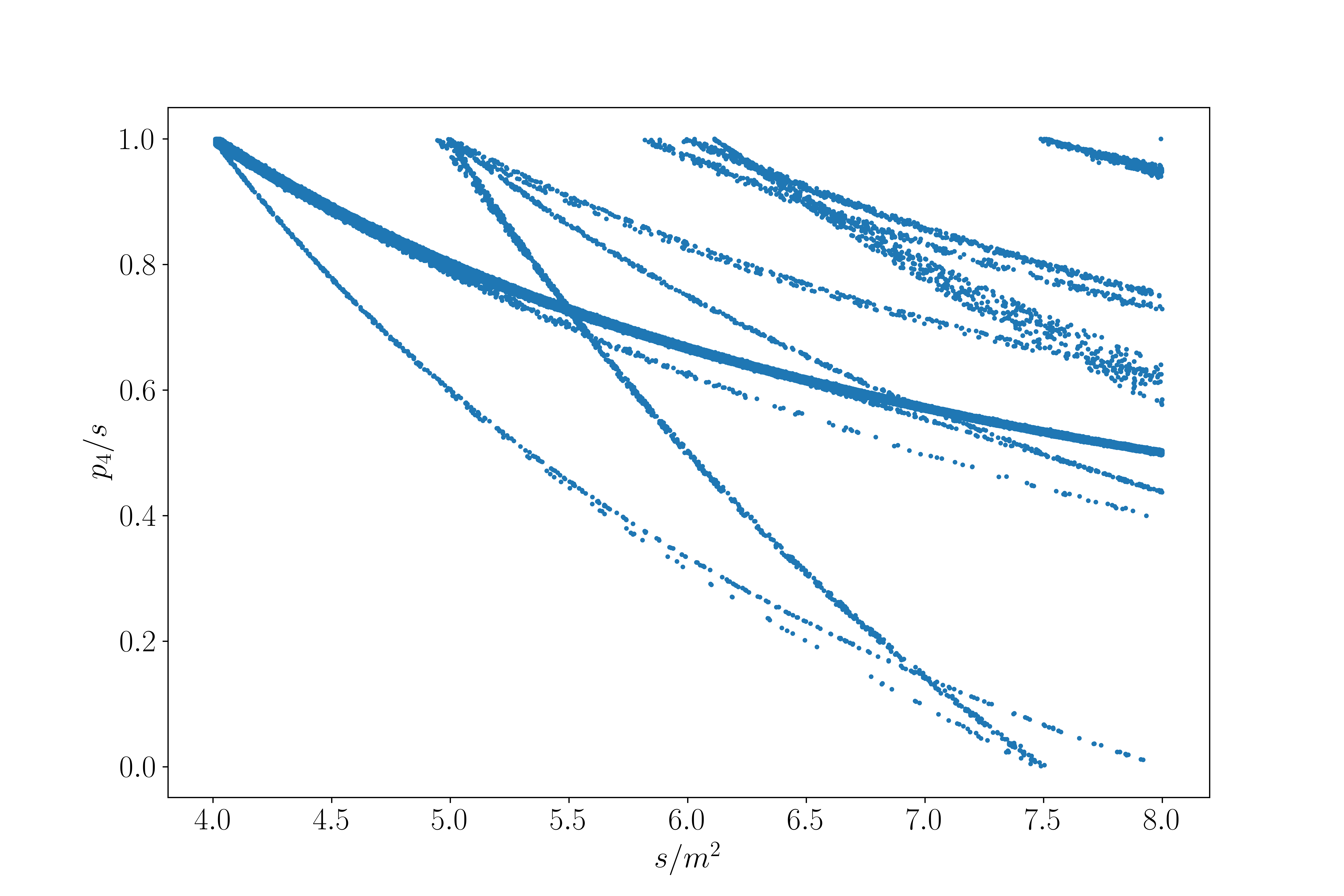}
\includegraphics[scale=0.5]{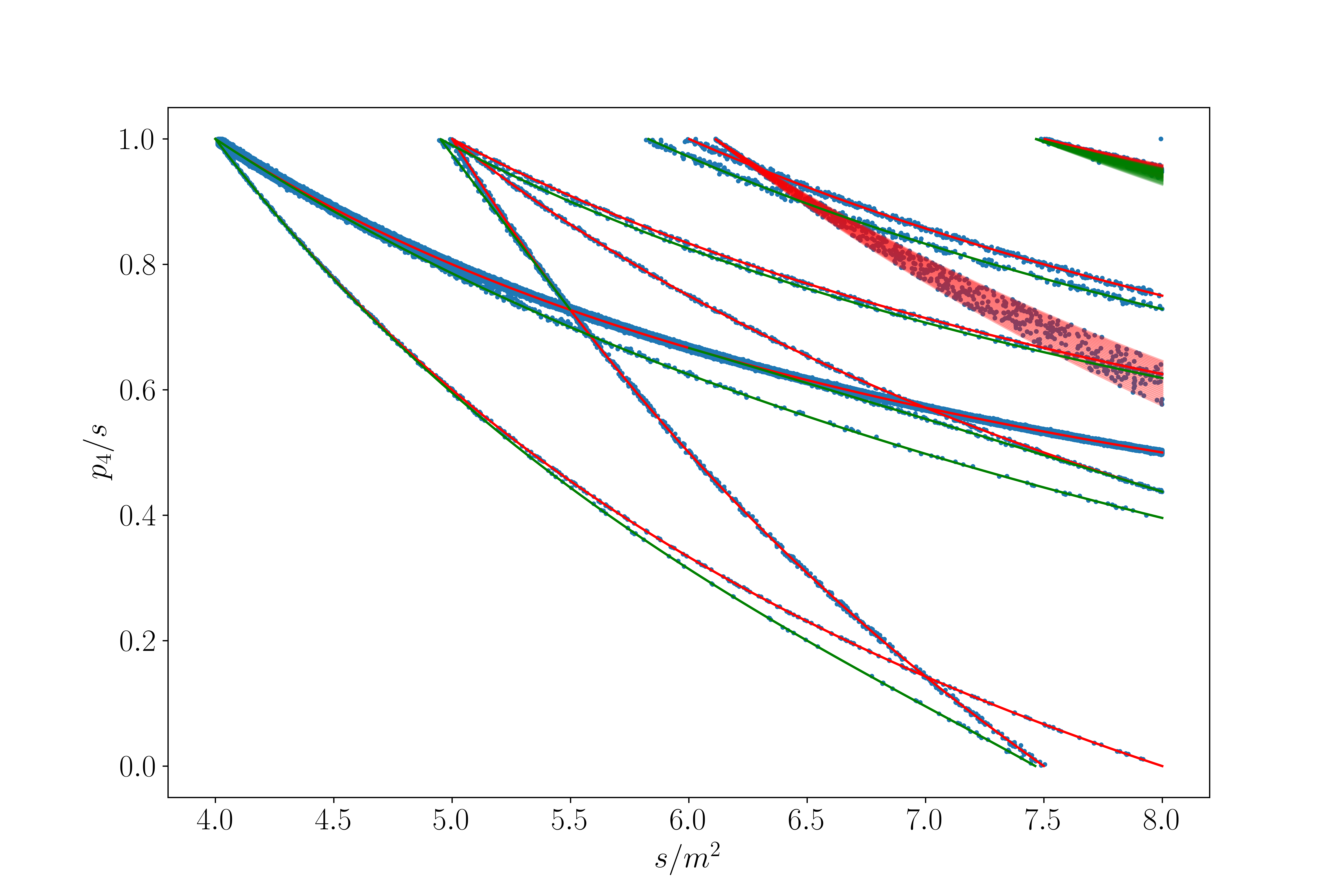}
\caption{Upper plot: Small-$T$ points identified empirically. Lower plot: Small-$T$ points with superimposed analytical curves for the zeros of the $T$ function. Red curves correspond to solutions with all $x$ variables set to either 0 or 1. Green curves correspond to solutions where all but one of the $x_i$ are either $0$ or $1$. The curves with a $\tau$ dependence are shown as shaded areas.}\label{fig:thresholdfinder}
\end{figure}

In this appendix, we illustrate the empirical detection of thresholds through the monitoring of the quantity 
\begin{align}
T=\frac{1}{i(\mathbf{x};\mathbf{s})} + \sum\limits_{i=1}^{N}x_i(1-x_i)\frac{\partial F}{\partial x_i}\;,
\end{align}
introduced in Section~\ref{sec:thresholds}. Sampling the integrand of Section~\ref{sec:ramon} and selecting points with $T<0.001$, we obtain the upper plot in Figure~\ref{fig:thresholdfinder}.

Examination of the data shows that most of the identified points correspond to corners of the $x_i$ phase space. These are integrable singularities which are not affected by the contour deformation as they vanish at the boundary of the integration region. The only singularity that can occur in the bulk of the $x$-space is the one corresponding to the physical threshold $p_4^2-4m^2$. 

With the knowledge gained from numerical sampling that low-$T$ points occur at $x$-space boundaries, the equation $T=0$ can be solved analytically at the boundaries detected by the numerical sampling as a cross-check. The lower plot of Figure~\ref{fig:thresholdfinder} shows the empirical data with superimposed analytical curves of zeros of $T$.  

\begin{figure}
\centering
\includegraphics[scale=0.5]{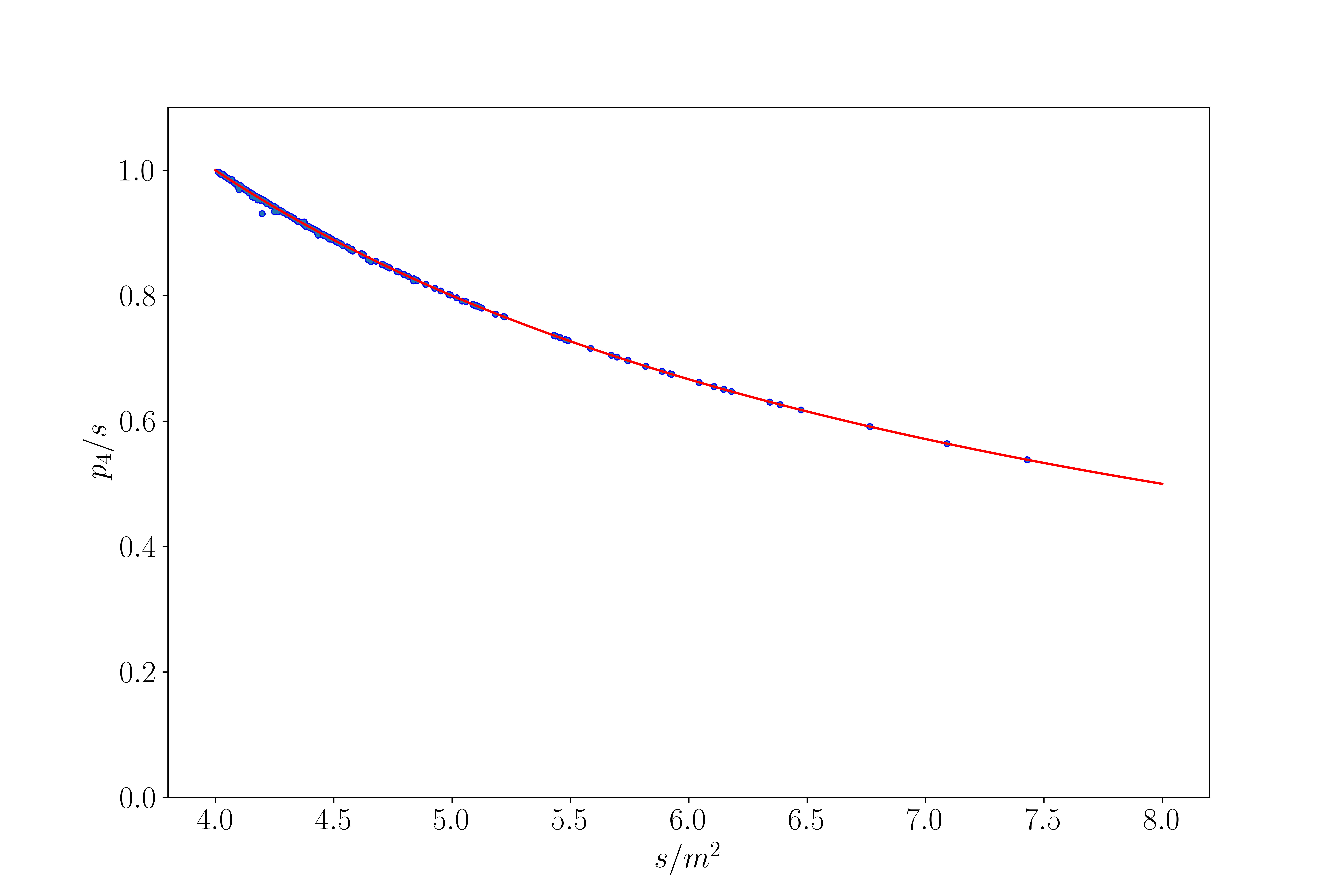}
\caption{Small-$T$ points identified empirically for the two-loop box integrand using the Korobov transform. The red curves corresponds to the physical threshold $p_4^2=4m^2$.}\label{fig:thresholdfinder_k3}
\end{figure}

A Korobov transform, as defined in Section~\ref{sec:montecarlo}, regularises the integrand at the boundary of the integration region, such that if such a transformation had been applied (we did not because the results we compared with did not) the only region to trigger the small-$T$ test is the physical threshold $p_4^2=4m^2$, as shown in Figure~\ref{fig:thresholdfinder_k3}. In practice, since we expect the behaviour of the optimal $\lambda$ to change at the first appearance of an imaginary part, we separate the training both according to the leftmost threshold, as derived in Appendix~\ref{sec:analytical_threshold} and the physical $p_4^2=4m^2$.

\end{document}

%% file: all.tikzstyles
\tikzstyle{blank}=[fill=white, shape=circle, draw=white, inner sep=0.8pt]
\tikzstyle{dot}=[fill=black, shape=circle, draw=black, inner sep=0.8pt]
\tikzstyle{fat}=[fill=blue, shape=circle, draw, line width=1pt, inner sep=0.8pt, minimum size=4mm]
\tikzstyle{blob}=[fill=gray, shape=circle, draw=black, line width=1pt, inner sep=0.8pt, minimum size=1cm]
\tikzstyle{star}=[fill=black, shape=star, inner sep=0.8pt, minimum size=5mm]
\tikzstyle{arrow}=[{-{Classical TikZ Rightarrow[length=2mm,width=1.5mm]}}, draw={rgb,255: red,61; green,171; blue,83}, line width=1pt, preaction={{draw=white,line width=2pt}}, line cap=rect]

\tikzstyle{gluoncoil}=[-, decorate, decoration={coil,aspect=1.4,segment length=2.5mm}]
\tikzstyle{fermion}=[-, draw={rgb,255: red,176; green,36; blue,39}, line width=1pt, preaction={{draw=white,line width=2pt}}, line cap=rect, postaction=decorate, decoration={markings,mark=at position .60 with {\arrow{Stealth[round,width=5pt]}}}]
\tikzstyle{scalar}=[-, line width=1pt, dashed, draw]
\tikzstyle{fermionarrow}=[-, postaction=decorate, decoration={markings,mark=at position .60 with {\arrow{Stealth[round,width=5pt]}}}]
\tikzstyle{ct scalar}=[-, line width=1pt, dashed, draw={rgb,255: red,102; green,102; blue,102}, postaction=decorate, decoration={markings,mark=at position .50 with {\node[style=star,minimum size=5mm]{};}}]
\tikzstyle{ct fermion}=[-, line width=1pt, preaction={{draw=white,line width=2pt}}, line cap=rect, postaction=decorate, decoration={markings,mark=at position 0.25 with {\arrow{Stealth[round,width=5pt]}}, mark=at position 0.75 with {\arrow{Stealth[round,width=5pt]}}, mark=at position .50 with {\node[style=star,minimum size=5mm]{};}}]

%% file: declarations.tex
\usepackage{bbding}
\usepackage{tikz}
\usetikzlibrary{arrows.meta}
\usetikzlibrary{calc}
\usetikzlibrary{decorations.markings}
\usetikzlibrary{decorations.shapes}
\usetikzlibrary{decorations.pathmorphing}
\usetikzlibrary{decorations.pathreplacing}
\usetikzlibrary{external}
\usetikzlibrary{fit}
\usetikzlibrary{positioning}
\usetikzlibrary{shapes.misc}
\usetikzlibrary{positioning}
\usetikzlibrary{external}
\usetikzlibrary{shapes.geometric}

\tikzset{baseline=-0.6ex}
\tikzset{inner sep=0}

\pgfdeclarelayer{nodelayer}
\pgfdeclarelayer{edgelayer}
\pgfsetlayers{edgelayer,main,nodelayer}
\tikzstyle{none}=[]

\tikzset{baseline=0pt}
\tikzset{inner sep=0}

\tikzset{every picture/.style={execute at end picture={\node[fit=(current bounding box),inner sep=0.1mm]{};}}}

\tikzstyle{none}=[]

\definecolor{SapGreen}{HTML}{3B7B3B}
\definecolor{PhtaloGreen}{HTML}{2b7a76}
\definecolor{EmeraldGreen}{HTML}{1ea78d}
\definecolor{EnglishRed}{HTML}{b02427}
\definecolor{IronOxideRed}{HTML}{a13634}
\definecolor{CadmiumRed}{HTML}{df2b3f}
\definecolor{Vermillion}{HTML}{e14235}
\definecolor{PropagatorColor}{HTML}{c0c0c0}
\definecolor{LegColor}{HTML}{8d6566}
\definecolor{TopPropagatorColor}{HTML}{1165ef}
\definecolor{TopLegColor}{HTML}{5571a2}
\definecolor{HiggsLegColor}{HTML}{77679b}

\tikzset{every loop/.style={min distance=10mm}}

\tikzstyle{blank}=[fill=white, shape=circle, draw=white, inner sep=0.8pt]
\tikzstyle{dot}=[fill=black, shape=circle, draw=black, inner sep=0.8pt]
\tikzstyle{fat}=[fill=white, shape=circle, draw={rgb,255: red,176; green,36; blue,39}, dashed, line width=1pt, inner sep=0.8pt]

\tikzstyle{edge}=[-, draw=PropagatorColor, line width=1.2pt, line cap=rect, preaction={{draw=white,line width=2.5pt}}]
\tikzstyle{incoming edge}=[line width=1pt, line cap=rect, draw={rgb,255: red,102; green,102; blue,102}, {|-}]
\tikzstyle{outgoing edge}=[line width=1pt, line cap=rect, draw={rgb,255: red,102; green,102; blue,102}, ->]
\tikzstyle{external edge}=[line width=1pt, line cap=rect, draw={rgb,255: red,102; green,102; blue,102}, -]
\tikzstyle{top}=[-, draw=TopPropagatorColor, line width=1.9pt, line cap=rect, preaction={{draw=white,line width=2.5pt}}]
\tikzstyle{edge dot1}=[-, postaction=decorate, decoration={markings,mark=at position .50 with {\node[style=dot]{};}}]
\tikzstyle{edge dot2}=[-, postaction=decorate, decoration={markings,mark=between positions 0.33 and 0.67 step 0.33 with {\node[style=dot]{};}}]
\tikzstyle{edge dot3}=[-, postaction=decorate, decoration={markings,mark=between positions 0.25 and 0.76 step 0.25 with {\node[style=dot]{};}}]
\tikzstyle{edge dot4}=[-, postaction=decorate, decoration={markings,mark=between positions 0.20 and 0.81 step 0.20 with {\node[style=dot]{};}}]
\tikzstyle{dot1}=[-, draw=none, postaction=decorate, decoration={markings,mark=at position .50 with {\node[style=dot]{};}}]
\tikzstyle{dot2}=[-, draw=none, postaction=decorate, decoration={markings,mark=between positions 0.33 and 0.67 step 0.33 with {\node[style=dot]{};}}]
\tikzstyle{dot3}=[-, draw=none, postaction=decorate, decoration={markings,mark=between positions 0.25 and 0.76 step 0.25 with {\node[style=dot]{};}}]
\tikzstyle{dot4}=[-, draw=none, postaction=decorate, decoration={markings,mark=between positions 0.20 and 0.81 step 0.20 with {\node[style=dot]{};}}]
\tikzstyle{incoming}=[line width=1pt, line cap=rect, draw=LegColor, {|-}]
\tikzstyle{outgoing}=[line width=1pt, line cap=rect, draw=LegColor, ->]
\tikzstyle{outgoing top}=[line width=1pt, line cap=rect, draw=TopLegColor, ->]
\tikzstyle{outgoing higgs}=[line width=1pt, line cap=rect, draw=HiggsLegColor, ->]
\tikzstyle{edge}=[-, draw={rgb,255: red,176; green,36; blue,39}, line width=1pt, preaction={{draw=white,line width=2pt}}, line cap=rect]
\tikzstyle{massive edge}=[-, draw={rgb,255: red,11; green,60; blue,191}, line width=2.0pt, preaction={{draw=white,line width=2.5pt}}, line cap=rect]
\tikzstyle{massive2 edge}=[-, draw={rgb,255: red,59; green,196; blue,114}, line width=2.0pt, preaction={{draw=white,line width=2.5pt}}, line cap=rect]
\tikzstyle{cut edge}=[-, draw={rgb,255: red,64; green,64; blue,64}, line width=0.5pt, densely dashed, line cap=rect]
\tikzstyle{xcut edge}=[-, draw={rgb,255: red,64; green,64; blue,64}, line width=0.5pt, densely dashed, line cap=rect, postaction={decorate, decoration={markings,mark=at position .50 with {\node[cross out,solid,draw=white,line width=2pt,inner sep=1.4pt,transform shape] {};}}}, postaction={decorate, decoration={markings,mark=at position .50 with {\node[cross out,solid,draw={rgb,255: red,176; green,36; blue,39},line width=1pt,inner sep=1.8pt,transform shape] {};}}}]
\tikzstyle{xcut edge 1/3}=[-, draw={rgb,255: red,64; green,64; blue,64}, line width=0.5pt, densely dashed, line cap=rect, postaction={decorate, decoration={markings,mark=at position .33 with {\node[cross out,solid,draw=white,line width=2pt,inner sep=1.4pt,transform shape] {};}}}, postaction={decorate, decoration={markings,mark=at position .33 with {\node[cross out,solid,draw={rgb,255: red,176; green,36; blue,39},line width=1pt,inner sep=1.8pt,transform shape] {};}}}]
\tikzstyle{cut}=[-, draw={rgb,255: red,61; green,171; blue,83}, line width=0.7pt, dotted, line cap=rect]
\tikzstyle{photon}=[-, draw={rgb,255: red,176; green,36; blue,39}, line width=1pt, preaction={{draw=white,line width=2pt}}, line cap=rect, decorate, decoration=snake]
\tikzstyle{gluon}=[-, draw={rgb,255: red,176; green,36; blue,39}, line width=1pt, preaction={{draw=white,line width=2pt}}, line cap=rect, decorate, decoration={coil,aspect=1.4,segment length=2.5mm}]
\tikzstyle{gluoncoil}=[-, decorate, decoration={coil,aspect=1.4,segment length=2.5mm}]
\tikzstyle{fermion}=[-, draw={rgb,255: red,176; green,36; blue,39}, line width=1pt, preaction={{draw=white,line width=2pt}}, line cap=rect, postaction=decorate, decoration={markings,mark=at position .60 with {\arrow{stealth[round]}}}]
\tikzstyle{ghost}=[-, style=fermion, line width=1pt, line cap=round, dash pattern={on 0pt off 3\pgflinewidth}]
\tikzstyle{scalar}=[-, line width=1pt, densely dashed, draw={rgb,255: red,102; green,102; blue,102}]
\tikzstyle{fermionarrow}=[-,postaction=decorate, decoration={markings,mark=at position .60 with {\arrow{stealth[round]}}}]
\tikzstyle{arrow}=[{-{Classical TikZ Rightarrow[length=2mm,width=1.5mm]}}, draw={rgb,255: red,61; green,171; blue,83}, line width=1pt, preaction={{draw=white,line width=2pt}}, line cap=rect]
\tikzstyle{brace}=[-,draw={rgb,255: red,61; green,171; blue,83}, line width=1pt, decorate, decoration={brace,amplitude=5pt}]

\def\pysecdec{py\textsc{SecDec}\xspace}
\def\lam{\lambda}
\def\lams{\lambda(\mathbf{s})}
\def\lamsx{\lambda(\mathbf{x};\mathbf{s})}

%% file: figs/triangle_fd.tex
\coordinate (x1) at (1.1340, 1.4999) ;
\coordinate (x2) at (2.8660, 1.5000) ;
\coordinate (x3) at (2,3) ;
\node (p1) at (0.2681, 0.9998) {};
\node (p2) at (3.7320, 1.0000) {};
\node (p3) at (2,3.8) {};
\node (s) at (2,4) {$s$};

\draw[color=Black] (x1) -- (p1);
\draw[ultra thick,color=Green] (x3) -- (p3);
\draw[color=Black] (x2) -- (p2);

\draw[ultra thick,color=Purple] (x1) -- (x2);
\draw[ultra thick,color=Purple] (x2) -- (x3);
\draw[ultra thick,color=Purple] (x3) -- (x1);

\path (x1) -- (x2) node [midway,yshift=-10pt,color=Black] {$x_2$};
\path (x2) -- (x3) node [midway,xshift=10pt,yshift=5pt,color=Black] {$x_3$};
\path (x3) -- (x1) node [midway,xshift=-10pt,yshift=5pt,color=Black] {$x_1$};

\draw[fill,thick,color=Blue] (x1) circle (1pt);
\draw[fill,thick,color=Blue] (x2) circle (1pt);
\draw[fill,thick,color=Blue] (x3) circle (1pt);

%% file: figs/triangle_np.tex
\begin{tikzpicture}[scale=1.5]
  \draw[very thin,color=gray] (-0.1,-0.1) grid (3.2,3.2);

  \draw[->] (-0.2,0) -- (3.2,0) node[right,xshift=5pt] {$x_1$};
  \draw[->] (0,-0.2) -- (0,3.2) node[above,yshift=5pt] {$x_2$};

  \draw[fill=gray,opacity=0.2]  (0,0) -- (0,2) -- (2,0) -- cycle;

  \draw[ultra thick,color=Black] (0,0) -- (2,0);
  \draw[ultra thick,color=Black] (0,0) -- (0,2);
  \draw[ultra thick,color=Black] (2,0) -- (0,2);

  \draw[->,thick] (1,0) -- (1,0.5);
  \draw[->,thick] (0,1) -- (0.5,1);
  \draw[->,thick] (1,1) -- (0.5,0.5);

  \draw[fill,thick,color=Red] (0,0) circle (2pt);
  \draw[fill,thick,color=Green] (2,0) circle (2pt);
  \draw[fill,thick,color=Yellow] (0,2) circle (2pt);

  \node at (1,-0.3) {$\mathbf{n}_2$};
  \node at (-0.3,1) {$\mathbf{n}_3$};
  \node at (1.3,1.3) {$\mathbf{n}_1$};

  \node at (0,-0.3) {$\mathbf{v}_1$};
  \node at (-0.3,2) {$\mathbf{v}_2$};
  \node at (2,-0.3) {$\mathbf{v}_3$};

\end{tikzpicture}

%% file: figs/triangle_nf.tex
\begin{tikzpicture}[scale=1.13,domain=0:4]
  \draw[very thin,color=gray,step=2.0] (-0.1,-0.1) grid (4.2,4.2);

  \draw[->] (-0.2,0) -- (4.2,0) node[right,xshift=5pt] {$x_1$};
  \draw[->] (0,-0.2) -- (0,4.2) node[above,yshift=5pt] {$x_2$};

  \draw[fill=Yellow,opacity=0.6]  (0,2) -- (2,2) -- (4,4) -- (0,4) -- cycle;
  \draw[fill=Green,opacity=0.6]  (2,2) -- (2,0) -- (4,0) -- (4,4) -- cycle;
  \draw[fill=Red,opacity=0.6]  (0,0) -- (2,0) -- (2,2) -- (0,2) -- cycle;

  \draw[->,thick] (2,2) -- (1,2);
  \draw[->,thick] (2,2) -- (2,1);
  \draw[->,thick] (2,2) -- (2.5,2.5);

  \draw[fill,thick,color=Black] (2,2) circle (1pt);

  \node at (2.5,1) {$-\mathbf{n}_2$};
  \node at (1,2.3) {$-\mathbf{n}_3$};
  \node at (3,2.65) {$-\mathbf{n}_1$};

  \node at (2,-0.3) {$1$};
  \node at (4,-0.3) {$\infty$};
  \node at (-0.3,2) {$1$};
  \node at (-0.3,4) {$\infty$};

  \node at (1.1,0.9) {$\mathbf{I}$};
  \node at (1.1,3.2) {$\mathbf{II}$};
   \node at (3.2,0.9) {$\mathbf{III}$};

  \draw[color=Blue,thick,smooth,samples=200,domain=0.4:3.5]   plot (\x,{-2+4*sqrt(abs(\x/2))-\x})    node[below left,xshift=9pt,yshift=0pt] {\small $\tilde{\mathcal{F}}(\mathbf{x};4)=0$};
  
  \draw[color=Blue,thick,smooth,samples=200,domain=0.15:3.8]   plot (\x,{-2+6*sqrt(abs(\x/2))-\x})    node[below,xshift=-25pt,yshift=-15pt] {\small $\tilde{\mathcal{F}}(\mathbf{x};9)=0$};

\end{tikzpicture}

%% file: figs/bubble.tex
\coordinate (x1) at (1, 2) ;
\coordinate (x2) at (3, 2) ;
\node (s) at (-0.2, 2) {$s$};
\node (p1) at (0, 2) {};
\node (p2) at (4, 2) {};
\draw[ultra thick,color=ForestGreen] (x1) -- (p1);
\draw[ultra thick,color=ForestGreen] (x2) -- (p2);

\draw[ultra thick,color=Purple] (x2) arc (0:180:1) node [midway,yshift=-6.5pt,color=Black] {$m$};
\draw[ultra thick,color=Purple] (x2) arc (0:-180:1) node [midway,yshift=+6.5pt,color=Black] {$m$};

\draw[fill,thick,color=Blue] (x1) circle (1pt);
\draw[fill,thick,color=Blue] (x2) circle (1pt);

%% file: figs/box.tex
  \begin{tikzpicture}[scale=1.3]
	\begin{pgfonlayer}{nodelayer}
		\node [style=dot] (0) at (-1, -0.75) {};
		\node [style=dot] (1) at (-1, 0.75) {};
		\node [style=dot] (2) at (0, -0.75) {};
		\node [style=dot] (4) at (1, -0.75) {};
		\node [style=dot] (5) at (1, 0.75) {};
		\node [style=none] (6) at (-1.75, -0.75) {};
		\node [style=none] (7) at (-1.75, 0.75) {};
		\node [style=none] (8) at (1.75, -0.75) {};
		\node [style=none] (9) at (1.75, 0.75) {};
		\node [style=none] (10) at (2, -0.75) {$p_4$};
		\node [style=none] (11) at (1.25, 0) {$m$};
		\node [style=none] (12) at (-1.25, 0) {$m$};
		\node [style=none] (13) at (0, 0.5) {$m$};
		\node [style=none] (14) at (-0.5, -0.5) {$m$};
		\node [style=none] (16) at (0.5, -0.5) {$m$};
		\node [style=none] (17) at (-2, 0.75) {$p_1$};
		\node [style=none] (18) at (-2, -0.75) {$p_2$};
		\node [style=none] (19) at (2, 0.75) {$p_3$};
	\end{pgfonlayer}
	\begin{pgfonlayer}{edgelayer}
		\draw [style=massive edge] (0) to (2);
		\draw [style=massive edge] (1) to (0);
		\draw [style=edge] (5) to (9.center);
		\draw [style=massive2 edge] (8.center) to (4);
		\draw [style=edge] (0) to (6.center);
		\draw [style=edge] (1) to (7.center);
		\draw [style=edge] (5) to (2);
		\draw [style=massive edge] (5) to (4);
		\draw [style=massive edge] (5) to (1);
		\draw [style=massive edge] (4) to (2);
	\end{pgfonlayer}
\end{tikzpicture}

%% file: tables/thresholds.tex
\begin{tabular}{c|c|c}
$s_{threshold}^{(k)}$ & numerical & sector $(k)$ \\ 
\hline
\hline
4 & 4 & 2,3,5,6,7,13,14,17,18 \\
$5/2+\sqrt{6}$ & $\simeq 4.95$ & 8,11 \\
5 & 5 & 10 \\
6 & 6 & 21,22 \\
$2/23 (37 + \sqrt{2001})$ & $\simeq7.1$ & 15,19 \\
8 & 8 & 1 \\
9 & 9 & 4,12,16 \\
28/3 & $\simeq 9.3$ & 9 \\
12 & 12 & 20 \\
\hline
\end{tabular}